\documentclass[12pt]{article}   
\usepackage{epsfig}
\usepackage[utf8]{inputenc} 
\usepackage{url}

\usepackage{amssymb}
\usepackage{rotating}
\usepackage{graphicx}
\usepackage{comment}
\usepackage{verbatim}
\usepackage{fancyvrb}
\usepackage{xcolor}
\usepackage{hyperref}
\usepackage{amsmath}

\VerbatimFootnotes

\newcommand{\bm}[1]{\mbox{\boldmath$#1$}}

\begin{document}

\title{Ratio of counts vs ratio of rates \\
  in Poisson processes
}
\author{Giulio D'Agostini \\
Universit\`a ``La Sapienza'' and INFN, Roma, Italia \\
{\small (giulio.dagostini@roma1.infn.it,
 \url{http://www.roma1.infn.it/~dagos})}
}
\date{}
\maketitle
\thispagestyle{empty}

\begin{abstract}
The often debated issue of `ratios of small numbers of events'
is approached from a probabilistic perspective,
making a clear distinction between the predictive problem
(forecasting numbers of events we might count under
well stated assumptions, and therefore of their ratios)
and inferential problem (learning about the
relevant parameters of the related probability distribution,
in the light of the observed number of events).
The quantities of interests and their relations are
visualized in a graphical model (`Bayesian network'),
very useful to understand how to approach the problem
following the rules of probability theory.
In this paper, written with didactic intent, we
discuss in detail the basic ideas, however giving
some hints of how real life complications,
like (uncertain) efficiencies and possible
background and systematics, can be included in the analysis,
as well as the possibility that the ratio of rates might depend
on some physical quantity. 
The simple models considered in this paper allow to obtain, 
under reasonable assumptions, closed expressions for the
rates and their ratios.
Monte Carlo methods are also used, both to cross check the exact results
and to evaluate by sampling the ratios of counts in the cases
in which large number approximation does not hold.
In particular it is shown how to make approximate inferences
using a Markov Chain Monte Carlo using JAGS/rjags.
Some examples of R and JAGS code are provided.
\end{abstract}

\section{Introduction}
Many measurements in Physics are based
on counting events belonging to a well defined `class'.
They could be the number
of electric pulses, registered within a given time interval,
exceeding a properly set threshold,
as in a Geiger counter;
or the number of events observed, for a given integrated
luminosity, in a region defined by properly
chosen `cuts' in the multi-dimensional space defined
on the basis of geometrical and kinematic variables
of the final state particles, a typical problem in Particle Physics.
However, the aim of physicists is not limited in counting
how many events {\em will} occur in each `class'
satisfying some detector related criteria, but rather
in {\em inferring} the {\em physical quantities} which are
related to them, as the intensity of radioactivity
or the production rate of a given physical final state
resulting from the collision of two particles, to continue
with our examples.
This also implies that the `experimentally defined class'
(`being inside cuts')
is only a proxy for the `physical class' of interest,
that might be a radioactive particle in a given energy range,
or a particular final state resulting from a collision.
 This is in analogy with the case when we are
interested in counting the number of individuals of a population
infected by a specific agent using as a proxy the number
of individuals tagged `positive' by suitable tests, by their nature 
imperfect.\footnote{This problem has been treated
in much detail in Ref.~\cite{sampling},
taking cue from questions related to the Covid-19 pandemic.
}

If we change the conditions of the experiment,
that is, going on with our examples,
we place the Geiger counter in a different place,
or we vary the initial energy of the colliding
particles (or we tag somehow the final state),
we usually register different numbers
of events in our reference class. This could
just be due to statistical fluctuations.
But it could (also) be due to a variation of
the related physical quantity.
It is then crucial, as well understood,
to associate  an uncertainty to the `measured'
variation.

If the observed numbers are `large', things
get rather easy, thanks to the Gaussian approximation
of the probability distributions of interest.
When, instead, the numbers are `small'
the question can be quite troublesome
(see, e.g., Refs.\,\cite{Nelson,JamesRoos,Coakley,Gu,SIMS,Coath}).
For example, Ref.\,\cite{JamesRoos} focus on the
{\em ``errors on ratios of small numbers of events''},
leading the readers astray: we are {\em usually} not interested in
the ratios of `counts', but rather on the ratios of radioactivity
levels or of production rates, and so on.

The aim of this paper is to review these 
questions following consistently  the rules of probability theory.
The initial, crucial point is to make a clear distinction
between the empirical observations
(the numbers of event of a given `experimentally defined class')
and the related physical quantities we are interested to infer,
although in a probabilistic way. We start playing with the Poisson
distribution in Sec.~\ref{sec:predictions},
referring to Appendix A for a reminder of how this distribution
is related
not only to the binomial (as well known),
but also to other important
distributions via the {\em Poisson process}, which has indeed its
roots in the {\em Bernoulli process}.
In Sec.~\ref{sec:inferring_lambda} we show how to
use the Bayes' rule to infer Poisson $\lambda$'s from the observed
number of counts and then how to get the probability distribution
of their ratio $\rho$
making an exact propagation of uncertainties, that is
$f(\lambda_1/\lambda_2)$ from $f(\lambda_1)$
and $f(\lambda_2)$.
Then in Sec.\,\ref{sec:inferring_r} we move to the
inference of {\em intensities of the Poisson processes}
(or `{\em rates}' $r$, in short), related to $\lambda$
by $\lambda\!=\!r\!\cdot\!T$, with $T$ being the `observation time' --
it can be replaced by `integrated luminosity' or other
quantities to which the Poisson parameter $\lambda$ is proportional.
In the same section the `anxiety-inducing'~\cite{PriorAnxiety}
question of the priors, assumed `flat' until
Sec.~\ref{ss:inference_r_flat}, is finally tackled
and the {\em conjugate priors} are introduced,
showing, in particular, how to apply them in sequential
measurements of the {\em same} rate. 
The technical question of getting the
probability distribution of the ratio of rates
is tackled  in  Sec.~\ref{sec:ratio_gamma}.
Again,  closed formulae are `luckily' obtained,
which can be extended to the more general problem
of getting the probability density function (pdf),
and its summaries, of a ratio of Gamma distributed variables.

When the game seems at the end, in Sec.~\ref{sec:inference_rho}
we modify  the `graphical' model (indeed a visualization of the
underlying logical {\em causal model}) and restart the analysis,
this time really {\em inferring directly} $\rho$, as it will be clear.
The implications of the different models and of the
priors appearing in each of them will be analyzed with some care.
Finally, in Sec.~\ref{sec:MCMC}
the same models are analyzed making use
of Markov Chain Monte Carlo (MCMC) methods, exploiting JAGS.
The purpose is twofold. First we want to cross-check the exact results
obtained in the previous section, 
although the latter were limited to uniform
priors of the `top parents' of the causal model.
Second this allows not only to take into account more realistic priors,
but also to enlarge the models including efficiencies
and background, for which examples of graphical model
are  provided. Another interesting question, that is
how to fit the ratio of rates as a function of another physical
question will be also addressed, showing how to modify the
causal  model, but without entering into the details.
The related issue of `combining ratios' is also discussed
and it shows once more the importance of the underlying model.

\section{Predicting numbers of counts, their difference and their ratio}
\label{sec:predictions}
The Poisson distribution hardly needs any introduction,
beside, perhaps, that it can be framed within the
{\em Poisson process},
which has indeed its roots in the {\em Bernoulli process}.
This picture makes the  Poissonian related
to other important distributions,
as reminded in Appendix A, which can be seen as a 
technical preface to the paper.

Using the notation introduced there,
the Poisson probability function is given by\footnote{I try,
  whenever it is possible,
    to stick to the convention
    of  capital letters for the name of a variable
    and small letters for its possible values.
    Exceptions are Greek letters and quantities
    naturally defined by a small letter, like $r$ for a `rate'. 
}
\begin{eqnarray}
  f(x\,|\,\lambda) \equiv P(X\!\!=\!x\,|\,\lambda)
  & = &\frac{\lambda^x}{x!}\cdot e^{-\lambda}
\hspace{1.0 cm}
\left\{ \begin{array}{l}   0 < \lambda < \infty \\
                           x = 0, 1, \ldots,  \infty\\
                           \end{array} \right.\,
\label{eq:poisson_distr}
\end{eqnarray}
and it quantifies how much we believe that $x$ counts will occur,
{\em if we assume} an exact value of the parameter
$\lambda$.\footnote{If, instead, we are uncertain about $\lambda$ 
and  quantify its uncertainty by the probability density function
  $f(\lambda\,|\,I)$, where $I$ stands for our status
  of information about that quantity, the distribution of the counts
  will be given by
  $ f(x\,|\,\lambda,I) = \int_0^\infty\!\!  f(x\,|\,\lambda,I)\cdot
  f(\lambda\,|\,I)\,\mbox{d}\lambda\,.$
  \label{fn:uncertain_lambda}
}
As well known,
expected value and standard deviation of $X$
are $\lambda$ and $\sqrt{\lambda}$. The most probable value
of $X$ (`mode') is equal to the integer just below $\lambda$
(`{\tt floor}$(\lambda)$') in the case $\lambda$ is not integer. 
Otherwise is equal to $\lambda$ itself, and also to $\lambda-1$
(remember that $\lambda$ cannot be null).

If we have two {\em independent} Poisson distributions
characterized
by $\lambda_1$ and $\lambda_2$, i.e. 
\begin{eqnarray*}
  X_1 &\sim& {\cal P}_{\lambda_1} \\
  X_2 &\sim& {\cal P}_{\lambda_2}\,,
\end{eqnarray*}
we `expect' their difference $D\!=\! X_1\!-\!X_2$ `to be'
$(\lambda_1\!-\!\lambda_2) \pm \sqrt{\lambda_1+\lambda_2}$,
as it results from well known theorems of probability
theory\footnote{In brief: the expected value of a linear combination
  is the linear combination of the expected values;
  the variance of a linear combination is the linear combination
  of the variances, with squared coefficients.}
(hereafter, unless indicated otherwise, the notation
`$xxx \pm yyy $' stands for `expected value of the quantity $\pm$
its {\em standard uncertainty}'\,\cite{ISO},
that is the standard deviation of the associated probability distribution).

\begin{figure}[t]
  \begin{center}
    \epsfig{file=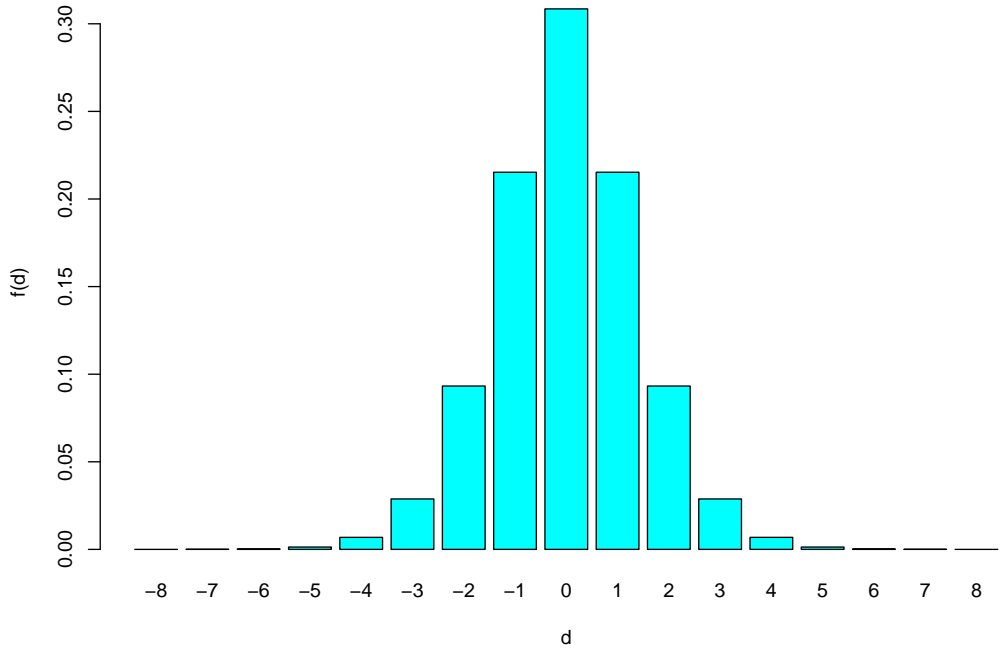,clip=,width=0.9\linewidth}
    \\  \mbox{} \vspace{-1.0cm} \mbox{}
    \end{center}
  \caption{\small \sf Distribution of the difference of counts
    resulting from two Poisson distributions with $\lambda_1=\lambda_2=1$.
  }
    \label{fig:diff_poisson} 
\end{figure}

The probability distribution of $D$ can be obtained
`from the inventory of the values of $X_1$ and $X_2$ that result
in each possible value of $D$', that is
\begin{eqnarray}
P(D\!=\!d\,|\,\lambda_1,\lambda_2) \equiv  f(d\,|\,\lambda_1,\lambda_2)& = &
 \!\! \sum_{\begin{array}{c}x_1,x_2 \\ x_1\!-\!x_2\!=\!d\end{array}}
  \!\! f(x_1\,|\,\lambda_1)\cdot
  f(x_2\,|\,\lambda_2)\,.
  \label{eq:sommatoria}
\end{eqnarray} 
For example, in the case of $\lambda_1=\lambda_2=1$,
the most probable contributions to $D$ are shown
in Tab.~\ref{tab:differenze}.
\begin{table}
  \begin{center}
{\small     
  \begin{tabular}{c|c|cccccc}
    & & \multicolumn{6}{c}{$X_2$} \\
    \hline
    & & 0 & 1 & 2 & 3 & 4 & 5 \\
    \hline
    & 0 & {\bf 0} & -1 & -2 & -3 & -4 & -5  \\
    && {\small[{\bf \em 0.135335}]} &
        {\small[{\em 0.135335}]} &  {\small[{\em 0.067668}]} &
        {\small[{\em 0.022556}]} &  {\small[{\em 0.005639}]} &
        {\small[{\em 0.001128}]}\\
        & 1 & 1 & {\bf 0} & -1 & -2 & -3 & -4  \\
        && {\small[{\em  0.135335}]} & {\small[{\bf\em 0.135335}]} &
        {\small[{\em 0.067668}]} &
       {\small[{\em  0.022556}]} & {\small[{\em  0.005639}]} &
       {\small[{\em 0.001128}]}  \\    
      & 2 & 2 & 1 & {\bf 0} & -1 & -2 & -3  \\
   $X_1$ & &{\small[{\em   0.067668}]} & {\small[{\em  0.067668}]} & {\small[{\bf\em  0.033834}]} & {\small[{\em  0.011278}]} & {\small[{\em  0.002819}]} & {\small[{\em  0.000564}]} \\      
       & 3 & 3 & 2 & 1 & {\bf 0} & -1 & -2  \\
   && {\small[{\em 0.022556}]} & {\small[{\em  0.022556}]} & {\small[{\em  0.011278}]} & {\small[{\bf \em 0.003759}]} & {\small[{\em  0.000940}]} & {\small[{\em  0.000188}]} \\        
       & 4 & 4 & 3 & 2 & 1 & {\bf 0} & -1  \\
   && {\small[{\em 0.005639}]} & {\small[ {\em 0.005639}]} & {\small[{\em  0.002819}]} & {\small[{\em  0.000940}]} & {\small[{\bf\em 0.000235}]} & {\small[{\em  0.000047}]} \\        
       & 5 & 5 & 4 & 3 & 2 & 1 & {\bf 0}    \\
       &&  {\small[{\em 0.001128}]} & {\small[{\em  0.001128}]} & {\small[{\em  0.000564}]} & {\small[{\em  0.000188}]} & {\small[{\em  0.000047}]} & {\small[{\bf\em 0.000009}]} 
  \end{tabular}
}
\\ \mbox{}\vspace{-0.65cm}\mbox{}
  \end{center}
  \caption{\sf \small Table of the most probable differences $D\!=\!X_1\!-\!X_2$ 
    for  $\lambda_1\!=\!\lambda_2\!=\!1$ (probability of each entry %
    in the table
    within square brackets).}
\label{tab:differenze}
\end{table}
For instance, the probability to get $D=0$ sums up to 30.9\%.
The probability decreases
symmetrically for larger absolute values of the difference.
Without entering into the question of getting
a closed form of
$f(d\,|\,\lambda_1,\lambda_2)$,\footnote{Such a distribution is known in
  the literature as Skellam distribution~\cite{Wiki_Skellam} and
  it is available in R~\cite{R} installing the
  homonym  package~\cite{R_Skellam}.
  The distribution of the differences corresponding to the cases of
  Tab.~\ref{tab:differenze} can be easily plotted by the
  following R commands, producing a bar plot similar to that
  of Fig.~\ref{fig:diff_poisson},\\
  \mbox{}\vspace{-0.55cm}\mbox{}
\begin{verbatim}
  library(skellam)
  d = -5:5
  barplot(dskellam(d,1,1), names=d, col='cyan')  
\end{verbatim}
\mbox{}\vspace{-0.5cm}\mbox{}
\label{fn:Skellam}
} 
it can be instructive to implement
Eq.~(\ref{eq:sommatoria}), although in an approximate
and rather inefficient way, in
a few lines of R code~\cite{R}:\footnote{This function,
  hopefully having a didactic value, is not
  optimized at all and it uses the fact that the R function
  {\tt dpois()} returns zero for negative values of the
variable.} 
\begin{verbatim} 
dPoisDiff <- function(d, lambda1, lambda2) {
   xmax = round(max(lambda1,lambda2)) + 20*sqrt(max(lambda1,lambda2))
   sum( dpois((0+d):xmax, lambda1) * dpois(0:(xmax-d), lambda2) )
}
\end{verbatim}
\mbox{}\vspace{-0.4cm}\mbox{}\\
This function is part of the code provided in Appendix B.1,
which produces  the plot of Fig.~\ref{fig:diff_poisson},
evaluating also expected value and standard deviation
(indeed approximated values, being {\tt xmax} not too large).

\begin{figure}
  \begin{center}
    \epsfig{file=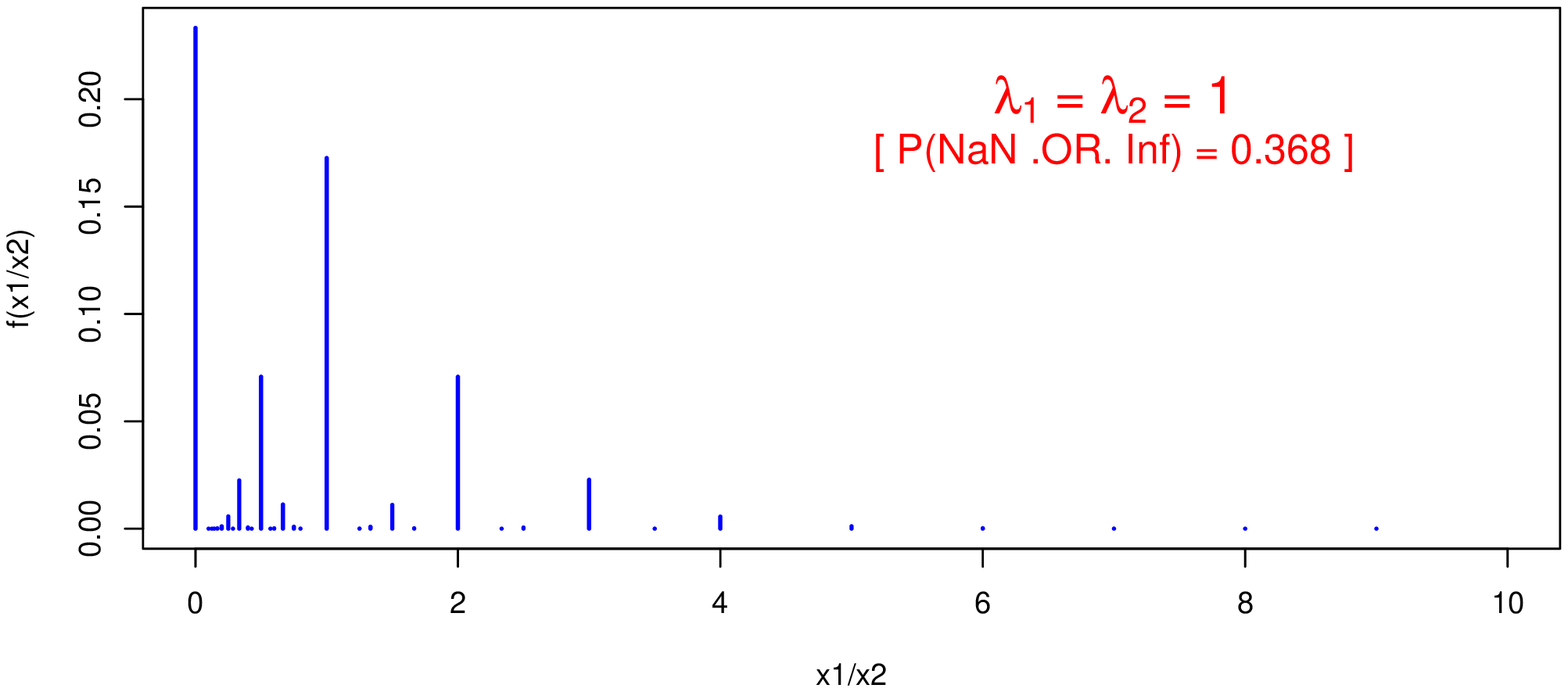,clip=,width=0.95\linewidth} \\
    \epsfig{file=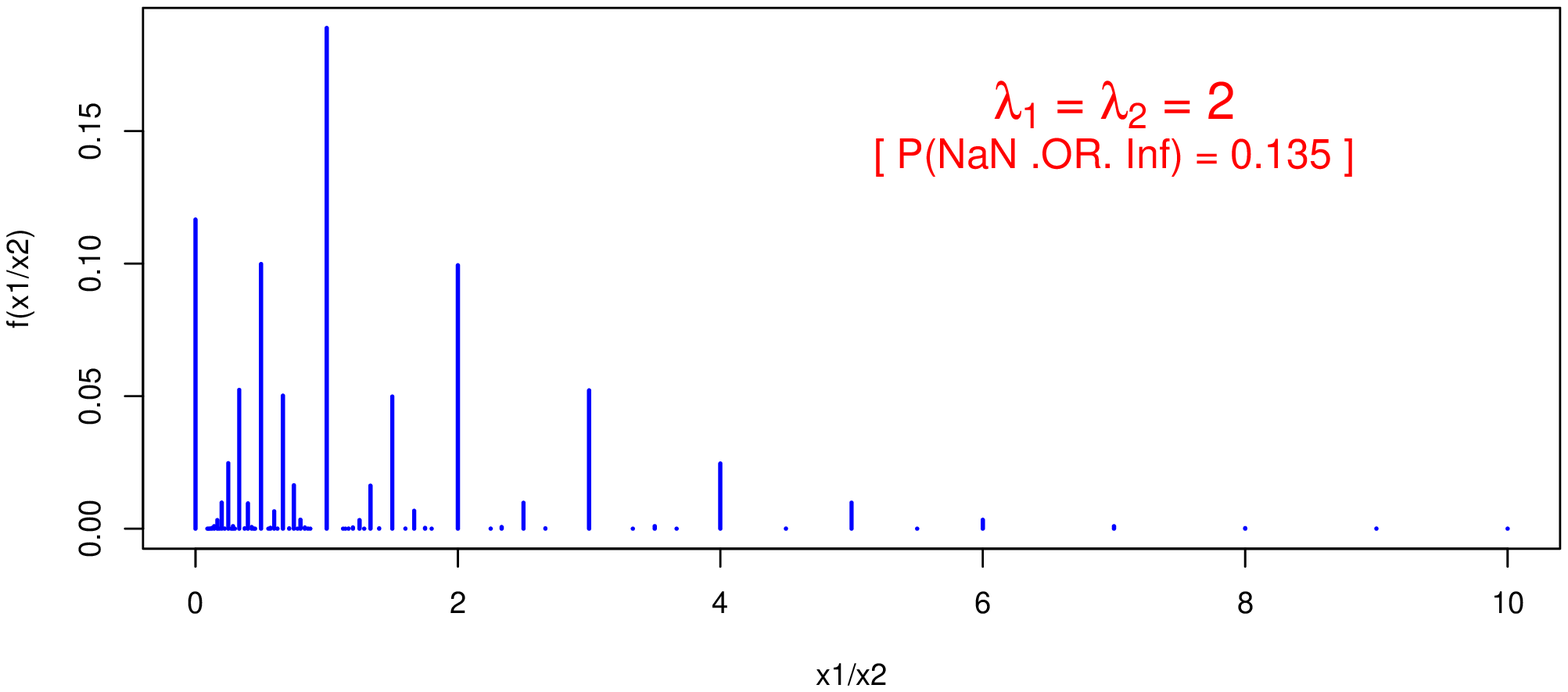,clip=,width=0.95\linewidth} \\
     \epsfig{file=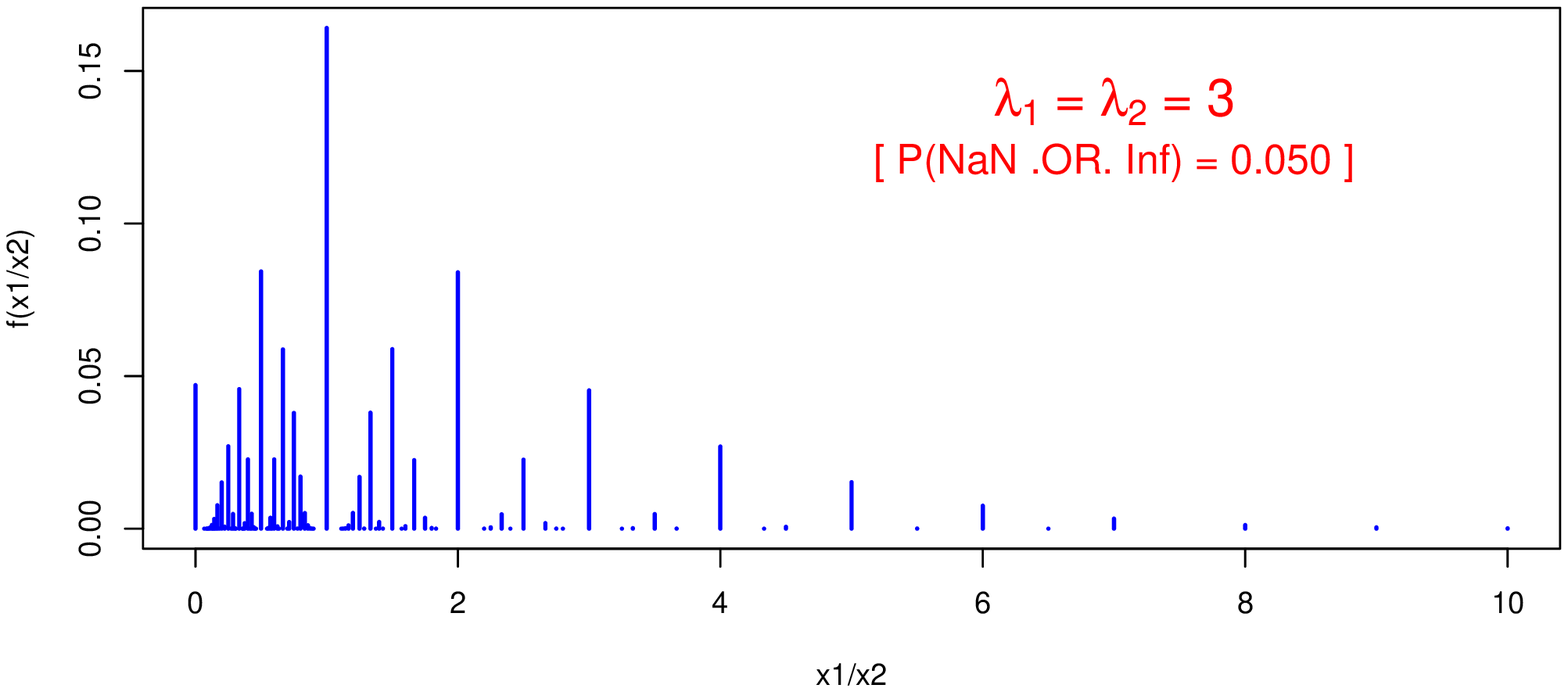,clip=,width=0.95\linewidth}
   \\  \mbox{} \vspace{-1.0cm} \mbox{}
    \end{center}
  \caption{\small \sf Monte Carlo distribution of the ratio of counts
    resulting from two Poisson distributions with $\lambda_1=\lambda_2$.
  }
    \label{fig:rapp_poisson} 
\end{figure}

\begin{figure}
  \begin{center}
    \epsfig{file=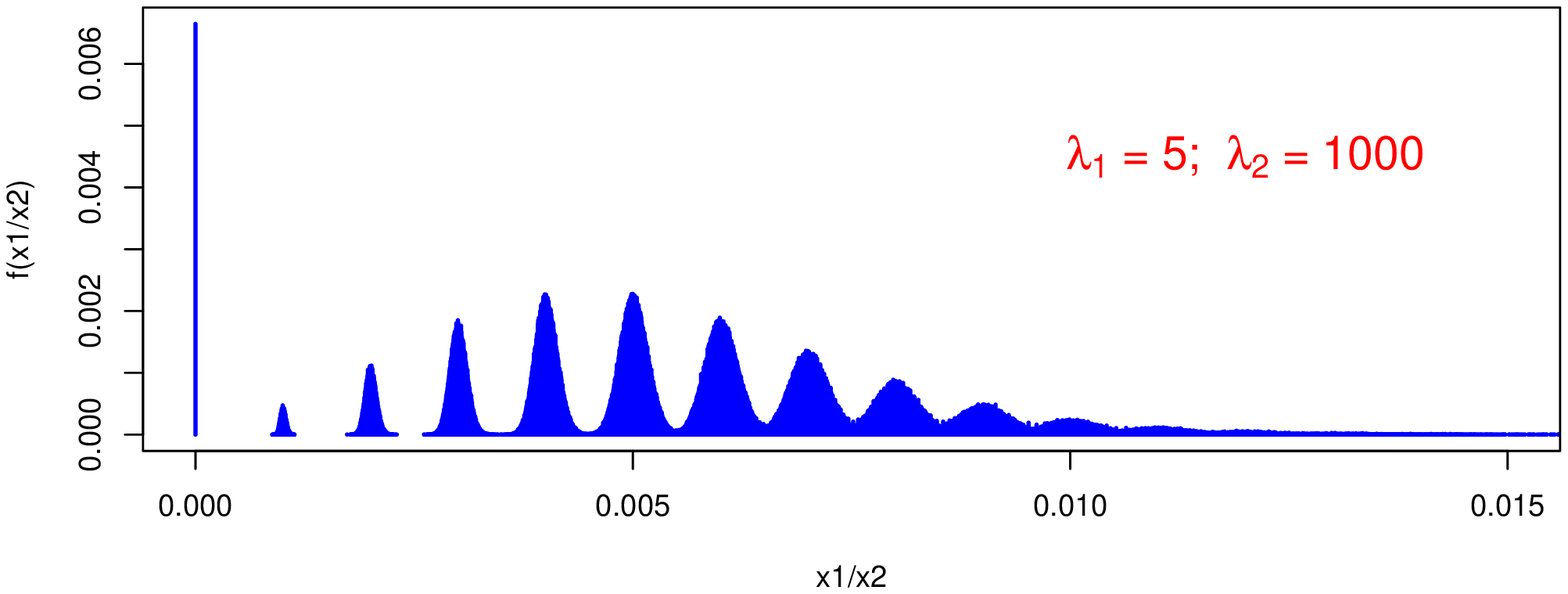,clip=,width=0.82\linewidth} \\
    \epsfig{file=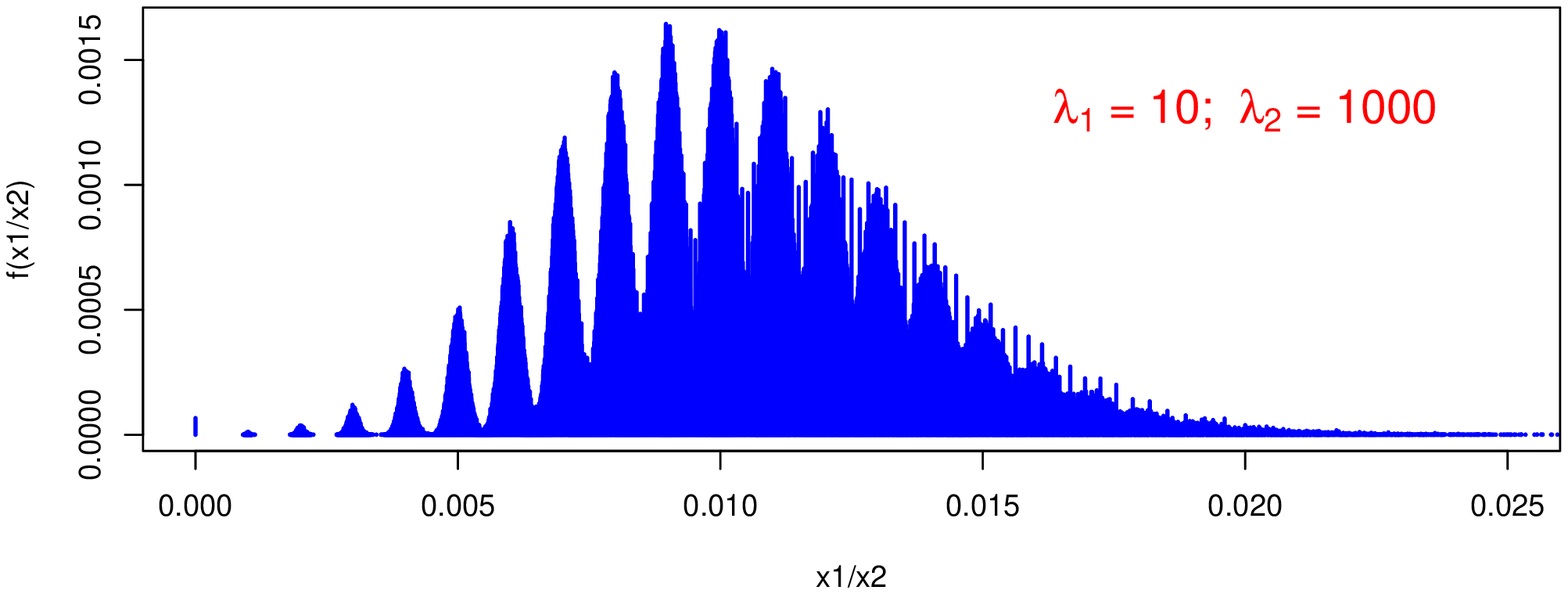,clip=,width=0.82\linewidth} \\
    \epsfig{file=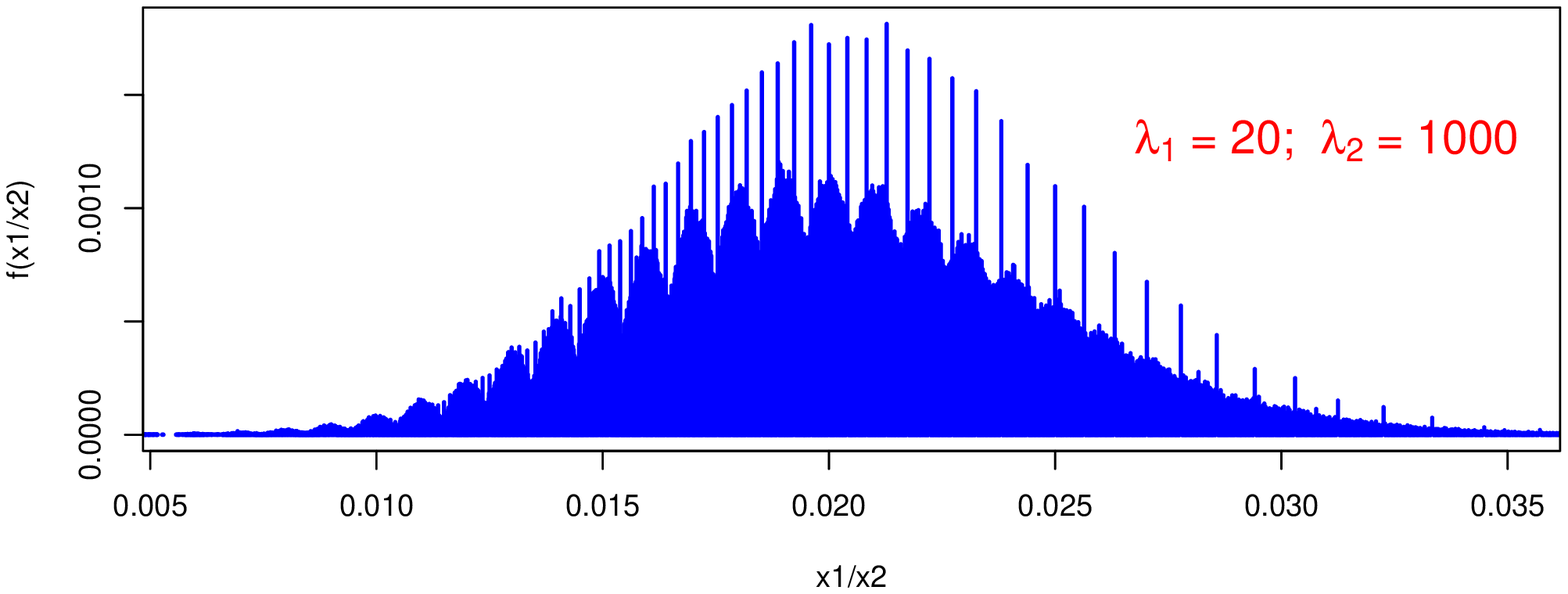,clip=,width=0.82\linewidth}\\
    \epsfig{file=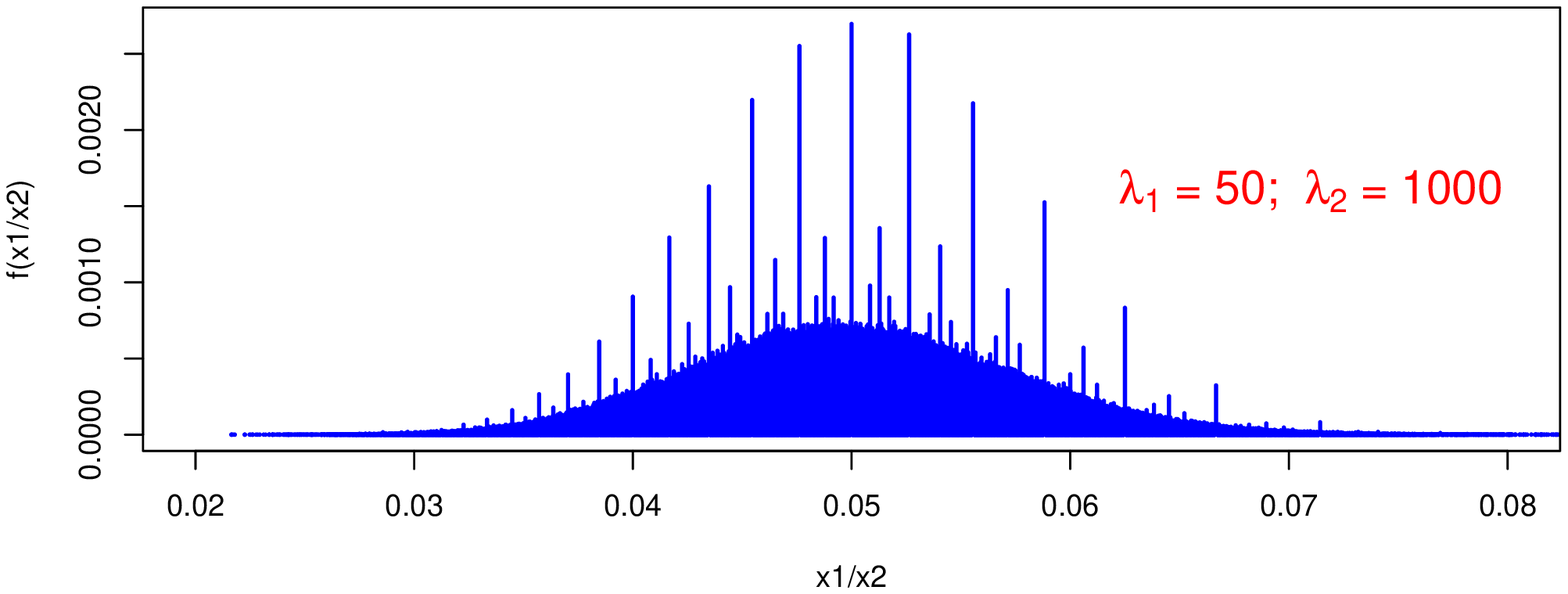,clip=,width=0.82\linewidth}
   \\  \mbox{} \vspace{-1.3cm} \mbox{}
    \end{center}
  \caption{\small \sf Monte Carlo distribution of the ratio of counts
    resulting from two Poisson distributions with $\lambda_1=5,10,20,50$
    and $\lambda_2=1000$.
  }
    \label{fig:rapp_poisson_2} 
\end{figure}

Moving to the ratio of counts, numerical problems might arise, 
as shown in Tab.~\ref{tab:rapporti}, analogue of Tab.~\ref{tab:differenze}.
\begin{table}[!h]
  \mbox{}
  \begin{center}
    {\small 
  \begin{tabular}{c|c|cccccc}
    & & \multicolumn{6}{c}{$X_2$} \\
    \hline
    & & 0 & 1 & 2 & 3 & 4 & 5 \\
    \hline
    & 0 & {\color{red} NaN} & 0 & 0 & 0 & 0 & 0  \\
    && {\small[  {\color{red}\em 0.135335}]} &
        {\small[{\em 0.135335}]} &  {\small[{\em 0.067668}]} &
        {\small[{\em 0.022556}]} &  {\small[{\em 0.005639}]} &
        {\small[{\em 0.001128}]}\\
        & 1 & {\color{red} Inf}  & 1 & 1/2 & 1/3 & 1/4 & 1/5  \\
        && {\small[{\em  \color{red} 0.135335}]} & {\small[{\em 0.135335}]} &
        {\small[{\em 0.067668}]} &
       {\small[{\em  0.022556}]} & {\small[{\em  0.005639}]} &
       {\small[{\em 0.001128}]}  \\    
      & 2 &  {\color{red} Inf} & 2 & 1 & 2/3 & 1/2 & 2/5  \\
   $X_1$ & &{\small[{\em  \color{red}  0.067668}]} & {\small[{\em  0.067668}]} & {\small[{\em  0.033834}]} & {\small[{\em  0.011278}]} & {\small[{\em  0.002819}]} & {\small[{\em  0.000564}]} \\      
       & 3 &  {\color{red} Inf} & 3 & 3/2 & 1 & 3/4 & 3/5  \\
   && {\small[{\em  \color{red} 0.022556}]} & {\small[{\em  0.022556}]} & {\small[{\em  0.011278}]} & {\small[{\em 0.003759}]} & {\small[{\em  0.000940}]} & {\small[{\em  0.000188}]} \\        
       & 4 &  {\color{red} Inf} & 4 & 2 & 4/3 & 1 & 4/5  \\
   && {\small[{\em  \color{red} 0.005639}]} & {\small[ {\em 0.005639}]} & {\small[{\em  0.002819}]} & {\small[{\em  0.000940}]} & {\small[{\em 0.000235}]} & {\small[{\em  0.000047}]} \\        
       & 5 &  {\color{red} Inf} & 5 & 5/2 & 5/3 & 5/4 & 1    \\
       &&  {\small[{\em  \color{red} 0.001128}]} & {\small[{\em  0.001128}]} & {\small[{\em  0.000564}]} & {\small[{\em  0.000188}]} & {\small[{\em  0.000047}]} & {\small[{\em 0.000009}]} 
  \end{tabular}
  }
  \end{center}
  \caption{\sf \small Table of the most probable ratios $X_1/X_2$ 
    for  $\lambda_1\!=\!\lambda_2\!=\!1$.
    `{\tt NaN}' and `{\tt Inf}' are the R symbols
    for undefined (`not a number') and infinity, resulting from
  a vanishing denominator.}
\label{tab:rapporti}
\end{table}
In fact for rather small values of $\lambda_2$
there is high chance
(exactly the probability of getting $X_2=0$) 
that the ratio results in
an undefined form or an infinite, reported 
in the table using the R symbols {\tt NaN} (`not a number')
and {\tt Inf}, respectively.
As we can see,
we have now quite a variety of possibilities
and the probability distribution of the ratios
is rather irregular. For this reason, in this case
we evaluate it by Monte Carlo methods using R.\footnote{The
  core of the R code is given,
  for the case of
  $\lambda_1=\lambda_2=1$, by\\
  \mbox{}\vspace{-0.7cm}\mbox{}
\begin{verbatim}
lambda1 = lambda2 = 1; n = 10^6
x1 = rpois(n,lambda1)
x2 = rpois(n,lambda2)
rx = x1/x2
rx = rx[!is.nan(rx) & (rx != Inf)]
barplot(table(rx)/n, col='cyan', xlab='x1/x2', ylab='f(x1/x2)')
\end{verbatim}
\mbox{}\vspace{-0.6cm}\mbox{}
}
Figure ~\ref{fig:rapp_poisson}
 shows the distributions
of the ratio for $\lambda_1\!=\!\lambda_2 = 1,2,3$.
The figure also reports the probability to get an infinite
or an undefined expression, equal to $P(X_2\!=\!0\,|\,\lambda_i)$. 
When $\lambda_2$ is very large the probability to get $X_2=0$,
and therefore of $X_1/X_2$ being equal to {\tt Inf} or {\tt NaN}, vanishes.
But the distribution of the ratio remains quite `irregular', if looked
into detail, even for $\lambda_1$ `not so small', as 
shown in Fig.~\ref{fig:rapp_poisson_2} for the cases
of $\lambda_1\!=\!5,10,20,50$ and  $\lambda_2\!=\!1000$.

However we should not be worried about
this kind of distributions, which are not more than entertaining
curiosities, as long as physics questions are concerned. Why should
we be interested in the ratio of counts that we might observe
for different $\lambda$'s? If we want to get an idea of how
much the counts could differ, we can just use the probability
distribution of their possible differences, which has a regular behavior,
without divergences or undefined forms.

The deep reason
for speculating about ``{\em ratios of small numbers of events}''
and their ``{\em errors}''\cite{JamesRoos} is due
to a curious ideology at the basis of a school of Statistics
which limits the applications of Probability Theory.
Indeed, we, as physicists,
are often interested in the {\em ratio of the rates of 
Poisson processes}, that is in $\rho=r_1/r_2$, being
this quantity related to some physical quantities having
a theoretical relevance. 
Therefore we aim to learn
which values of $\rho$ are more or less probable in the light
of the experimental observations. Stated in this terms,
we are interested in evaluating `somehow' (not always in closed form)
the probability density function (pdf) $f(\rho\,|\,x_1,T_1,x_2,T_2,I)$,
given the observations of $x_1$ counts during $T_1$ and
of $x_2$ counts  during $T_2$ (and also conditioned
on the  background state of
knowledge, generically indicated by $I$).
But there are statisticians who maintain that we can only talk about
the probability of $X$ counts, assuming $\lambda$, and
not of the probability distribution
of $\lambda$ having observed $X\!=\!x$, and even less
of $\lambda_1/\lambda_2$ (same as $r_1/r_2$, if $T_1\!=\!T_2$)
having observed $x_1$ and $x_2$ counts.\footnote{Ref.~\cite{JamesRoos}
  is a kind of `masterpiece' of the kind of convoluted reasoning involved.
  For example, the paper starts with the following {\em incipit}
  (quote marks original):
  \begin{quote}
    {\sl When the result of the measurement of a physical quantity
      is published as $R=R_0\pm \sigma_0$ without further explanation, it is
      implied that $R$ is a gaussian-distributed measurement with mean $R_0$
      and variance $\sigma_0^2$. This allows one to calculate
      various confidence intervals of given ``probability'',
      i.e., the ``probability'' $P$ that the true value of $R$ is within
    a given interval.}
  \end{quote}
  However, nowhere in the paper is explained why {\em probability}
  is within quote marks. The reason is simply because the authors
  are fully aware that frequentist ideology, to which they overtly
  adhere, refuses to attach the concept of probability to
  {\em true values}, as well as to {\em model parameters}, and so on
  (see e.g. Ref.~\cite{BR}). But authoritative statements of this
  kind might contribute to increase the confusion of
  practitioners~\cite{maxent98}, who then tend to take
  frequentist `confidence levels' as if they were
  probability values~\cite{WavesSigmas}.
  \label{fn:quote_JR}
}

If, instead, we follow an approach closer to that innate
in the human mind, which naturally attaches the concept of probability to
whatever is considered uncertain\,\cite{Hume},
 there is no need to go through contorted reasoning.
For example,
if we observe $X=1$, then we tend to believe that this {\em effect}
has been  {\em caused}
more likely by a value of $\lambda$ around 1 rather than around 10 or 20, or larger,
although
we cannot rule out with {\em certainty} such values.
Similarly, sticking to the observation of $X=1$,
we tend to believe to $\lambda\approx 1$ much more
than to $\lambda \approx 10^{-2}$, or smaller.
In particular, {\em $\lambda=0$ is definitely ruled out},
because it could only yield 0 counts (this is just a limit case,
since the Poisson $\lambda$ is defined positive).

It is then clear that, as far as the ratio
of $\lambda_1/\lambda_2$ is concerned, there are
no divergences, no matter how small the numbers
of counts might be, obviously with two exceptions.
The first is when $X_2$ is observed to be exactly 0 (but in this case
  we could turn our interest in $\lambda_2/\lambda_1$,
  assuming $X_1>0$). The second is
when $X_2$ is not zero, but there could be some background,
such that $r_2=0$ is not excluded with certainty\,\cite{BR,conPia}.
The effect of possible background is not going to be treated
in detail in this paper,
and only some hints on how to include it into the model
will be given.
  
\section{Inferring Poisson \texorpdfstring{$\lambda$'s}{lambdas} and
  then {\em deducing} their ratio}\label{sec:inferring_lambda}
We are now faced to the inference of $r_1$ and $r_2$ from the observed
numbers of counts and the observation times $T_1$ and $T_2$.
For simplicity we start assuming
$T_1=T_2$, so that we can focus on $\lambda_1$, $\lambda_2$
and their ratio. The extension to the general case will be
straightforward, as we shall see from Sec.~\ref{sec:inferring_r} on.

\subsection{Inference of \texorpdfstring{$\lambda$}{lambda} given
  \texorpdfstring{$x$}{x}, assuming
  \texorpdfstring{$X\sim {\cal P}_\lambda$}{X Pois(lambda)}}
\label{ss:inference_lambda}
\begin{figure}[t]
  \begin{center}
    \epsfig{file=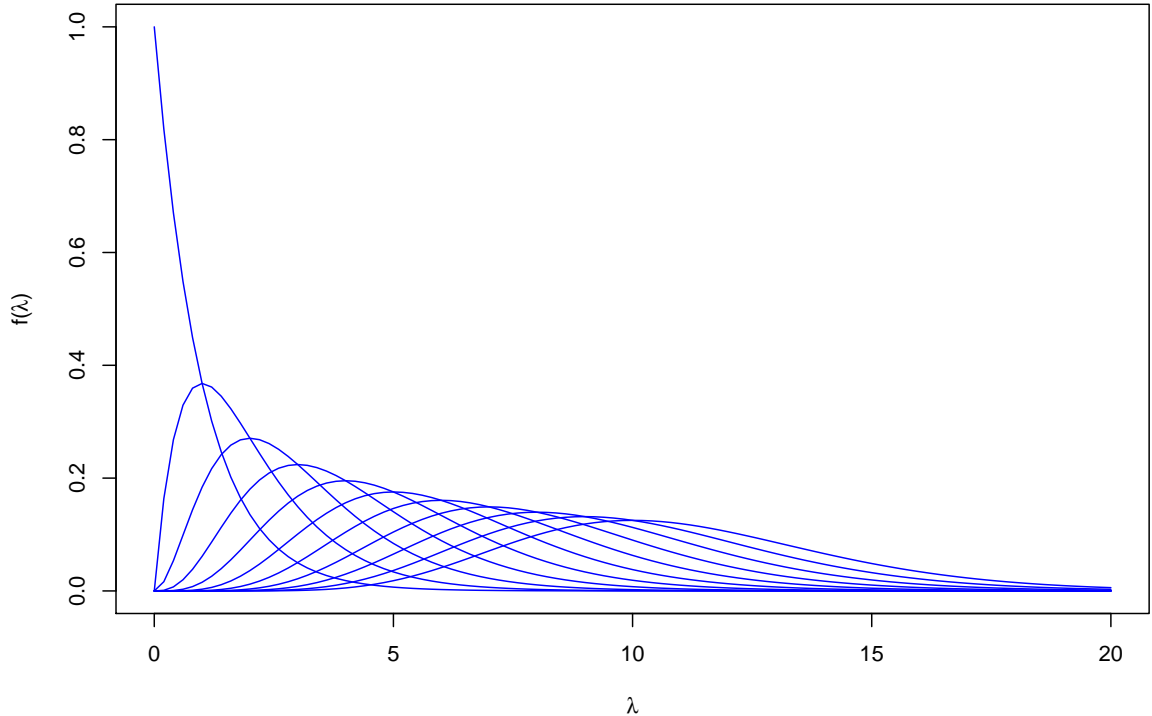,clip=,width=\linewidth}
   \\  \mbox{} \vspace{-1.0cm} \mbox{}
    \end{center}
  \caption{\small \sf Inferred $f(\lambda\,|\,x)$, using a flat prior,
    for $x=0,1,\ldots,10$.
  }
    \label{fig:esempi_pdf_lambda} 
\end{figure}

The probability density function of $\lambda$ is evaluated
from the so called {\em Bayes' rule}:
\begin{eqnarray}
  f(\lambda\,|\,x) &\propto & f(x\,|\,\lambda)\cdot f_0(\lambda)\\
   &\propto & \frac{\lambda^{x}\cdot e^{-\lambda}}{x!}\cdot f_0(\lambda) \,,
\end{eqnarray}
where $f_0(\lambda)$ is the so called
`prior'.\footnote{This name is somehow unfortunate, because
  it might induce people think to time order, as discussed
  in Ref.~\cite{sampling}, in which it is shown how,
  instead, the `prior' can be applied in a second step,
in particular by someone else, if a `flat prior' used.}
 Assuming for the moment 
 a `flat' prior, that is $f_0(\lambda)=k$,
 and neglecting 
 all factors non depending on $\lambda$, we get
 \begin{eqnarray}
 f(\lambda\,|\,x) & \propto & \lambda^{x}\cdot e^{-\lambda} \\
 & \propto & \lambda^{(x+1)-1}\cdot e^{-\lambda}\,,
\end{eqnarray}
 in which we recognize a Gamma pdf with $\alpha=x+1$
 and $\beta=1$ (see Appendix A -- for a detailed derivation
 see e.g. Ref.~\cite{BR}), and therefore
  \begin{eqnarray}
    f(\lambda\,|\,x) & = & \frac{1}{\Gamma(x+1)}\cdot
    \lambda^{(x+1)-1}\cdot e^{-\lambda} \\
    &=& \frac{\lambda^x\cdot e^{-\lambda}}{x!}\,. \label{eq:pdf_lambda_given_x}
\end{eqnarray}
 Expected value, standard deviation and mode are
 $x+1$, $\sqrt{x+1}$ and $x$, respectively.
 The advantage of having expressed the distribution of $\lambda$
 in terms of a Gamma is that we can use the 
 probability distributions made available from programming languages, e.g.
 in R, which usually include also useful random generators
 (e.g. {\tt rgamma()} in R). For example,
 making use of the R function {\tt dgamma()} 
 we can draw
 Fig.~\ref{fig:esempi_pdf_lambda},
 which shows
 $f(\lambda\,|\,x)$, for $x=1,2,\ldots,10$, with the following
 few lines of code:
\begin{verbatim}
for (x.o in 0:10) {
  curve(dgamma(x,x.o+1,1),xlim=c(0,20),ylim=c(0,1),col='blue',add=x.o>0,
        xlab=expression(lambda),ylab=expression(paste('f(',lambda,')')))  
}
\end{verbatim}

\subsection{Distribution of the ratio of Poisson
  \texorpdfstring{$\lambda$'s}{lambdas} by sampling}
\begin{figure}
  \begin{center}
    \epsfig{file=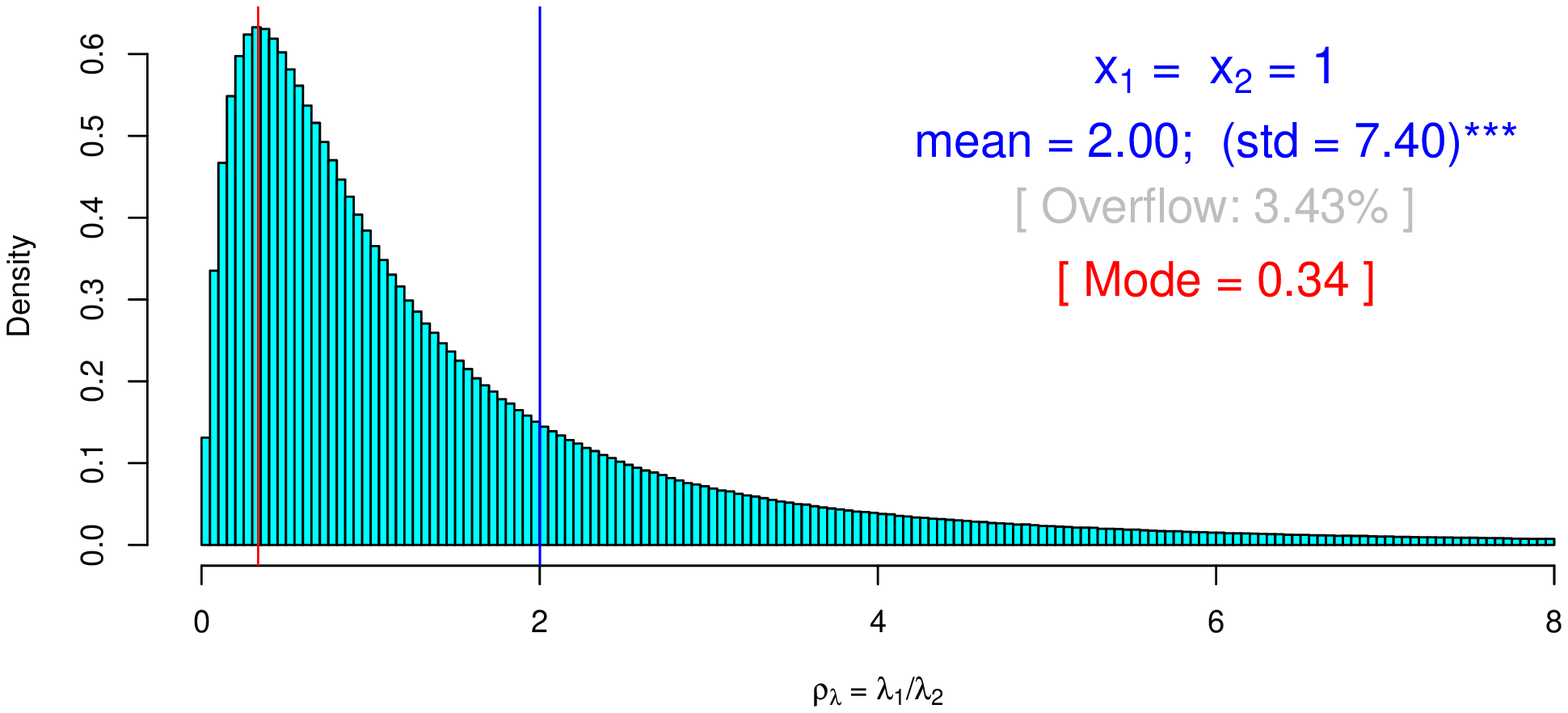,clip=,width=0.97\linewidth}
    \epsfig{file=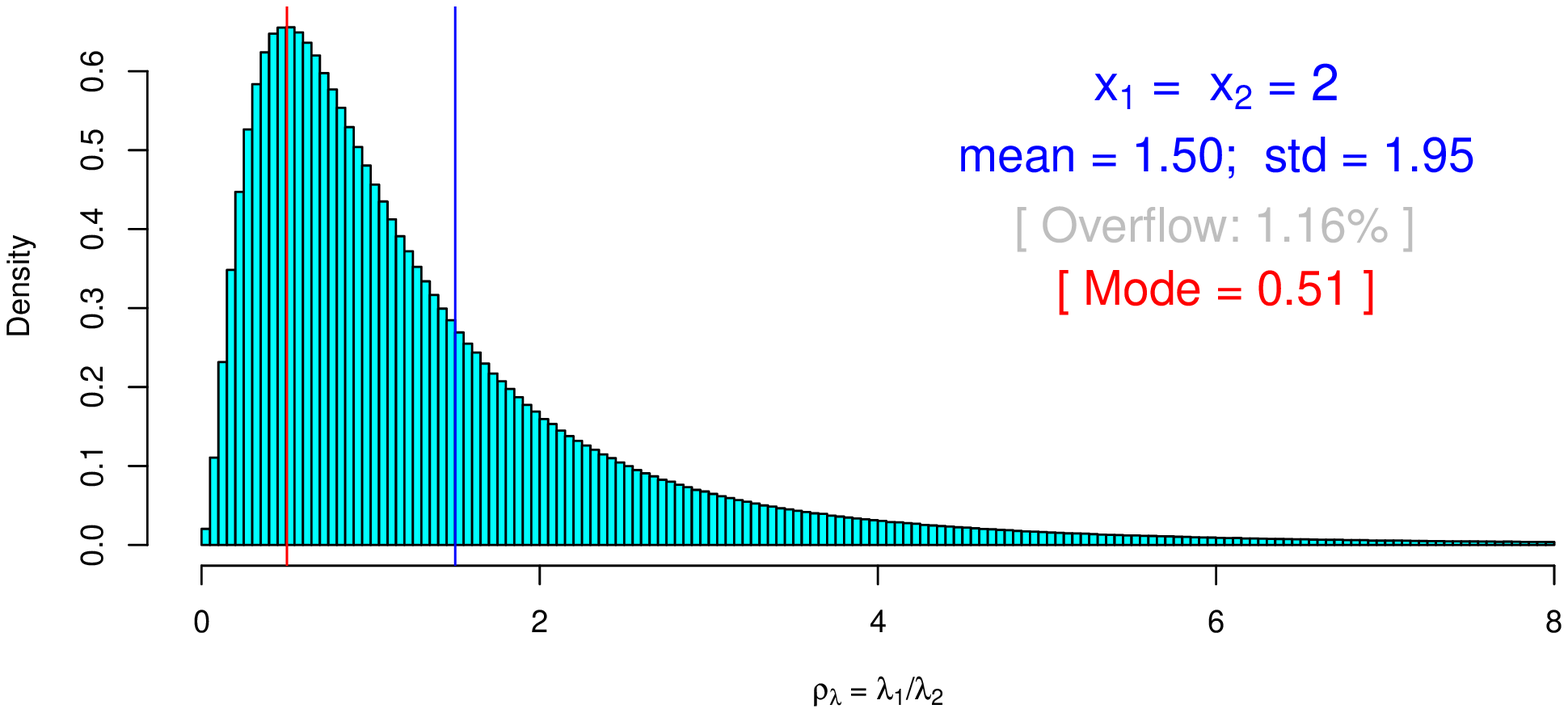,clip=,width=0.97\linewidth}
     \epsfig{file=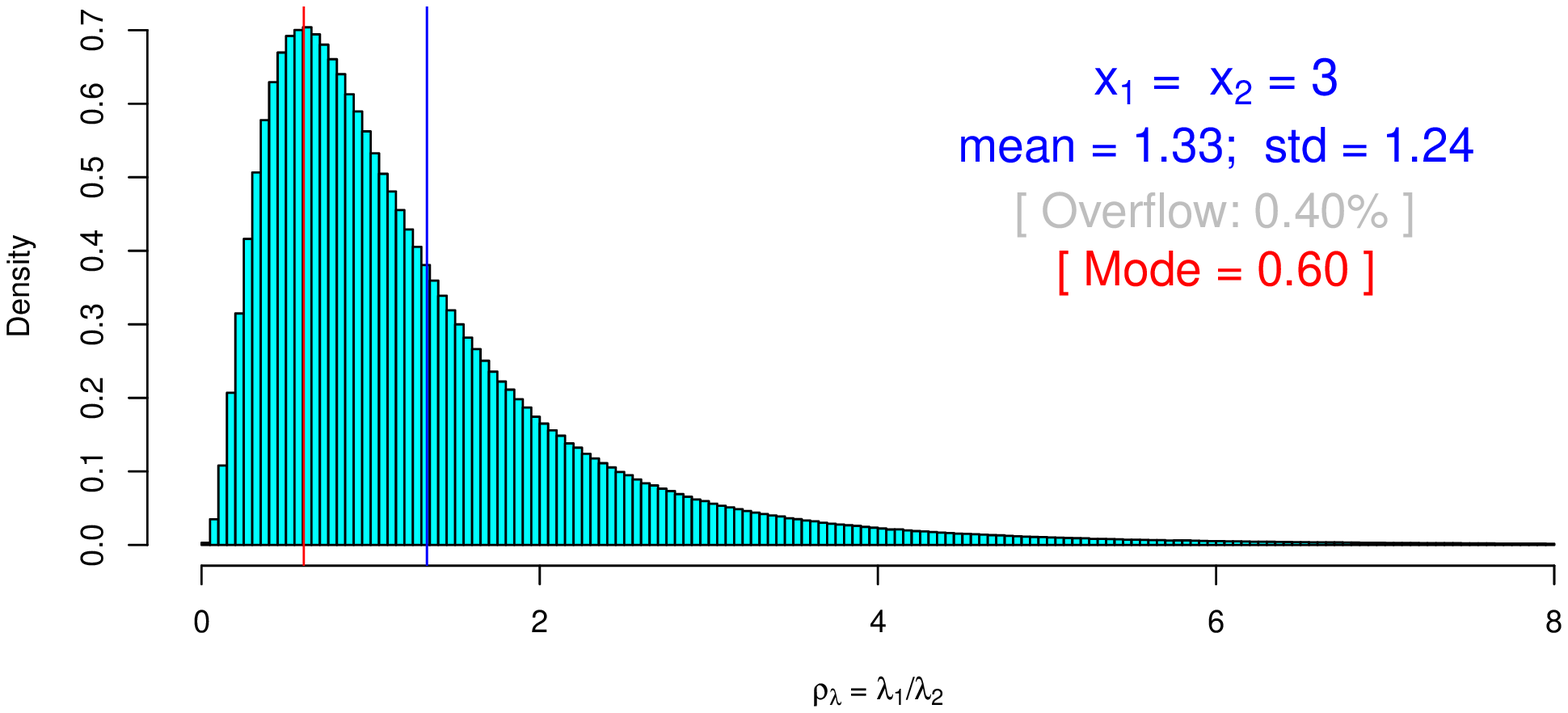,clip=,width=0.97\linewidth}
   \\  \mbox{} \vspace{-1.2cm} \mbox{}
    \end{center}
  \caption{\small \sf Estimate by sampling ($n\!=\!10^7$) of
    $f(\rho_\lambda\!=\!\lambda_1/\lambda_2)$ for some `observed' counts.
    (For the (non) meaning of standard deviation for $x_1=x_2=1$
    see Sec.~\ref{ss:pdf_rho_closed}.)
  }
    \label{fig:pdf_lambda_x1_x2_MC_low} 
\end{figure}

Once we have learned that the pdf of  $\lambda$,
in the light of the observation of $x$ count and assuming
a flat prior, is a Gamma distribution, the easiest way to
evaluate the distribution of $\lambda_1/\lambda_2$, for  
$x_2>0$, is by sampling. For example, using the following
lines of R code,
\begin{verbatim}
x1 = 1
x2 = 1
lambda1 = rgamma(n, x1+1, 1)
lambda2 = rgamma(n, x2+1, 1)
rho = lambda1/lambda2
\end{verbatim}
Then, varying {\tt x1} and {\tt x2}
we can get the plots of Figs.~\ref{fig:pdf_lambda_x1_x2_MC_low}
and \ref{fig:pdf_lambda_x1_x2_MC_high}.\footnote{The
  complete script is provided in Appendix B.2.} 
\begin{figure}
  \begin{center}
    \epsfig{file=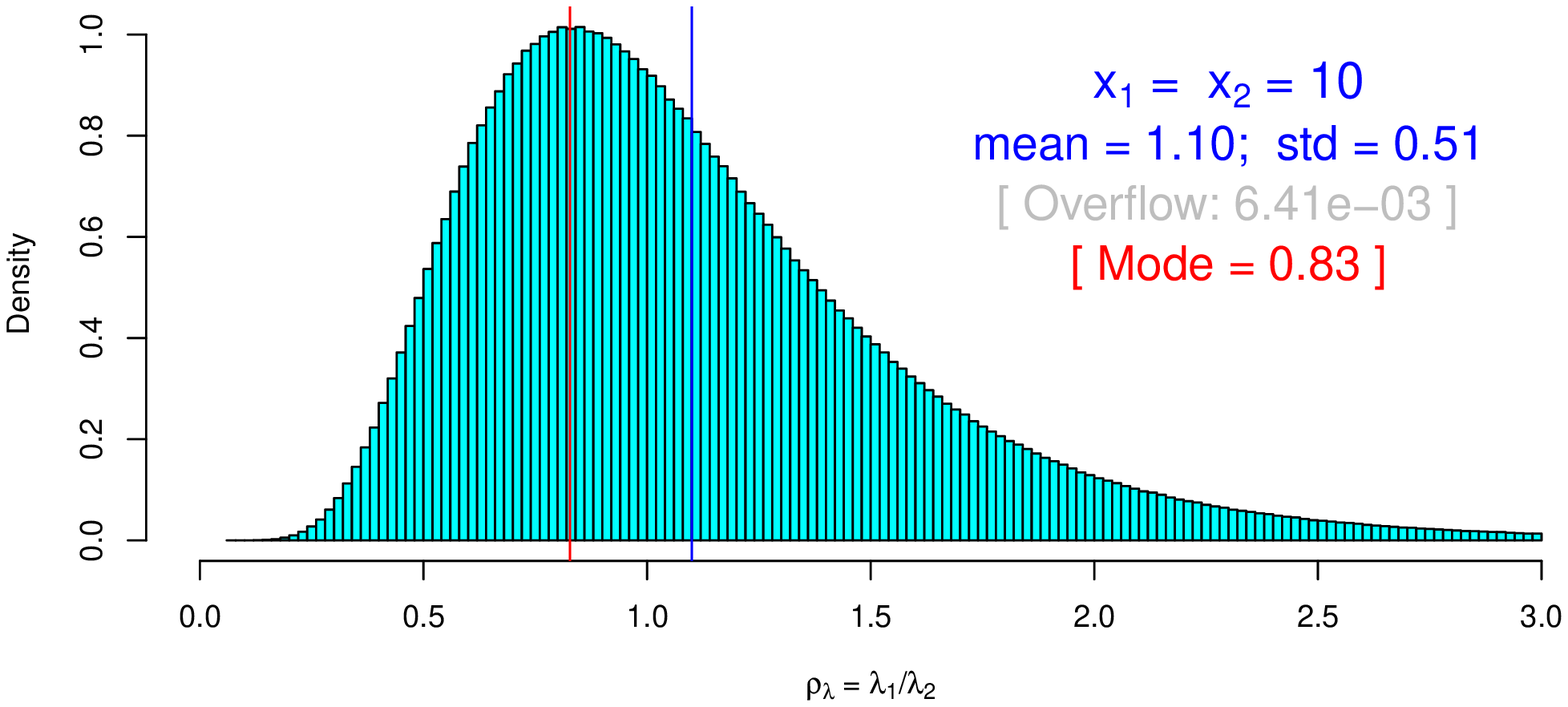,clip=,width=0.97\linewidth}
    \epsfig{file=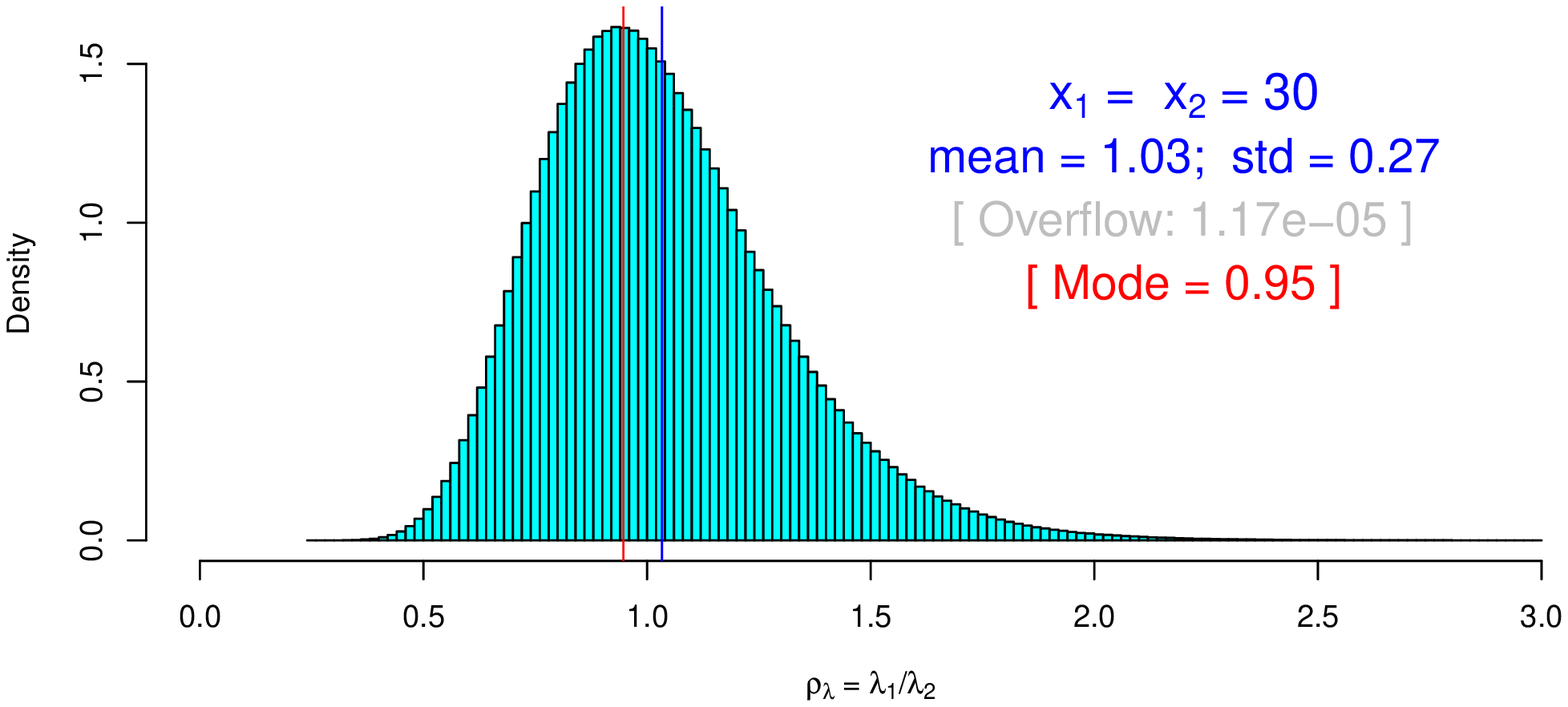,clip=,width=0.97\linewidth}
     \epsfig{file=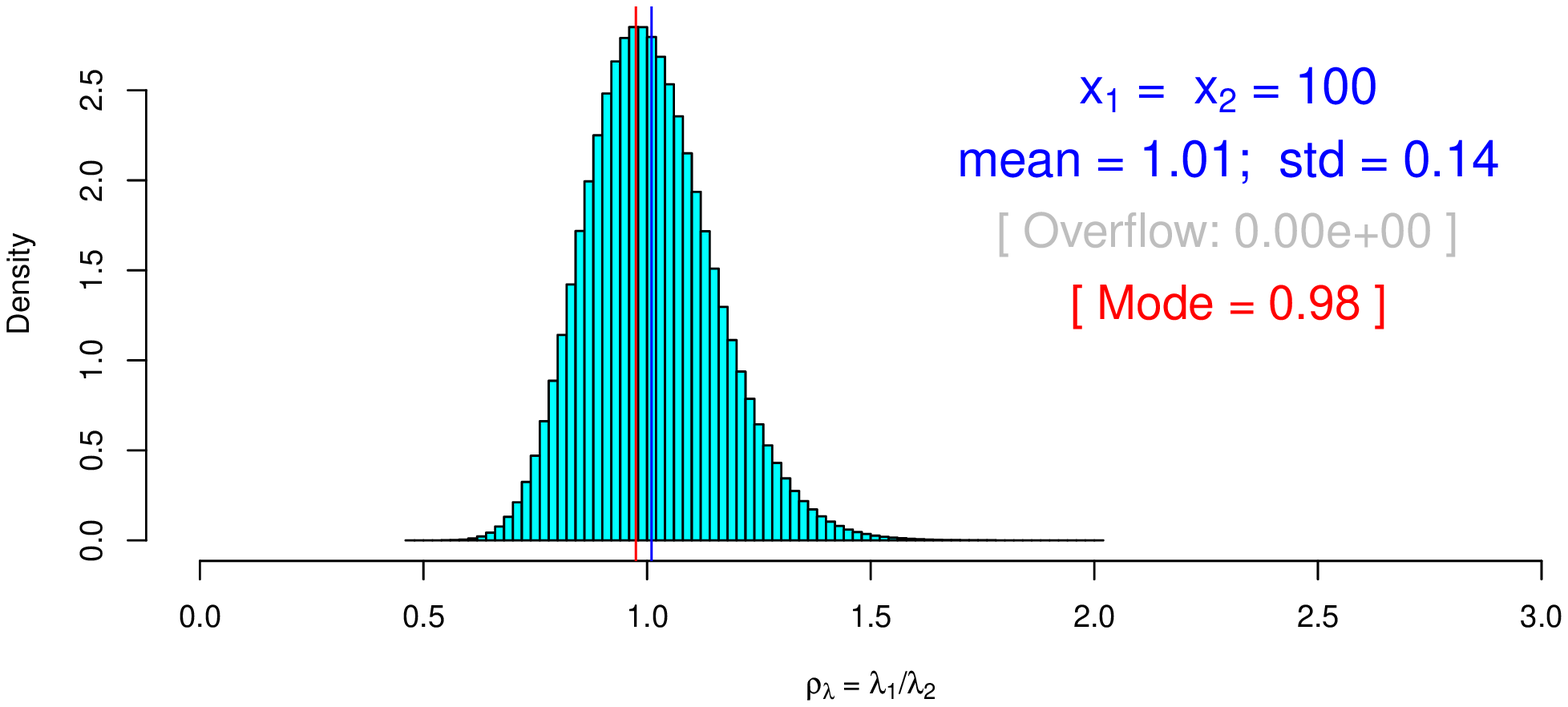,clip=,width=0.97\linewidth}
   \\  \mbox{} \vspace{-1.0cm} \mbox{}
    \end{center}
  \caption{\small \sf As Fig.~\ref{fig:pdf_lambda_x1_x2_MC_low}
    for larger values of the `observed' counts.
  }
    \label{fig:pdf_lambda_x1_x2_MC_high} 
\end{figure}
We immediately
observe that the histograms are very regular, without
divergences,
although for small values
of $x_2$ there is quite a long tail up to
infinity, which is however reached with vanishing probability
(the figures report also the proportions of overflows,
having chosen the horizontal scale of the plots
in order to show the most interesting part
of each probability distribution). 
The mean and standard deviation
(`std') shown on each plot are calculated from the Monte Carlo samples. 

The effect of the long tails is that there is quite a big difference
between mean value and the most probable one, located around the
highest bar of the histogram. This
is not a surprise (the famous exponential distribution
has modal value equal to zero independently of its parameter!),
but it should sound as a {\em warning for those who use analysis methods
  which provide, as `estimator',
  ``the most probable value''}\footnote{For
the reason of the quote marks see footnote \ref{fn:quote_JR}.}\,\cite{BR}.
Moreover, 
for very small $x_2$ the tails do not seem to
go very fast to zero (in comparison e.g. to the exponential), 
leading to not-defined moments of the distribution
(of the theoretical one, obviously,
since in most cases mean and standard deviation
of the Monte Carlo distribution have finite values).
This question will be investigated
in the next subsection, after the derivation
of closed expressions.

When $x_1$ and $x_2$ become `quite large' the Gamma distribution
tends ({\em slowly} -- think to the $\chi^2$, that is
indeed a particular Gamma, as
reminded in Appendix A) to a Gaussian, and likewise does
({\em a bit slower}) the ratio of two Gamma variables 
(as $\lambda_1$ and $\lambda_2$ are), as we can see for the case
$x_1\!=\!x_2\!=\!100$ of
Fig.~\ref{fig:pdf_lambda_x1_x2_MC_high},\footnote{Zero overflow
  in that plot is only due to the `limited' number of sampled,
  chosen to be $10^7$, and to the fact that the same script
  of the other plot has been used.}
in which  some skewness is still visible and the mode is
about $3\%$ smaller than the mean value.  

\subsection{Distribution of the ratio of Poisson
  \texorpdfstring{$\lambda$'s}{lambdas} in closed form}
\label{ss:pdf_rho_closed}
Being the evaluation of ratio of rates (to which the ratio of
$\lambda$'s is related)
an important issue in Physics, it is worth trying to get
an analytic expression for its pdf. 
This can be done extending to the continuum
Eq.~(\ref{eq:sommatoria}),\footnote{ for a different
  approach to get Eq.~(\ref{eq:propagazione_delta})
  see  footnote \ref{eq:propagazione_delta} ,
  in which the integrand of  Eq.~(\ref{eq:propagazione_delta})
  is interpreted as the joint pdf of $\rho_\lambda$, $\lambda_1$ and $\lambda_2$.
 }
that is replacing the sums
by integrals, and applying the constraint between the
two variables by a Dirac delta~\cite{BR}:
\begin{eqnarray}
  f(\rho_\lambda\,|\,x_1,x_2) &=& \int_0^\infty\!\!\!\int_0^\infty
  \!\delta\left(\rho_\lambda-\frac{\lambda_1}{\lambda_2}\right)\cdot
  f(\lambda_1\,|\,x_1)\cdot f(\lambda_2\,|\,x_2)\,
  \mbox{d}\lambda_1 \mbox{d}\lambda_2\,.
  \label{eq:propagazione_delta}
\end{eqnarray}  
Making use of the properties of the $\delta()$, we can rewrite it as
\begin{eqnarray}
  \delta\left(\rho_\lambda-\frac{\lambda_1}{\lambda_2}\right) &=&
  \frac{\delta(\lambda_1-\lambda_1^*)}
       {\left|\frac{\mbox{d}}{\mbox{d}\lambda_1}
         \left(\rho_\lambda-\frac{\lambda_1}{\lambda_2}\right)
         \right|_{\lambda_1=\lambda_1^*}} \\
       &=& \lambda_2\cdot \delta(\lambda_1-\lambda_1^*)\,,
\end{eqnarray}
with $\lambda_1^*$ root of the equation $\rho_\lambda-\lambda_1/\lambda_2=0$,
and therefore equal to $\rho_\lambda\cdot\lambda_2$. Equation
(\ref{eq:propagazione_delta}) becomes then
\begin{eqnarray}
  f(\rho_\lambda\,|\,x_1,x_2) &=& \int_0^\infty\!\!\!\int_0^\infty
  \!\lambda_2\cdot \delta(\lambda_1-\rho_\lambda\cdot\lambda_2)
  \cdot f(\lambda_1\,|\,x_1)\cdot f(\lambda_2\,|\,x_2)\,
  \mbox{d}\lambda_1 \mbox{d}\lambda_2 \\
  &=& \int_0^\infty\!\lambda_2\cdot
    \frac{(\rho_\lambda\cdot \lambda_2)^{x_1}\cdot e^{-\rho_\lambda\cdot\lambda_2}}
         {x_1!}\cdot
         \frac{\lambda_2^{x_2}\cdot e^{-\lambda_2}}{x_2!}
         \,\mbox{d}\lambda_2 \\
         &=& \frac{\rho_\lambda^{x_1}}{x_1!\,x_2!}\cdot
         \int_0^\infty\!\lambda_2^{x_1+x_2+1}\cdot e^{-(\rho_\lambda+1)\cdot\lambda_2}
         \,\mbox{d}\lambda_2\,. \label{eq:integrale_quasi_gamma}
\end{eqnarray}  
Once more we recognize in the integrand something related
to the Gamma distribution. In fact, identifying the power of $\lambda_2$
with `$\alpha-1$' of a Gamma pdf, and `$(1+\rho_\lambda)$' at the exponent
with the `rate parameter' $\beta$, that is
\begin{eqnarray}
   \alpha-1 &=& x_1+x_2+1 \\
  \beta &=& \rho_\lambda+1\,,
\end{eqnarray}
the integrand in Eq.~(\ref{eq:integrale_quasi_gamma}) can be rewritten as
\begin{eqnarray}
  \lambda_2^{\alpha-1}\cdot e^{-\beta\,\lambda_2} &=&
  \frac{\Gamma(\alpha)}{\beta^{\,\alpha}}\cdot
  \left(\frac{\beta^{\,\alpha}}{\Gamma(\alpha)}\cdot
  \lambda_2^{\alpha-1}\cdot e^{-\beta\,\lambda_2} \right)
\end{eqnarray}
in order to recognize within parentheses a Gamma pdf
in the variable $\lambda_2$, whose integral over
$\lambda_2$ is then equal to one because of normalization. We get then
\begin{eqnarray}
  f(\rho_\lambda\,|\,x_1,x_2) &=&
  \frac{\rho_\lambda^{\,x_1}}{x_1!\,x_2!}\cdot \frac{\Gamma(\alpha)}{\beta^{\,\alpha}}
  \cdot \int_0^\infty\!\frac{\beta^{\,\alpha}}{\Gamma(\alpha)}\cdot
  \lambda_2^{\alpha-1}\cdot e^{-\beta\lambda_2}
  \,\mbox{d}\lambda_2
  \label{eq:integrale_norm_gamma}\\
  &=&  \frac{\rho_\lambda^{x_1}}{x_1!\,x_2!}\cdot \frac{\Gamma(\alpha)}{\beta^{\,\alpha}} \\
  &=&   \frac{\Gamma(x_1+x_2+2)}{\Gamma(x_1+1)\,\Gamma(x_2+1)}\cdot
  \frac{\rho_\lambda^{\,x_1}}{(\rho_\lambda+1)^{\,x_1+x_2+2}} \\
   &=&   \frac{(x_1+x_2+1)!}{x_1!\,x_2!}\cdot
  \rho_\lambda^{\,x_1}\cdot (\rho_\lambda+1)^{\,-(x_1+x_2+2)}
  \label{eq:pdf_rho_closed}
\end{eqnarray}    
The mode of the distribution can be easily obtained
finding the maximum (of the log) of the pdf, thus getting
\begin{eqnarray}
  \mbox{mode}(\rho_\lambda) &=& \frac{x_1}{x_2+2}\,,
  \label{eq_mode_rho}
\end{eqnarray}
in agreement with what we have got in Figs.~\ref{fig:pdf_lambda_x1_x2_MC_low}
and \ref{fig:pdf_lambda_x1_x2_MC_high} by Monte Carlo
(indeed, done there in a fast and rather rough way -- see Appendix B.2).

\begin{figure}
  \begin{center}
    \epsfig{file=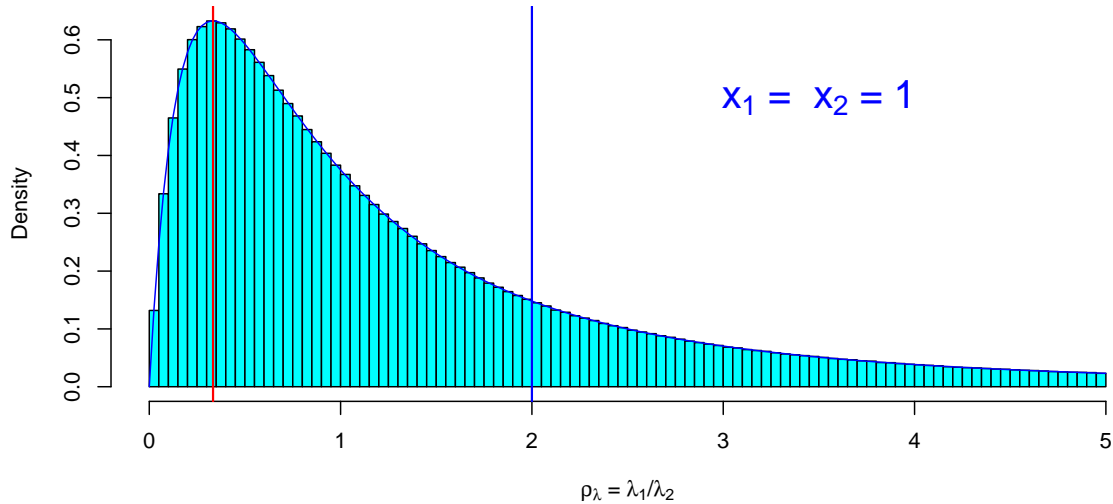,clip=,width=\linewidth}
   \\  \mbox{} \vspace{-1.2cm} \mbox{}
    \end{center}
  \caption{\small \sf Comparison of the distribution of
    $\rho_\lambda=\lambda_1/\lambda_2$ obtained by the closed expression
    (\ref{eq:pdf_rho_closed}) with that 
    estimated by Monte Carlo (same as top plot of
    Fig.~\ref{fig:pdf_lambda_x1_x2_MC_low}). The vertical
    lines indicate mode and expected value, evaluated
    using Eqs.~(\ref{eq_mode_rho}) and
    (\ref{eq:expected_value_rho}), equal to 1/3 and 2, respectively.
    (Note that none of these values is close to 1, that is what
    one would naively expect for the ratio of the rates -- indeed, only for
    $x_1$ and $x_2$ above ${\cal O}(100)$ mode, expected value
    and ratio of the observed counts become approximately equal).
  }
    \label{fig:pdf_rho_1_1_esatta} 
\end{figure}

In order to get expected value and standard deviation, we need
to evaluate the relevant integrals\footnote{Work
  done on a Raspberry Pi3, thanks to Mathematica 12.0 generously
offered by Wolfram Inc. to the Raspbian system.}
\begin{itemize}
\item First we check that $ f(\rho_\lambda\,|\,x_1,x_2) $ is properly
  normalized. Indeed the integral
  $\int_0^\infty\! f(\rho_\lambda\,|\,x_1,x_2)\,\mbox{d}\rho_\lambda$
  is equal to unity
  for `all possible' $x_1$ and $x_2$.\footnote{To be precise,
    the condition is $x_1>-1$
    and $x_2>-1$, but, given the role of the two variables in our context,
    it means for all possible counts (including
    $x_1=x_2=0$, for which the pdf becomes $1/(1+\rho_\lambda)^2$, having
    however infinite mean and variance.)
  }
\item The expected value is equal to
  \begin{eqnarray}
    \mbox{E}(\rho_\lambda\,|\,x_1,x_2) &=& \frac{x_1+1}{x_2}
    \hspace{0.7cm}(\mathbf{x_2>0})\,,
    \label{eq:expected_value_rho}
  \end{eqnarray}
  in perfect agreement with what we can read
  from the Monte Carlo results of Figs.~\ref{fig:pdf_lambda_x1_x2_MC_low}
  and \ref{fig:pdf_lambda_x1_x2_MC_high}.
\item The expected value of $\rho_\lambda^2$ is given by
   \begin{eqnarray}
     \mbox{E}(\rho_\lambda^2\,|\,x_1,x_2) &=& \frac{(x_1+1)\cdot
       (x_1+2)}{x_2\cdot (x_2-1)}
     \hspace{3.0cm}(\mathbf{x_2>1})\,,
   \end{eqnarray}
   from which we evaluate
   (subtracting to it the square of the expected value)
   \begin{eqnarray}
     \mbox{Var}(\rho_\lambda\,|\,x_1,x_2) &=&
     \frac{x_1+1}{x_2}\cdot \left(\frac{x_1+2}{x_2-1} -   \frac{x_1+1}{x_2}\right)
     \hspace{0.8cm}(\mathbf{x_2>1})\,,
   \end{eqnarray}
   from which the standard deviation follows,
   that we rewrite in a more compact form
   as
  \begin{eqnarray}
    \sigma(\rho_\lambda) &=& \sqrt{\mu_{\rho_\lambda}\cdot
      \left(\frac{x_1+2}{x_2-1} - \mu_{\rho_\lambda}\right)}
     \hspace{4.0cm}(\mathbf{x_2>1})\,, \label{eq:eq:sigma_rho}
  \end{eqnarray}
  having indicated by $\mu_{\rho_\lambda}$ the expected value
  of $\rho_\lambda$.
  For the values $x_1$ and $x_2$ used in Figs.~\ref{fig:pdf_lambda_x1_x2_MC_low}
  and \ref{fig:pdf_lambda_x1_x2_MC_high}, we get, {\bf starting from
  $\mathbf{x_1=x_2=2}$} in increasing order, the following
  standard deviations: 1.936, 1.247, 0.507, 0.269 and 0.143,
  in agreement with the Monte Carlo results (or the other way around). 
\end{itemize}
The detailed comparison between closed expression of the pdf
and the Monte Carlo outcome is shown in
Fig.~\ref{fig:pdf_rho_1_1_esatta} for the toughest case
we have met, that is $x_1=x_2=1$. 

\section{Inferring \texorpdfstring{$r_1$}{r1} and
  \texorpdfstring{$r_2$}{r2} (\texorpdfstring{$T_1$}{T1}
  possibly different from
  \texorpdfstring{$T_2$}{T2})}\label{sec:inferring_r}
After having been playing with $\lambda$'s and their
ratios, from which we started for simplicity, let us 
move now to the rates $r_1$ and $r_2$ of the two Poisson processes,
i.e. to the case in which the observation times $T_1$
and $T_2$ might be different.

\begin{figure}[t]
\begin{center}
  \epsfig{file=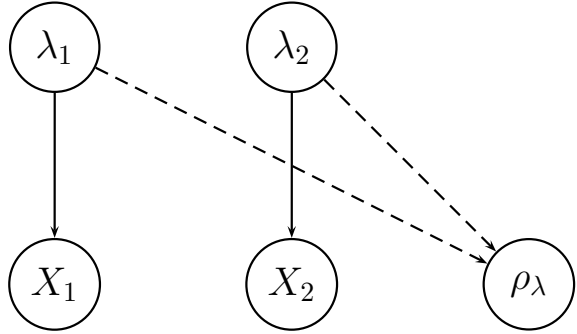,clip=,width=0.5\linewidth}
  \\ \mbox{}\vspace{-0.8cm}\mbox{}
\end{center}
\caption{\small \sf Graphical model showing the model underlying
  the inference of $\lambda_1$ and $\lambda_2$ from the observed
numbers of counts $X_1$ and $X_2$, followed by the deduction of $\rho$.}
\label{fig:BN_lambdas_x_rho}
\end{figure}

But, before doing that, let us spend a few words on the reason
of the word `deducing', appearing 
in the title of the previous section.
Let us start framing
what we have been doing in the past section in the 
graphical model
of Fig.~\ref{fig:BN_lambdas_x_rho},
known as a {\em Bayesian network}
(the reason for the adjective will be clear in the sequel).
The {\em solid} arrows from the {\em nodes} $\lambda_i$ to
the {\em nodes} $X_i$
indicate that the {\em effect} $X_i$ is {\em caused} by $\lambda_i$,
although in a probabilistic way (more properly, $X_i$ is
{\em conditioned} by $\lambda_i$, since, as it is well understood
{\em causality is a tough concept}\footnote{For
  a historical review and modern developments,
  with implication on Artificial Intelligence application, 
  see Ref.~\cite{PearlCausality}, an influential book on which I have
  however
  reservations when the author talks about causality in Physics
  (I have the suspicion he has never really read Newton
  or Laplace~\cite{Laplace_PC}  or Poincaré~\cite{Poincare'},
  and perhaps not even Hume~\cite{Hume}).}).
The dashed arrows indicate, instead,
{\em deterministic} links (or deterministic
`cause-effect' relations, if you wish).
For this reason we have been talking about `deduction': each couple of values
$(\lambda_1,\lambda_2)$ provides a unique value of $\rho_\lambda$,
equal to $\lambda_1/\lambda_2$, and any uncertainty about
the $\lambda$'s is `directly propagated' into
uncertainty about $\rho_\lambda$. The same will happen with $\rho=r_1/r_2$.

Navigating back along the solid arrows, that is from $X_i$ to
$\lambda_i$, is often called a problem of `inverse probability',
although nowadays
many experts do not like this expression, which however
gives an idea of what is going on. More precisely, it is
an {\em inferential problem}, notoriously tackled for the first time
in mathematical terms  by  Thomas Bayes~\cite{Bayes}
and Simon de Laplace, who was indeed talking about
``{\em  la probabilité des causes par les événements}''\cite{Laplace_PC}.
Nowadays the probabilistic tool to perform this `inversion'
goes under the name of {\em Bayes' theorem}, or {\em Bayes' rule},
whose essence, in terms
of the possible causes $C_i$ of the
observed effect $E$,
is given, besides normalization, by this very simple formula 
\begin{eqnarray}
  P(C_i\,|\,E,I) &\propto & P(E\,|\,C_i,I)\cdot P(C_i\,|\,I)\,,
  \label{eq:BayesRule}
\end{eqnarray}
having indicated again by $I$ the background state of information.
$P(C_i\,|\,I)$ quantifies our belief that $C_i$ is true taking into account
all available information, \underline{except} for $E$.
If \underline{also} the {\em hypothesis that $E$ is occurred}
is considered to be true,\footnote{Usually we say
  `if $E$ has occurred', but, indeed, in probability theory
  there are, in general, `hypotheses', to which we associate
  a degree of belief, being the states TRUE and FALSE just the
limits, mapped into $P=0$ and $P=1$.}
then the `prior' $P(C_i\,|\,I)$ is turned into the `posterior'
$ P(C_i\,|\,E,I)$.

We shall came back on the question of the priors,
but now let us move to infer $r_1$, $r_2$ and their ratio $\rho=r_1/r_2$.

\subsection{Inferring \texorpdfstring{$r$}{r},
  having observed \texorpdfstring{$x$}{x} counts in
the measuring time \texorpdfstring{$T$}{T}}\label{ss:inference_r_flat}
Being $r$ equal to $\lambda/T$, we can obtain its pdf
by a simple change of variables.\footnote{Starting from $f(\lambda\,|\,x)$
  given by Eq.~(\ref{eq:pdf_lambda_given_x}) we get
\begin{eqnarray*}
  f(\lambda\,|\,x)\,\mbox{d}\lambda\, =\,
  \frac{\lambda^x\cdot e^{-\lambda}}{x!}\,\mbox{d}\lambda &=&
  \frac{(r\cdot T)^x\cdot e^{-r\,T}}{x!}\cdot T\,\mbox{d}r \,=\,
  f(r\,|\,x,T)\,\mbox{d}r \\
  f(r\,|\,x,T) &=& \frac{T^{x+1}\cdot r^x\cdot e^{-T\,r}}{x!}\,.
\end{eqnarray*}
in which we recognize a Gamma pdf with  $\alpha=x+1$ and $\beta=T$.
}
But, having practiced a bit with the Gamma distribution, we can reach
the identical result observing that, using again a {\em flat prior} and
neglecting irrelevant factors, the pdf of $r$ is given by
\begin{eqnarray}
  f(r\,|\,x,T) &\propto&  f(x\,|\,r,T) \\
  &\propto&  (r\cdot T)^x\cdot e^{-r\,T}\\
                &\propto& r^x\cdot  e^{-T\,r}, 
\end{eqnarray}
in which we recognize, besides the normalization factor,
a Gamma pdf for the variable $r$ with
$\alpha=x+1$ and $\beta=T$, and hence
\begin{eqnarray}
  f(r\,|\,x,T) &=& \frac{\beta^{\,\alpha}\cdot r^{\,\alpha-1}\cdot e^{-\beta\,r}}{\Gamma(\alpha)}\\
  &=&  \frac{T^{\,x+1}\cdot r^{\,x}\cdot e^{-T\,r}}{x!}\,.
  \label{eq:f_r_|_x_T}
\end{eqnarray}
Mode, expected value and standard deviation of $r$ are then
(see Appendix A)
\begin{eqnarray*}
   \mbox{mode}(r) &=& \frac{\alpha-1}{\beta} = \frac{x}{T} \\
  \mbox{E}(r) &=&  \frac{\alpha}{\beta} = \frac{x+1}{T} \\
  \sigma(r) &=&  \frac{\sqrt{\alpha}}{\beta} =  \frac{\sqrt{x+1}}{T}\,,
\end{eqnarray*}
as also expected from the `summaries' of $f(\lambda\,|\,x)$
and making use of $r=\lambda/T$.\\
$[$\,Note that the pdf (\ref{eq:f_r_|_x_T})
  assumes, as explicitly written in the condition, a precise
  value of $T$. If this is not the case and $T$ is uncertain, then,
  similarly to what we have seen in footnote
  \ref{fn:uncertain_lambda},
  the pdf of $r$ is evaluated as
  $f(r\,|\,x,I) = \int_{0}^\infty
  f(r\,|\,x,T,I)\cdot f(T\,|\,I)\,\mbox{d}T$.\,$]$

\subsection{Role of the priors and sequential update of
  \texorpdfstring{$f(r)$}{f(r)} as new observations are considered}
The very essence of the so called probabilistic
inference (`Bayesian inference')
is given by Eq.~(\ref{eq:BayesRule}). 
The rest is just a question of normalization and of
extending it to the continuum, that in our case of interest is
\begin{eqnarray}
  f(r\,|\,x,T,I) &\propto &
  f(x\,|\,r,T,I)\cdot f(r\,|\,I)\,.
  \label{eq:Bayes_continuo}
\end{eqnarray}
It is evident the symmetric role of $f(x\,|\,r,T,I)$
and $f(r\,|\,I)$, if the former is seen
as a mathematical function of $r$ for a given (`observed')
$x$, that is $x$ playing the role of a parameter.
This function is known as {\em likelihood} and commonly indicated by
${\cal L}(r\,;\, x,T)$.\footnote{The real issue
  with the `likelihood' is not just replacing in Eq.~(\ref{eq:Bayes_continuo})
  $f(x\,|\,r,T,I)$ by ${\cal L}(r\,;\, x,T)$, but rather the fact,
  that, being this a function of $r$, it is perceived
  as `the likelihood of $r$'.
  The result is that it is often (almost always) turned by practitioners
  into  `probability of $r$', being `likelihood' and 'probability'
  used practically as synonyms in the spoken language.
  It follows, for example, that the value that maximizes the likelihood
  function is perceived as the `most probable' value, in the
light of the observations.}
Indicating the second factor of
Eq.~(\ref{eq:Bayes_continuo}), that is the `infamous' prior
that causes so much anxiety in practitioners\,\cite{PriorAnxiety},
by $f_0(r)$, we get (assuming $I$ implicit, as it is
usually the case)
\begin{eqnarray}
  f(r\,|\,x,T,I) &\propto &
  {\cal L}(r;x,T)\cdot f_0(r)\,,
  \label{eq:Bayes_lik-prior}
\end{eqnarray}
which makes it clear that we have two mathematical functions of
$r$ playing {\bf symmetric} and
{\bf peer} roles. Stated in different words,
{\em each of the two has the role of  `reshaping'
  the other}~\cite{sampling}.
In usual `routine' measurements
(as watching your weight on a balance) the information provided
by ${\cal L}(\ldots)$ is so much narrower, with
respect to $f_0(\ldots)$,\footnote{This is true
  unless the balance shows a `clear anomaly',
  and then you stick to what you believed your weight should
  be. But you still learn something from the measurement,
  indeed: the balance is broken~\cite{BR}.}
that we can neglect the latter and absorb it
in the proportionality factor, as we have done above
in Sec.~\ref{ss:inference_lambda}. Employing a uniform
`prior' is then usually a good idea to start with,
unless $f_0(\ldots)$ arises from previous measurements
or from strong theoretical prejudice on the quantity
of interest. It is also very important to understand that
 {\em `the reshaping' due to the priors can be done 
  in a second step}, as it has been pointed out, with
practical examples, in Ref.~\cite{sampling}.

Let us now see what happens when, in our case, the
Bayes rule is applied in sequence in order to account for
several results on the \underline{same}
rate $r$, that is assumed to be stable.   
Imagine we start from rather vague ideas about the value of $r$, 
such that $f_0(r)=k$ is, in practice, the best
practical choice we can do. After
the observation of $x_1$ counts during $T_1$ we get,
as we have learned above, 
\begin{eqnarray}
  f(r\,|\,x_1,T_1) &=&  \frac{T_1^{\,x_1+1}\cdot r^{\,x_1}\cdot e^{-T_1\,r}}{x_1!}.
  \label{eq:inferenza_r-flat_prior}
\end{eqnarray}
Then we perform a {\em new} campaign of observations
and record $x_2$ counts in $T_2$. It is clear now that in the second
inference we \underline{have to} use as `prior'
the piece of knowledge derived from the first inference.
So, we have, all together, besides irrelevant factors,
\begin{eqnarray}
  f(r\,|\,x_1,T_1,x_2,T_2) &\propto& f(x_2\,|\,r,T_2)\cdot f(r\,|\,x_1,T_1) \\
  &\propto& r^{\,x_2}\cdot e^{-T_2\,r}\cdot r^{\,x_1}\cdot e^{-T_1\,r}\\
  &\propto& r^{\,x_1+x_2}\, e^{-(T_1+T_2)\,r}\,,
\end{eqnarray}
that is exactly as we had done a single experiment,
observing $x_{tot} = x_1+x_2$ counts in $T_{tot}=T_1+T_2$. 
The only real physical {\em strong} assumption is that the intensity
of Poisson process was the same during the two measurements,
i.e. {\em we have being measuring the same thing}.

This teaches us immediately how to `combine the results',
an often debated subject
within experimental teams, if we have sets of counts $x_i$ during times $T_i$
(indicated all together by '$\underline x$' and `$\underline T$'):
\begin{eqnarray}
  f(r\,|\,\underline{x},\underline{T}) &=&
  \frac{(\sum_i T_i)^{\,\sum_ix_i+1}\cdot
    r^{\,\sum_ix_i}\cdot e^{-(\sum_iT_i)\,r}}{(\sum_ix_i)!}\,,
  \label{eq:combination_xi_Ti}
\end{eqnarray}
without imaginative averages or fits. But this does not mean
that we can blindly sum up counts and measuring times.
Needless to say, it is important, whenever it is possible, 
to make a detailed study of the
behavior of $f(r\,|\,x_i\,T_i)$ in order to
be sure that the intensity $r$ is compatible with being constant 
during the measurements. But, once we are confident about its constancy
(or that there is no strong evidence against that hypothesis),
the result is provided by Eq.~(\ref{eq:combination_xi_Ti}),
from which  all summaries of interest can be derived.\footnote{It is perhaps
  important to remind that in probability theory
  the {\em  full result} of the inference is the
  probability distribution (typically a pdf, for continuous quantities)
  of the quantity of interest as inferred from the data,
 the model and all other pertinent pieces of information.
  Mode, mean, median, standard deviation and probability
  intervals are just useful numbers to summarize
  with few numbers the distribution,
  with should always be reported, unless it is
  (with some degree of approximation) as
  simple as a Gaussian, so that mean and standard
  deviation provide the complete information.
  For example, the shape of a not trivial pdf can be
  expressed with coefficients of a suitable fit made in the
  region of interest. Or one can provide several moments
  of a distribution, from which the pdf can be reobtained
  (see e.g. Ref.~\cite{KendalStuart}).
}

\subsection{Relative belief updating ratio}
Let us consider again Eq.~(\ref{eq:Bayes_lik-prior})
and focus on the role of the likelihood to reshape $f_0(r)$.
Being multiplicative factor irrelevant, it can be useful
to rewrite that equation as
\begin{eqnarray}
  f(r\,|\,x,T,I) &\propto &
  \frac{{\cal L}(r\,;\, x,T)}{{\cal L}(r_R\,;\, x,T)}\cdot f_0(r)
  \label{eq:Bayes_lik-prior_relative}
\end{eqnarray}
with $r_R$ a reference value, in principle arbitrary, but
conceptually very interesting if
properly chosen. In fact, the ratio in the above formula
acquires the meaning of
{\em relative belief update
  factor}\,\cite{BR,conPia,conPeppe},\footnote{Note how
  at that time we wrote
  Eq.~(\ref{eq:Bayes_lik-prior_relative})
  in a more expanded
  way, but the essence of this {\em factor}
  is given by Eq.~(\ref{eq:RBUF}).
  For recent developments and applications see
  Refs.~\cite{Gariazzo,Salas_et_al,Grilli_Messina_Piacentini}.
}
and the updating Bayes' rule can be rewritten as
\begin{eqnarray}
  f(r\,|\,x,T,I) &\propto &  {\cal R}(r\,;\,x,T,r_R)\cdot f_0(r) \,.
  \label{eq:RBUF}
\end{eqnarray}
This way of rewriting the Bayes' rule is particularly convenient
when the likelihood is not 'closed', that is it does not
go to zero when the quantity of interest is `very large'
or `very small'.

\begin{figure}[t]
  \begin{center}
    \epsfig{file=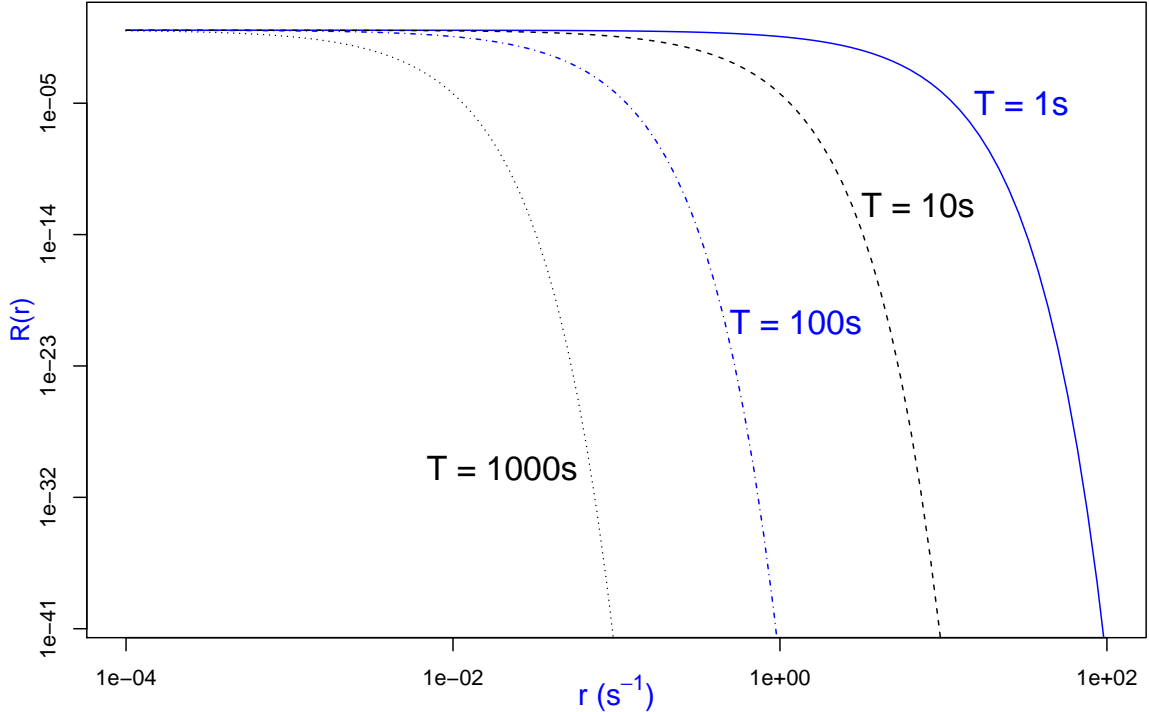,clip=,width=\linewidth}
   \\  \mbox{} \vspace{-1.2cm} \mbox{}
    \end{center}
  \caption{\small \sf Relative belief updating factor
    ${\cal R}(r;x=0,T,r_0=0)$ for different observation times.
  }
    \label{fig:esempi_RBUF} 
\end{figure}

To be clear, let us make the example of
{\em having \underline{observed} zero counts}, that is
the experiment was indeed performed, but no event of interest
was found during the measurement time $T$.
If we use a flat prior and
only stick to the summaries, we have that the most probable
value is zero, with $\mbox{E}(r)=\sigma(r)=1/T$:
the larger is the measuring time, the more the distribution
of $r$ is squeezed towards zero.
But this does not give a complete picture of what is going on.
Since ${\cal L}(r; x=0,T)$ goes to 1 for $r\rightarrow 0$,
the likelihood is {\em opened in the left side}.
Figure \ref{fig:esempi_RBUF} shows   ${\cal R}$
functions for this case, for different $T$, although in this very
simple case ${\cal R}$ is mathematically equivalent to the
likelihood.\footnote{The more interesting case, originally
  taken into account in Refs.~\cite{conPia,conPeppe},
  is when some events are observed, which could be, however, also
  attributed to irreducible background.}
If our beliefs about $r$ were above {\cal O}($100\,\mbox{s}^{-1}$),
the observation of zero events {\em practically} rule them out,
even with $T=1\,$s (`1\,s' is arbitrarily chosen in this hypothetical example,
just to remind that both $T$ and $r$ have physical
dimensions).

If we run the experiment longer and longer,
keeping observing zero events, the possible values of $r$
gets smaller and smaller. What is mostly interesting,
in this plot, is the region in which ${\cal R}$ is flat: it means
that if our beliefs are concentrate there, then
the experiment does not teach us more than what we already believed:
{\em the experiment looses sensitivity} in that region
and then {\em reporting `probabilistic' upper limits makes no sense}
and it can be highly misleading (even more reporting `C.L.
upper limits`)\,\cite{BR,CLW2000}.

\subsection{Conjugate priors}
At this point a technical remark is in order. The reason
why the Gamma appears so often is that the expression of
the Poisson probability function, seen as a function of $\lambda$
 and neglecting multiplicative factors,
that is $f(\lambda)\propto \lambda^x\cdot\exp(-\lambda)$,
has the same structure of a Gamma pdf.
The same is true if the variable $r$ is considered,
that is $f(r)\propto r^x\cdot\exp(-T\cdot r)$. If
then we have a Gamma distribution as prior, with parameters
$\alpha_0$ and $\beta_0$, the `final' distributions is still a
Gamma:
\begin{eqnarray}
  f(\lambda\,|\,x) = \lambda^x\cdot e^{-\lambda} \cdot
  \lambda^{\,\alpha_0-1}\cdot e^{-\beta_0\lambda} & = &
  \lambda^{\,\bm{\footnotesize \alpha_0+x-1}}\cdot
  e^{-\bm{(\small\beta_0+1)}\,\cdot\, \lambda}    \\
  &\propto&  \lambda^{\,\bm{\small \alpha_f-1}}
  \cdot e^{-{\footnotesize \bm{\beta_f}}\,\cdot\, \lambda}  \\
  f(r\,|\,x,T) = r^x\cdot e^{-T\cdot r} \cdot
  r^{\,\alpha_0-1}\cdot e^{-\beta_0r} & = & 
  r^{\,\bm{\alpha_0+x-1}}\cdot e^{-\bm{(\beta_0+T)}\,\cdot\, r} \\
  &\propto &  r^{\,\bm{\alpha_f-1}}\cdot e^{-\bm{\beta_f}\,\cdot\, {\large r}}
\end{eqnarray}
This kind of distributions, such that the `posterior' belongs to 
the same family of the `prior', with {\em updated parameters},
are called {\em conjugate priors} for obvious reasons,
as it is rather obvious how convenient they are
in applications,
{\em provided they are flexible enough to describe `somehow'
  the prior belief}.\footnote{Remember that, as Laplace used to say,
  {\em ``the theory of probabilities is basically
    just common sense reduced to calculus''}, that
  {\em ``All models are wrong, but some are useful''} (G.Cox)
  and that even Gauss was `sorry' because `his' error function
  could not be strictly true~\cite{Gauss}
  (see quote in footnote 9 of Ref.~\cite{SkepticalJags}).
} This was particularly important at the times
when the monstrous computational power nowadays available
was not even imaginable
(also the development of logical and mathematical
tools has a strong relevance).
Therefore a quite rich
collection of conjugate priors
is available in the literature (see e.g. Ref.~\cite{ConjugatePriors}).

In sum, these are the {\em updating rules} of the Gamma parameters
for our cases of interest (the subscript '$f$' is to remind
that is the parameter of the `final' distribution):
\mbox{}\vspace{-0.3cm}\mbox{}
\begin{eqnarray}
  \mbox{{\bf Inferring $\bm{\lambda}$:}}\hspace{2.0cm}
    \alpha_f &=& \alpha_0 + x \label{eq:update_alpha_lambda}\\
    \beta_f &=& \beta_0 + 1 \label{eq:update_alpha_beta}\\
     \mbox{{\bf Inferring $\bm{r}$:}}\hspace{2.0cm}
    \alpha_f &=& \alpha_0 + x\label{eq:update_alpha_r}\\
    \beta_f &=& \beta_0 + T \label{eq:update_beta_r}
  \end{eqnarray}
  (Note that in the case of $r$ the parameter $\beta$ has the dimension
  of a time, being $r$ a rate, that is counts per unit of time.)
A flat prior distribution is recovered for
$\alpha_0=1$ and $\beta_0\rightarrow 0$.
Technically,
  for $\alpha=1$ a Gamma distribution turns into a negative exponential:
  if then the `rate parameter' $\beta$ is chosen to be
  very small, the exponential
becomes `essentially flat' in the region of interest.

Once we have learned the updating rules
(\ref{eq:update_alpha_lambda})-(\ref{eq:update_alpha_beta})
and (\ref{eq:update_alpha_r})-(\ref{eq:update_beta_r}),
it might be convenient to turn a prior expressed in terms
of mean $\mu_0$ and standard deviation $\sigma_0$
into $\alpha_0$
and $\beta_0$, inverting the expressions of
expected value and standard deviation of a Gamma distributed
variable (see Appendix A), thus getting
\begin{eqnarray}
  \alpha_0 &=& {\mu_0^2}/{\sigma_0^2} \\
  \beta_0 &=& {\mu_0}/{\sigma_0^2}\,.
\end{eqnarray}
For example, if we have good reason to think that
$r$ should be $(5\pm 2)\,\mbox{s}^{-1}$\!, the parameters
of our initial Gamma distribution are $\alpha_0=6.25$
and $\beta_0=1.25\,\mbox{s}$. This is equivalent to having
started from a flat prior and having observed (rounding the numbers)
5 counts in about 1.2 seconds. This gives a clear idea of the
`strength' of the prior -- not much in this case, but it certainly
excludes the possibility of $r=0$. This happens in fact
as soon as $\alpha_0$ is larger then 1, implying $r^{\alpha_0-1}$
vanishing at $r=0$.
This observation can be a used as a trick to forbid a vanishing value
of $\lambda$ or of $r$, if we have good physical reason
to believe that they cannot be zero, although we are
highly uncertain about  even  their order of magnitude: 
just choose a prior $\alpha_0$ slightly larger than one.

\section{Ratio of Gamma distributed variables}\label{sec:ratio_gamma}
\begin{figure}[t]
\begin{center}
  \epsfig{file=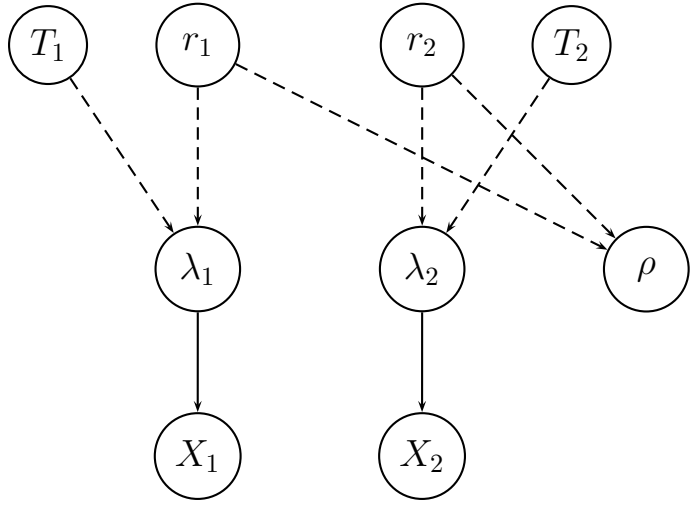,clip=,width=0.6\linewidth}
  \\ \mbox{}\vspace{-0.8cm}\mbox{}
\end{center}
\caption{\small \sf Graphical model relating the physical
  quantities (rates and measurement times) to the observed
  numbers of events.}
\label{fig:BN_r_x_rho}
\end{figure}

Having inferred the two rates, we can now evaluate
the distribution of $\rho=r_1/r_2$, which
is technically just a problem of `direct probabilities',
that is getting the pdf $f(\rho\,|\,x_1,T_1,x_2,T_2)$
from $f(r_1\,|\,x_1,T_1)$ and $f(r_2\,|\,x_2,T_2)$
(the Bayesian network that relates the variables
of interest is shown in Fig.~\ref{fig:BN_r_x_rho}).
We just need to repeat what it has been
done in Sec.~\ref{ss:pdf_rho_closed}, taking the advantage
of having understood that $f(\lambda_1\,|\,x_1)$
and $f(\lambda_2\,|\,x_2)$ appearing in Eq.~(\ref{eq:propagazione_delta})
are indeed Gamma distributions.
Therefore, we start evaluating the probability distribution of the ratio
of generic Gamma variables, denoted as $Z_1$ and $Z_2$
(and their possible occurrences $z_1$ and $z_2$) in order
to avoid confusion with $X$'s, associated so far to measured counts:
\begin{eqnarray}
  Z_1\!\! &\sim&\!\! \mbox{Gamma}(\alpha_1,\beta_1)\\
  Z_2\!\! &\sim&\!\! \mbox{Gamma}(\alpha_2,\beta_2)\,.
\end{eqnarray}
The pdf of $Z_1/Z_2$ is the given by
\begin{eqnarray}
  f(\rho_z\,|\,\alpha_1,\beta_1,\alpha_2,\beta_2)\!\! &=&\!\!
  \int_0^\infty\!\!\!\int_0^\infty\!
  \delta\!\left(\rho_z-\frac{z_1}{z_2}\right)\!\cdot\!
 f(z_1\,|\,\alpha_1,\beta_1)\!\cdot\! f(z_2\,|\,\alpha_2,\beta_2)\,
  \mbox{d}z_1 \mbox{d}z_2\,,\ \ \ \ \ \ 
  \label{eq:propagazione_delta_z1_z2}
\end{eqnarray}
in which we have indicated by $\rho_z$ their ratio.
In detail, taking benefit of what we have
learned in Sec.~\ref{ss:pdf_rho_closed},
\begin{eqnarray}
  f(\rho_z\,|\,\ldots)\!\! &=&\!\!  \int_0^\infty\!\!\!\int_0^\infty\!\!\!
  z_2\cdot \delta(z_1\!-\!\rho_z\!\cdot\! z_2)\!\cdot\!
  \frac{\beta_1^{\,\alpha_1}\cdot z_1^{\,\alpha_1-1}\cdot e^{-\beta_1\,z_1}}{\Gamma(\alpha_1)}
  \!\cdot\!
  \frac{\beta_2^{\,\alpha_2}\cdot z_2^{\,\alpha_2-1}\cdot e^{-\beta_2\,z_2}}{\Gamma(\alpha_2)}
 \, \mbox{d}z_1 \mbox{d}z_2 \nonumber \\
  \!\! &=&\!\!
  \frac{\beta_1^{\,\alpha_1}\beta_2^{\,\alpha_2}}
       {\Gamma(\alpha_1)\cdot \Gamma(\alpha_2)}\cdot 
       \int_0^\infty\!\!\!z_2\cdot 
       {(\rho_z\! \cdot\! z_2)^{\,\alpha_1-1}\cdot e^{-\beta_1\,(\rho_z\cdot z_2)}}
  \cdot {z_2^{\,\alpha_2-1}\cdot e^{-\beta_2\,z_2}}
  \, \mbox{d}z_2 \\
   \!\! &=&\!\!
  \frac{\beta_1^{\,\alpha_1}\cdot \beta_2^{\,\alpha_2}}
       {\Gamma(\alpha_1)\cdot \Gamma(\alpha_2)}\cdot
       \rho_z^{\,\alpha_1-1}
       \int_0^\infty\!\!\! z_2^{\,\alpha_1+\alpha_2-1}\cdot
       e^{-(\beta_2+\rho_z\,\beta_1)\cdot z_2}
       \, \mbox{d}z_2\,.   
\end{eqnarray}
Writing $\alpha_1\! +\! \alpha_2$ as $\alpha_*$
and $\beta_2\!+\!\rho_z\!\cdot\! \beta_1$ as $\beta_*$, we get
\begin{eqnarray}
  f(\rho_z\,|\,\alpha_1,\beta_1,\alpha_2,\beta_2)\!\!  &=& \!\!
   \frac{\beta_1^{\,\alpha_1}\cdot \beta_2^{\,\alpha_2}}
       {\Gamma(\alpha_1)\cdot \Gamma(\alpha_2)}\cdot
       \rho_z^{\,\alpha_1-1}
       \int_0^\infty\!\!\! z_2^{\alpha_*-1}\cdot
       e^{-\beta_*\cdot z_2}
       \, \mbox{d}z_2 \\
    &=& 
   \frac{\beta_1^{\,\alpha_1}\cdot \beta_2^{\,\alpha_2}}
       {\Gamma(\alpha_1)\cdot \Gamma(\alpha_2)}\cdot
       \rho_z^{\,\alpha_1-1}\cdot \frac{\Gamma(\alpha_*)}{\beta_*^{\alpha_*}}
       \label{eq:step_to_pdf_rho_closed_general}\\
       &=& \frac{\Gamma(\alpha_1+\alpha_2)}
           {\Gamma(\alpha_1)\cdot \Gamma(\alpha_2)}
       \cdot \beta_1^{\,\alpha_1}\cdot \beta_2^{\,\alpha_2}\cdot
       \frac{ \rho_z^{\,\alpha_1-1}}{(\beta_2+\rho_z\!\cdot\beta_1)^{\,\alpha_1+\alpha_2}}\\
       &=& \frac{\Gamma(\alpha_1+\alpha_2)}
            {\Gamma(\alpha_1)\cdot \Gamma(\alpha_2)}
       \cdot \beta_1^{\,\alpha_1}\cdot \beta_2^{\,\alpha_2}
      \cdot \rho_z^{\,\alpha_1-1}\!\cdot\!  (\beta_2+\rho_z\!\cdot\beta_1)^{\,-(\alpha_1+\,\alpha_2)}\!. \ \ \ \ \ \ \label{eq:pdf_rho_closed_general}
\end{eqnarray}
\subsection{Application to
\texorpdfstring{$\rho_\lambda=\lambda_1/\lambda_2$}{rho\_lambda=lambda1/lambda2}
and to \texorpdfstring{$\rho=r_1/r_2$}{rho=r1/r2}}
In the case of ratio of $\lambda$'s, and starting from uniform prior,
as done in   Sec.~\ref{ss:pdf_rho_closed}, we get,
applying Eq.~(\ref{eq:pdf_rho_closed_general}) and
writing the conditions in terms of the Gamma parameters,
\begin{eqnarray}
  f(\rho_\lambda\!=\!\frac{\lambda_1}{\lambda_2}\,|\,x_1+1,1,x_2+1,1)\!\!  &=& \!\!
  \frac{\Gamma(x_1+x_2+2)}{\Gamma(x_1+1)\cdot \Gamma(x_2+1)}\cdot
  \rho_\lambda^{\,x_1}\cdot (1+\rho_\lambda)^{\,-(x_1+\,x_2+2)} \nonumber \\
  &=& \frac{(x_1+x_2+1)!}{x_1!\,x_2!} \cdot
  \rho_\lambda^{\,x_1}\cdot (1+\rho_\lambda)^{\,-(x_1+\,x_2+2)}\,,
  \label{eq:pdf_rho_closed_again}
\end{eqnarray}   
re-obtaining exactly Eq.~(\ref{eq:pdf_rho_closed}).

\begin{figure}
  \begin{center}
    \epsfig{file=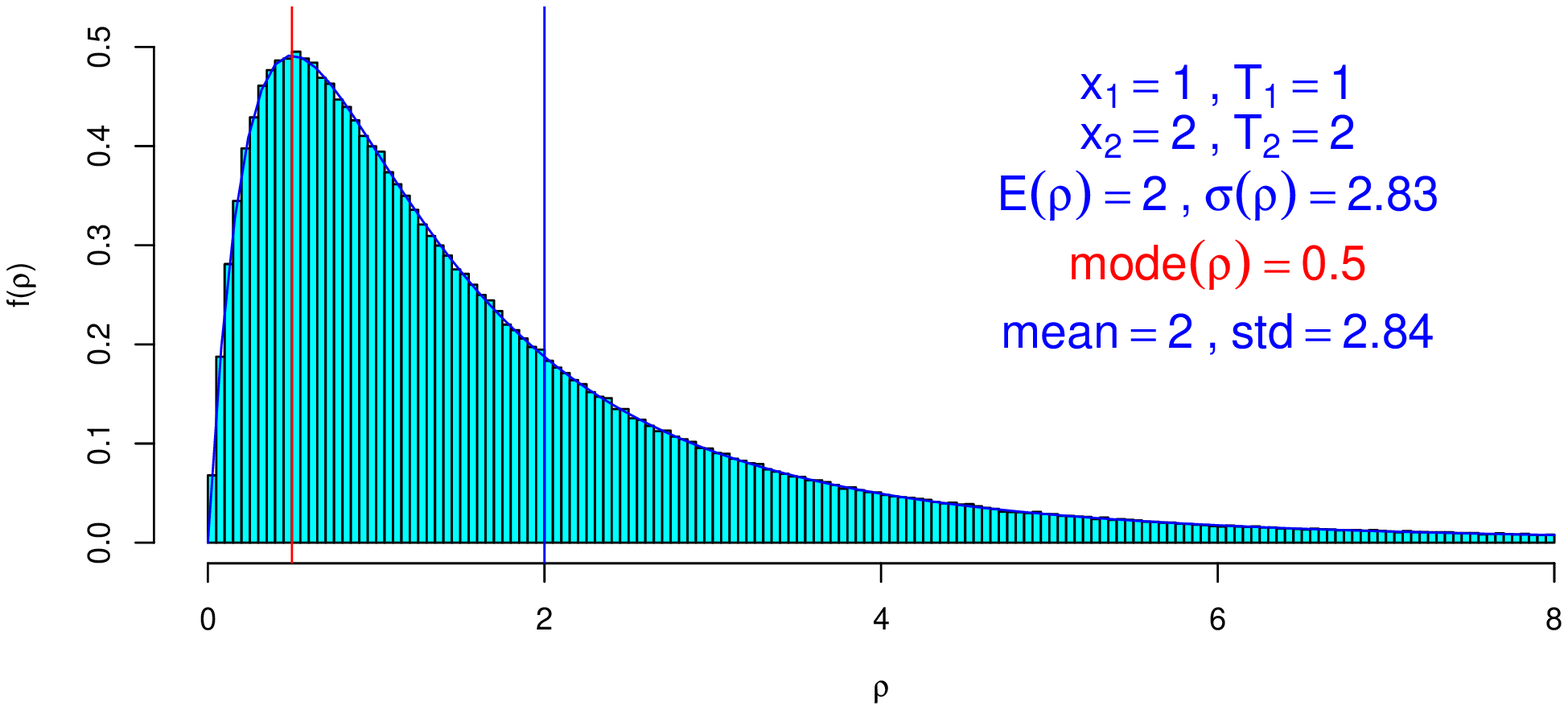,clip=,width=0.97\linewidth}
    \epsfig{file=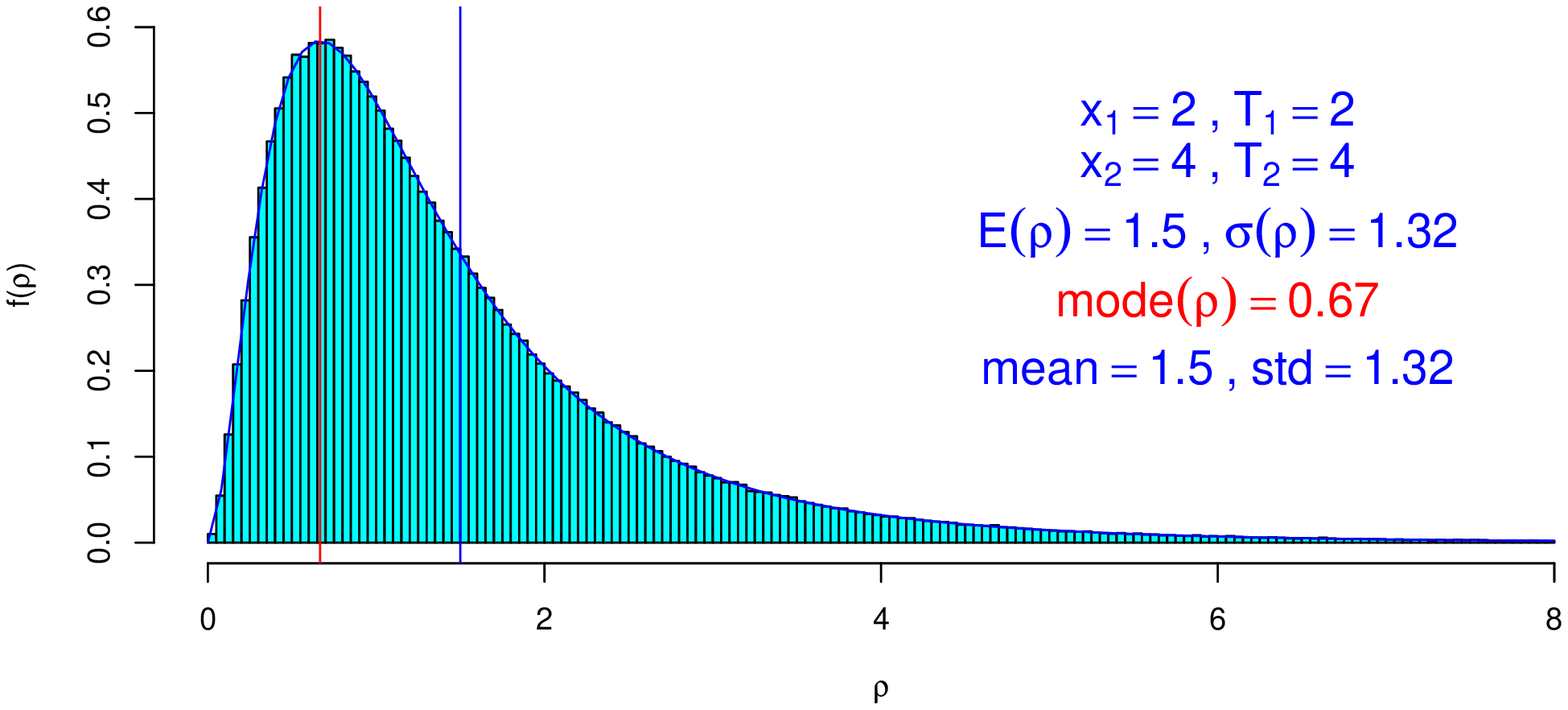,clip=,width=0.97\linewidth}
     \epsfig{file=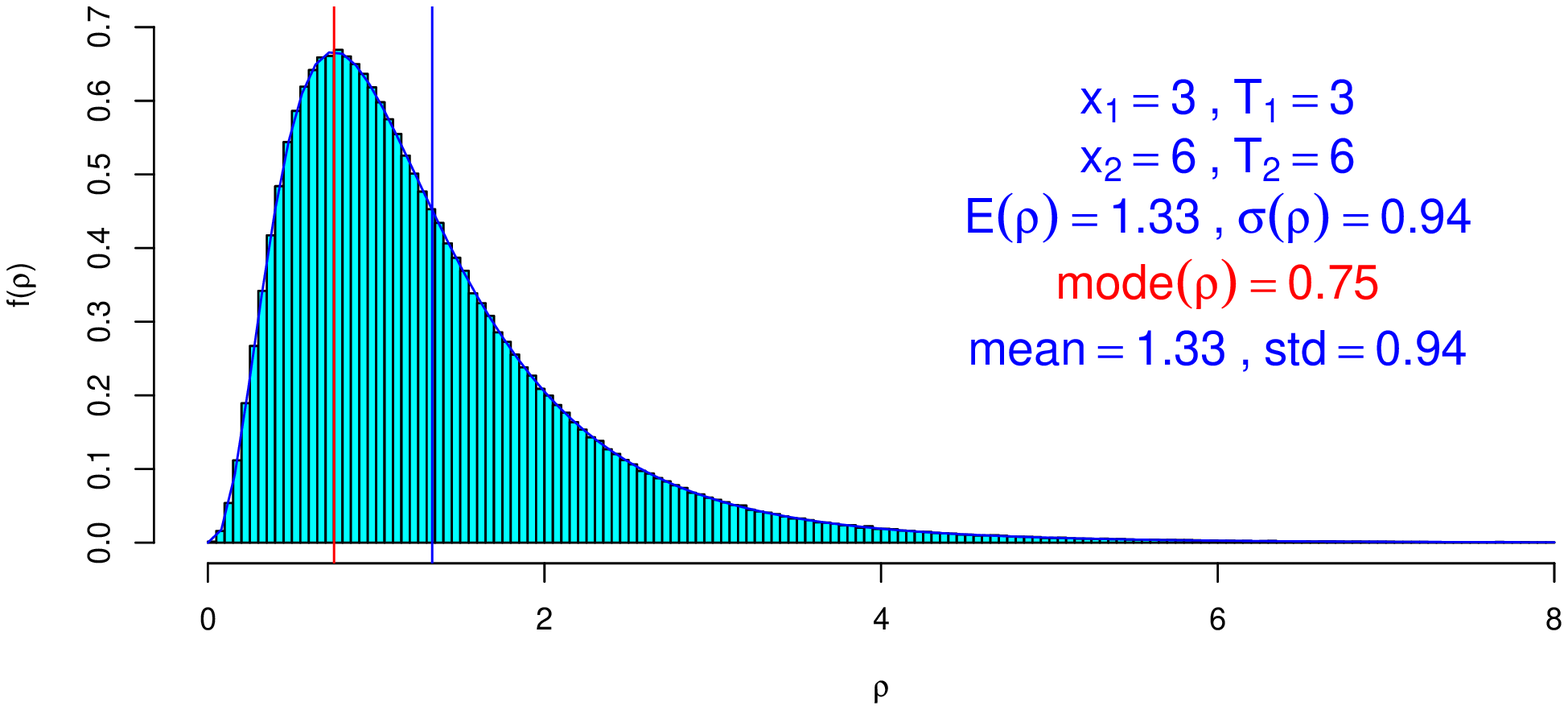,clip=,width=0.97\linewidth}
   \\  \mbox{} \vspace{-1.2cm} \mbox{}
    \end{center}
  \caption{\small \sf Ratios of rates, given counts and times.
  }
    \label{fig:pdf_rho_r_low} 
\end{figure}

\begin{figure}
  \begin{center}
    \epsfig{file=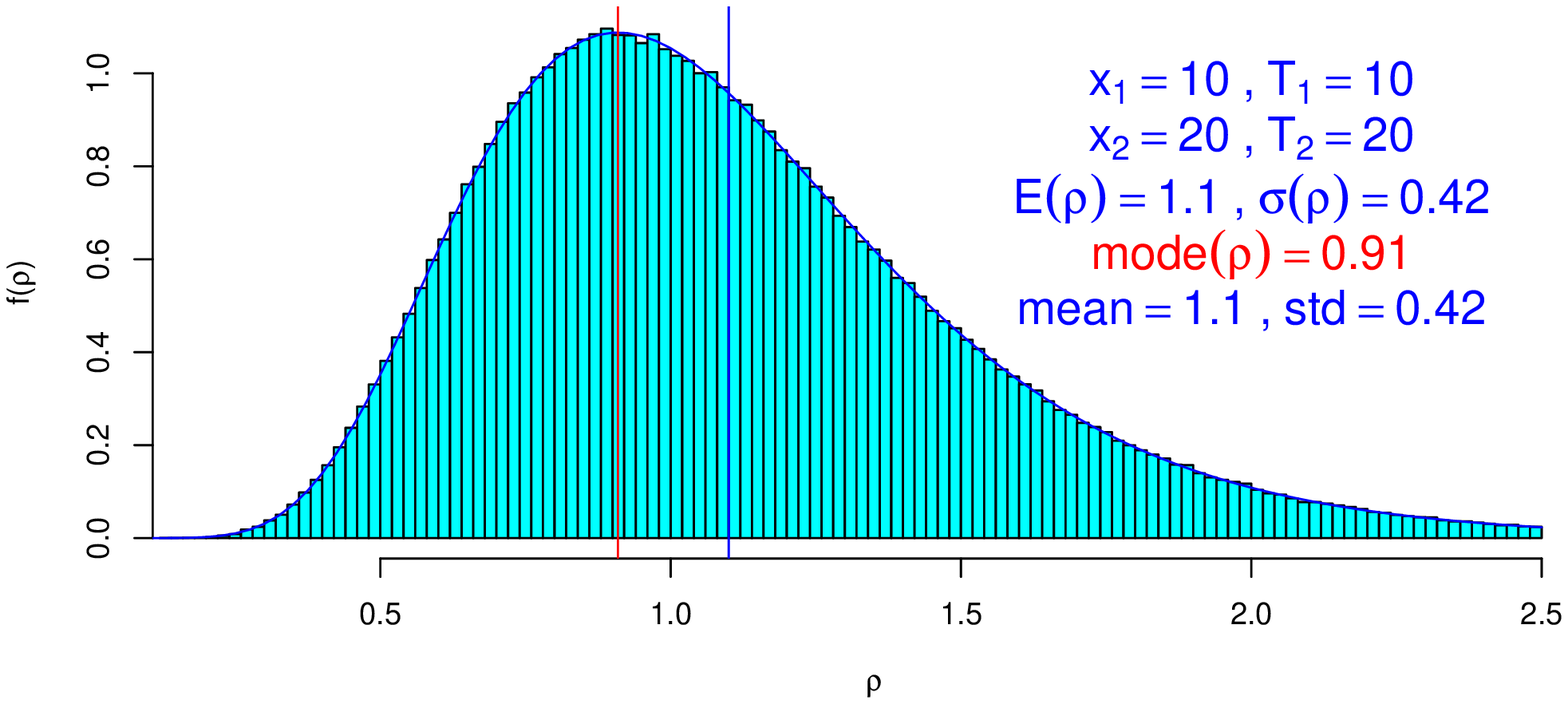,clip=,width=0.97\linewidth}
    \epsfig{file=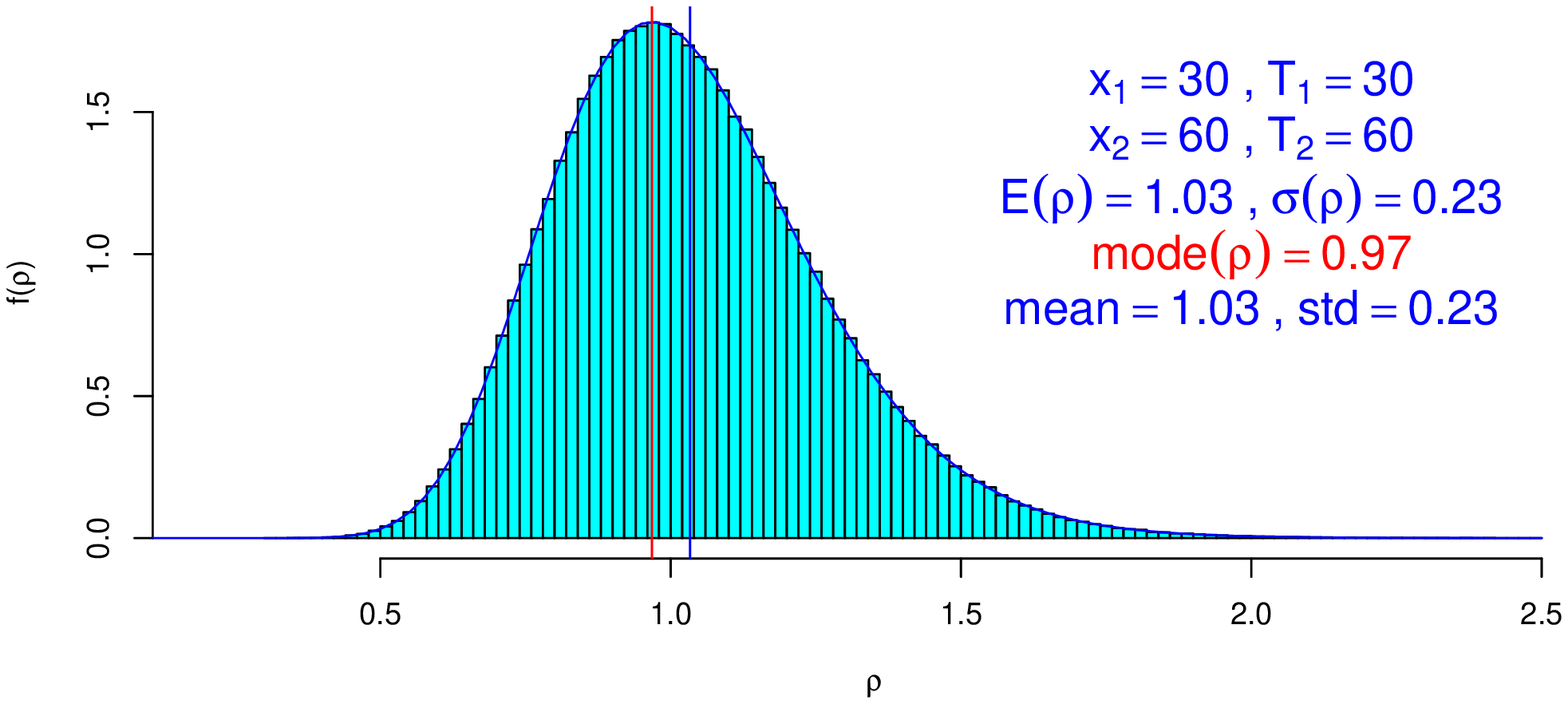,clip=,width=0.97\linewidth}
     \epsfig{file=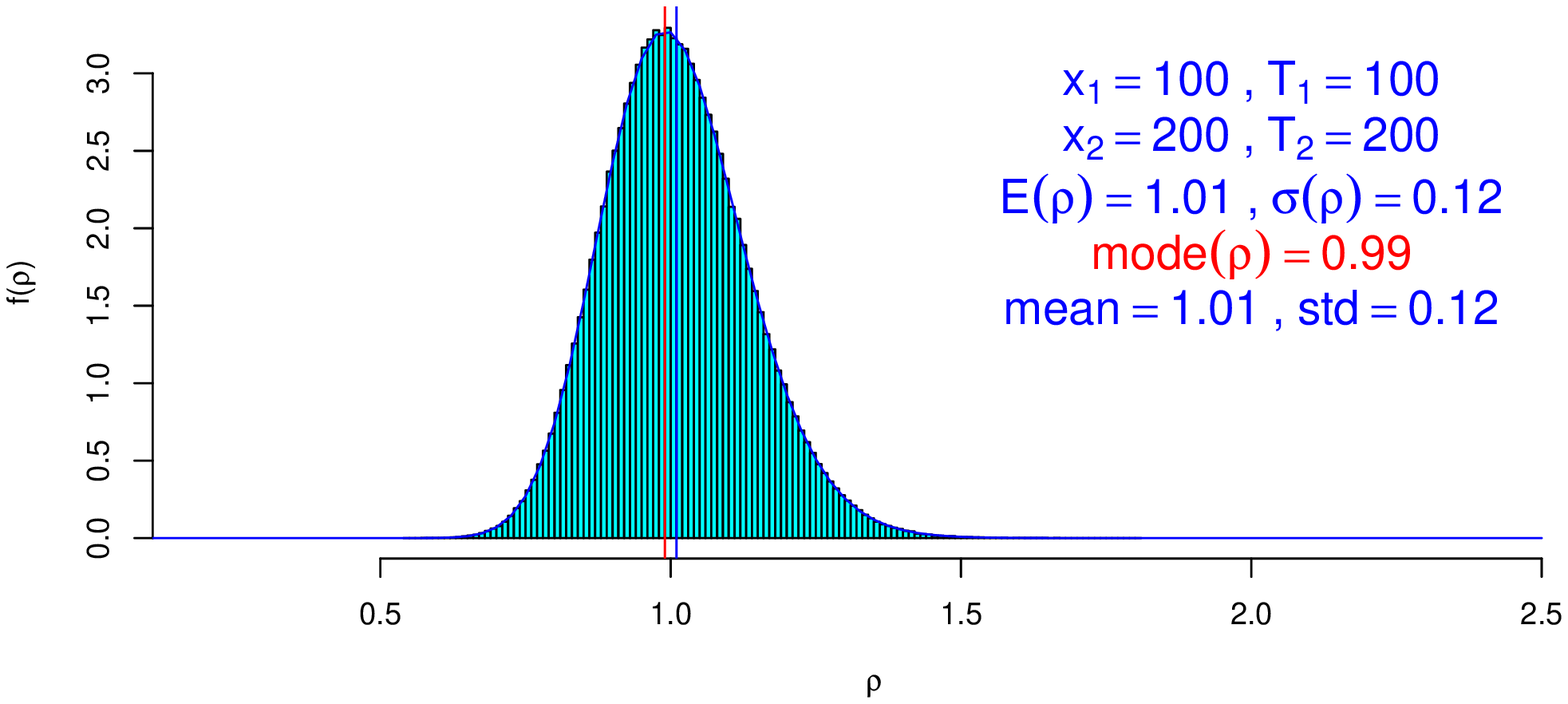,clip=,width=0.97\linewidth}
   \\  \mbox{} \vspace{-1.0cm} \mbox{}
    \end{center}
  \caption{\small \sf As Fig.~\ref{fig:pdf_rho_r_low}
    for larger values of counts, observed in proportionally larger times.
  }
    \label{fig:pdf_rho_r_high} 
\end{figure}

As far as the ratio of rates, starting again from a uniform prior,
implying then  $\alpha_i=x_i+1$ and $\beta_i=T_i$,
we get, 
writing, as in Eq.~(\ref{eq:pdf_rho_closed_again}),
the conditions in terms of the Gamma parameters,
\begin{eqnarray}
  f(\rho\!=\!\frac{r_1}{r_2}\,|\,x_1\!+\!1,T_1,x_2\!+\!1,T_2)\!\!  &=& \!\!
  \frac{(x_1+x_2+1)!}{x_1!\,x_2!} \!\cdot\!  T_1^{\,x_1+1}\!\cdot\!
  T_2^{\,x_2+1}\!\cdot\!
  \rho^{\,x_1}\!\cdot\! (T_2+ T_1\cdot \rho)^{\,-(x_1+\,x_2+2)}, \nonumber 
\end{eqnarray}
that is, without redundant details,\footnote{Another
  way to arrive to Eq.~(\ref{eq_pdf_rho_r}) is to start
  from Eq.~(\ref{eq:pdf_rho_closed}), applying
  the transformation of variables $\rho=\rho_\lambda\cdot T_2/T_1$:
\begin{eqnarray*}
  f(\rho_\lambda\,|\,\ldots)\,\mbox{d}\rho_\lambda
  &=& \frac{(x_1+x_2+1)!}{x_1!\,x_2!} \cdot
  \rho_\lambda^{\,x_1}\cdot (1+\rho_\lambda)^{\,-(x_1+\,x_2+2)}\, \mbox{d}\rho_\lambda\\
  &=& \frac{(x_1+x_2+1)!}{x_1!\,x_2!} \cdot
  \left(\frac{T_1}{T_2}\right)^{x_1}\!\cdot \rho^{x_1}\cdot
  \left(1+\frac{T_1}{T_2}\cdot \rho\right)^{\,-(x_1+\,x_2+2)}\cdot
  \frac{T_1}{T_2} \, \mbox{d}\rho \\
  &=&  \frac{(x_1+x_2+1)!}{x_1!\,x_2!} \cdot
  T_1^{x_1}\cdot  T_2^{-x_1}\cdot T_2^{x_1+x_2+2}
  \cdot \rho^{x_1}\cdot
  \left(T_2 + T_1\!\cdot\! \rho\right)^{\,-(x_1+\,x_2+2)}\cdot
  \frac{T_1}{T_2} \, \mbox{d}\rho \\
  &=&  \frac{(x_1+x_2+1)!}{x_1!\,x_2!} \cdot
   T_1^{x_1+1}\cdot  T_2^{x_2+1}  \cdot \rho^{x_1}\cdot
   \left(T_2 + T_1\!\cdot\! \rho\right)^{\,-(x_1+\,x_2+2)}  \, \mbox{d}\rho\\
 f(\rho\,|\,\ldots) &=&  \frac{(x_1+x_2+1)!}{x_1!\,x_2!} \cdot
   T_1^{x_1+1}\cdot  T_2^{x_2+1}  \cdot \rho^{x_1}\cdot
   \left(T_2 + T_1\!\cdot\! \rho\right)^{\,-(x_1+\,x_2+2)}\,.   
\end{eqnarray*}
}
\begin{eqnarray}
  f(\rho\,|\,x_1,T_1,x_2,T_2)\!\!  &=& \!\!
  \frac{(x_1+x_2+1)!}{x_1!\,x_2!} \!\cdot\!  T_1^{\,x_1+1}\!\cdot\!
  T_2^{\,x_2+1}\!\cdot\!
  \rho^{\,x_1}\!\cdot\! (T_2+ T_1\cdot \rho)^{\,-(x_1+\,x_2+2)}\!,\ \ \ \ \ \ 
\label{eq_pdf_rho_r}
\end{eqnarray}
from which we re-obtain Eqs.~(\ref{eq:pdf_rho_closed}) and
(\ref{eq:pdf_rho_closed_again})
in the special case $T_1=T_2$, as it has to be.
Mode, expected value and standard deviation can be obtained
quite easily from Eqs.~(\ref{eq_mode_rho})-(\ref{eq:eq:sigma_rho}),
just noting that
\begin{eqnarray*}
  \rho = \frac{r_1}{r_2} &=& \frac{\lambda_1/T_1}{\lambda_2/T_2}
  = \frac{\lambda_1}{\lambda_2}\cdot \frac{T_2}{T_1} 
  = \frac{T_2}{T_1} \cdot \rho_\lambda\,,
\end{eqnarray*}
and then 
\begin{eqnarray}
  \mbox{mode}(\rho) &=&  \frac{T_2}{T_1} \cdot  \mbox{mode}(\rho_\lambda) =
  \frac{T_2}{T_1} \cdot \frac{x_1}{x_2+2}
  = \frac{x_1/T_1}{(x_2+2)/T_2} \\
  && \nonumber \\
  \mbox{E}(\rho) &=&  \frac{T_2}{T_1} \cdot  \mbox{E}(\rho_\lambda) =
  \frac{T_2}{T_1} \cdot \frac{x_1+1}{x_2} =
  \frac{(x_1+1)/T_1}{x_2/T_2}   \hspace{2.1cm}(\mathbf{x_2>0})
  \label{eq:E_rho_ModelA}\\
  \sigma(\rho) &=& \frac{T_2}{T_1}\cdot \sigma(\rho_\lambda) 
  = \frac{T_2}{T_1}\cdot
  \sqrt{ \frac{x_1+1}{x_2}\cdot\left(\frac{x_1+2}{x_2-1} -
    \frac{x_1+1}{x_2}\right)}  \hspace{0.6cm}(\mathbf{x_2>1})\ \ \
   \label{eq:sigma_rho_ModelA}
\end{eqnarray}
that we can rewrite in a more compact form, in terms of
 $\mu_\rho\equiv \mbox{E}(\rho)$, as
\begin{eqnarray}
 \sigma(\rho) &=& \sqrt{\mu_\rho\cdot
    \left(\frac{T_2}{T_1}\cdot\frac{x_1+2}{x_2-1} - \mu_\rho\right)}\,,
\hspace{2.1cm} (\mathbf{x_2>1}) 
\end{eqnarray}  
Some examples are provided in Figs.~\ref{fig:pdf_rho_r_low} and
\ref{fig:pdf_rho_r_high},
for low and relatively high
numbers of counts, respectively.
Each plot shows
both the curve of the pdf, calculated with the closed formulae
just derived,
and the histogram of Monte Carlo simulation
(the script to reproduce these plots is given in Appendix B.3).
The value of mode, expected value and standard
deviation calculated from exact formulae are reported too,
together with `mean' and `std' (`empirical standard deviation')
evaluated from the sampling. The excellent agreement can be
considered a cross check of the exact formulae, derived above for
the purpose.

The counts and the measuring times have been chosen such that
$(x_1/T_1)/(x_2/T_2)$ are equal to one in all cases.
Therefore the plots are comparable to those of
Figs.~\ref{fig:pdf_lambda_x1_x2_MC_low}
and \ref{fig:pdf_lambda_x1_x2_MC_high} reporting $\rho_\lambda$
for several values of $\lambda_1=\lambda_2$
(but in that case all summaries were evaluated  from sampling,
having, at that stage of the work, not yet derived
the closed formulae of interest). As we can again see,
for small numbers of counts the distribution of the ratio of rates
is strongly asymmetric, with mode and expected value
systematically below and above, respectively,
the ratio calculated naively as $(x_1/T_1)/(x_2/T_2)$.
This value is reached asymptotically, as we see in
Fig.~\ref{fig:pdf_rho_r_high}, and as expected by the fact
that for  high numbers of counts we get
\begin{eqnarray}
  \mbox{mode}(\rho) = \frac{x_1/T_1}{(x_2+2)/T_2}\,
 & \xrightarrow[\ x_2\gg 2\ ]{} &  
  \frac{x_1/T_1}{x_2/T_2} 
  \\
  && \nonumber \\
  \mbox{E}(\rho) =
  \frac{(x_1+1)/T_1}{x_2/T_2}\,
  & \xrightarrow[\ x_1\gg 1\ ]{} & 
  \frac{x_1/T_1}{x_2/T_2} \\
  \sigma(\rho) =  \frac{T_2}{T_1}\cdot
  \sqrt{ \frac{x_1+1}{x_2}\cdot\left(\frac{x_1+2}{x_2-1} -
    \frac{x_1+1}{x_2}\right)}
 &  \xrightarrow[\ x_1,x_2\rightarrow\infty \ ]{} & 0\,.
\end{eqnarray}  

\subsection{More on the ratio of Gamma distributed variables}
Keeping the notation $\rho_z$ for the ratio of the generic
Gamma distributed variables $Z_1$ and $Z_2$, the pdf
of Eq.~(\ref{eq:pdf_rho_closed_general}) can be further 
simplified reminding that the {\em beta}
special function (or {\em Euler integral of the
first kind}~\cite{WikiBetaFunction}), defined as
\begin{eqnarray}
\mbox{ B}(r,s) &=& \int_0^1 t^{\,r-1} \cdot (1-t)^{s-1}\,\mbox{d}t\,, 
\end{eqnarray}
can be written as
\begin{eqnarray}
\mbox{ B}(r,s) &=& \frac{\Gamma(r)\cdot \Gamma(s)}{\Gamma(r+s)}\,.
\end{eqnarray}
We can then rewrite the combination of three gamma
functions appearing in Eq.~(\ref{eq:pdf_rho_closed_general})
as $1/\mbox{B}(\alpha_1,\alpha_2)$,
thus getting 
\begin{eqnarray}
f(\rho_z\,|\,\alpha_1,\beta_1,\alpha_2,\beta_2)  \!\! &=&\!\!
\frac{1}{\mbox{B}(\alpha_1,\alpha_2)}
       \cdot \beta_1^{\,\alpha_1}\cdot \beta_2^{\,\alpha_2}
      \cdot \rho_z^{\,\alpha_1-1}\!\cdot\!
      (\beta_2+\rho_z\!\cdot\beta_1)^{\,-(\alpha_1+\,\alpha_2)}\,.
\end{eqnarray}
As far as mode, expected value and variance are concerned,
they can be obtained, without direct calculations,
just transforming those of $\rho=r_1/r_2$, seen above,
remembering that, starting from a flat prior,
$r_i \sim \mbox{Gamma}(\alpha_i=x_i+1, \beta_i=T_i)$.
We get then
\begin{eqnarray}
  \mbox{mode}(\rho_z) &=&  \frac{\beta_2}{\beta_1} \cdot
  \frac{\alpha_1-1}{\alpha_2+1} \\
  && \nonumber \\
  \mbox{E}(\rho_z) &=&  
  \frac{\beta_2}{\beta_1} \cdot \frac{\alpha_1}{\alpha_2-1}
  \hspace{5.3cm}(\mathbf{\alpha_2>1}) \\
  && \nonumber \\
  \mbox{Var}(\rho_z) &=& 
 \frac{\beta_2^2}{\beta_1^2}\cdot
  \left[ \frac{\alpha_1}{\alpha_2-1}\cdot\left(\frac{\alpha_1+1}{\alpha_2-2} -
    \frac{\alpha_1}{\alpha_2-1}\right)\right]
    \hspace{0.85cm}(\mathbf{\alpha_2>2}).
\end{eqnarray} 
Moreover, just for completeness, let us mention 
the special case $\beta_1=\beta_2=1$, that
written for the generic variable $X$, becomes
\begin{eqnarray}
f(x\,|\,\alpha_1,\alpha_2)  \!\! &=&\!\!
\frac{1}{\mbox{B}(\alpha_1,\alpha_2)}
      \cdot \rho_z^{\,\alpha_1-1}\!\cdot\!
      (1+\rho_z)^{\,-(\alpha_1+\,\alpha_2)},
\end{eqnarray}
`known' (certainly not to me before I was
attempting to write these subsection)
as {\em Beta prime distribution}\,\cite{WikiBetaPrime},
with parameters $\alpha$ and $\beta$:
\begin{eqnarray}
f(x\,|\,\alpha,\beta)  \!\! &=&\!\!
\frac{1}{\mbox{B}(\alpha,\beta)}
      \cdot x^{\,\alpha-1}\!\cdot\!
      (1+x)^{\,-(\alpha+\,\beta)}\,.
\end{eqnarray}
The name of the distribution is clearly due to the special function
resulting from normalization. It is `prime'
in order to distinguish it 
from the more famous (and more important as far as
practical applications are concerned)
{\em Beta distribution} which arises quite `naturally' when inferring
the parameter $p$ of the Bernoulli trials, in the light of
$x$ successes in $n$ trials (essentially the original problem
tackled by Bayes~\cite{Bayes} and Laplace~\cite{Laplace_PC}),
and then used as conjugate
prior of the binomial distribution
(see e.g. Ref.~\cite{ConjugatePriors} as well as
Ref.~\cite{sampling} for practical applications).
The Beta prime distribution
is actually what has been independently derived in
Sec.~\ref{ss:pdf_rho_closed} to describe $\rho_\lambda$, although
the beta special function was not used there,
nor in Sec.~\ref{sec:ratio_gamma}.

\section{Direct inference of the rate ratio
  \texorpdfstring{$\rho$}{rho}
  (and of \texorpdfstring{$r_2$}{r2})}\label{sec:inference_rho}
\begin{figure}[t]
\begin{center}
  \epsfig{file=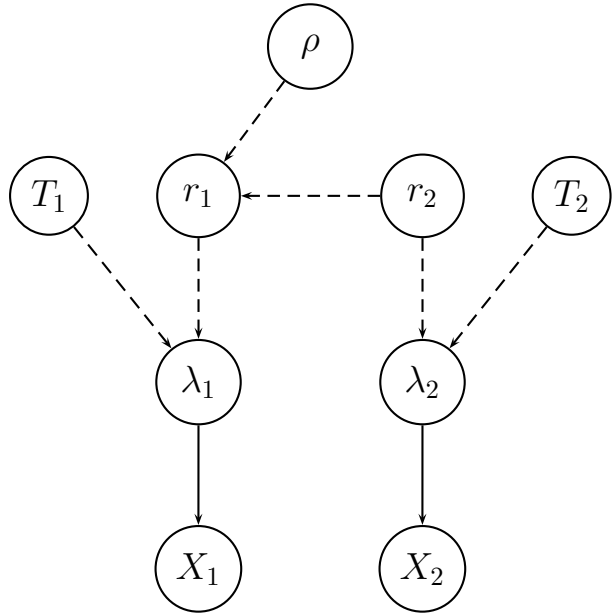,clip=,width=0.53\linewidth}
    \\ \mbox{}\vspace{-0.8cm}\mbox{}
\end{center}
\caption{\small \sf Alternative graphical model to that of  Fig.~\ref{fig:BN_r_x_rho}.}
\label{fig:inf_r2_rho}
\end{figure}

We have remarked several times  that $r_1$ and $r_2$ are inferred
from the observed numbers of events $X_1$ and $X_2$ (we assume $T_1$
and $T_2$ can be exactly known), and that
the possible values of their ratio $\rho$ are
successively evaluated (`deduced')
from each possible pair of values of the rates. 
This logical scheme is represented by the graphical model
of Fig.~\ref{fig:BN_r_x_rho}. But this is not the only way to approach
the problem.
An alternative model is shown in Fig.~\ref{fig:inf_r2_rho},
in which the node $\rho$ appears `at the top' of the network
and it is then really {\em inferred}\,\footnote{For those who
  have doubts about the meaning of `deduction' and  `induction',
  Ref.~\cite{PlatoPlatypus} is highly recommended
  (and they will discover that Sherlock Holmes
  was indeed not {\em deducing} explanations).}
(indeed also $r_2$ is `at the top', having above it
no {\em parents nodes}
from which to depend).

Writing one diagram or another one is not just a question
of drawing art. Indeed, the network reflects the supposed causal model
(`what depends from what') and therefore 
the choice of the model can have an effect on the results.
It is therefore important to understand in what they differ.
In the model of Fig.~\ref{fig:BN_r_x_rho} the rates $r_1$
and $r_2$ assume a primary role. We infer their values and,
{\em as a byproduct}, we get $\rho$. In this new model, instead,
it is $\rho$ to have a primary role, together with one of the
two rates (they cannot be both at the same level because
there is a constraint between the three quantities).
Our choice to make $r_1$ depend on $r_2$ is due to the fact
that $r_2$, appearing at the denominator,
 can be seen as a 
`baseline' to which the other rate is referred
(obviously, here $r_1$ and $r_2$ are just names, and
therefore the choice of their role depend on their meaning).

The strategy to get $f(\rho\,|\,\ldots)$ is then different, being
this time $\rho$ directly inferred using the Bayes theorem applied to the
entire network. 
A strong advantage of this second model
is  that, as we shall see, its prior can be factorized
(see also Ref.~\cite{sampling}, especially
Appendix A there, in which there is a summary
of the formulae we are going to use). 

In analogy to what has been done
in detail in Ref.~\cite{sampling}, the pdf of $\rho$
is obtained in two steps: first infer
$f(\rho,r_1,r_2\,|\,x_1,x_2,T_1,T_2)$;
then get the pdf of $\rho$ by marginalization.
For the first step we need to write down the joint distribution of all
variables in the network (apart from $T_1$ and $T_2$ which we consider
just as fixed parameters, having usually negligible uncertainty)
using the most convenient {\em chain rule},
obtained navigating  bottom up the graphical model.
Indicating, as in Ref.~\cite{sampling}, with $f(\ldots)$ the joint pdf
of all relevant variables, we obtain from the chain rule
\begin{eqnarray}
  f(\ldots) &=&
  f(x_2\,|\,r_2,T_2)\cdot f_0(r_2)\cdot f(x_1\,|\,r_1,T_1)
  \cdot f(r_1\,|\,r_2,\rho)\cdot f_0(\rho)\,\,
  \label{eq:chain_rule_model_B}
\end{eqnarray}
from which we can get, besides a normalization constant,
the pdf's of interest as
\begin{eqnarray}
  f(\rho\,|\,x_1,T_1,x_2,T_2)  &\propto&
  \int_0^\infty\!\!\!\int_0^\infty\!   f(\ldots) \,
  \mbox{d}r_1\, \mbox{d}r_2 \\
    f(r_2\,|\,x_1,T_1,x_2,T_2)  &\propto&
  \int_0^\infty\!\!\!\int_0^\infty\!   f(\ldots) \,
  \mbox{d}\rho\, \mbox{d}r_1\,, 
\end{eqnarray}
or the joint pdf $f(r_2,\rho\,|\,x_1,T_1,x_2,T_2)$,
integrating only over $r_1$.
Using explicit expressions of the pdf's, of which
$f(r_1\,|\,r_2,\rho)$ is just the Dirac delta
$\delta(r_1\!-\!\rho\cdot r_2)$,\footnote{It is interesting
  to note that there is an alternative 
  way to get Eq.~(\ref{eq:propagazione_delta}),
  starting from the joint distribution
  $f(\rho_\lambda,\lambda_1,\lambda_2\,|\,x_1,x_2)$
  and then marginalizing it.
  In fact, using the chain rule, 
  we get
  \begin{eqnarray*}
    f(\rho_\lambda,\lambda_1,\lambda_2\,|\,x_1,x_2) &=&
     f(\rho_\lambda\,|\,\lambda_1,\lambda_2)\cdot
     f(\lambda_1\,|\,x_1,x_1)\cdot  f(\lambda_2\,|\,x_1,x_2)\\
     &=& 
    f(\rho_\lambda\,|\,\lambda_1,\lambda_2)\cdot
    f(\lambda_1\,|\,x_1)\cdot  f(\lambda_2\,|\,x_2)\,.
  \end{eqnarray*}
  But, being $\rho_\lambda$ deterministically related
  to $\lambda_1$ and $\lambda_2$,
  $f(\rho_\lambda\,|\,\lambda_1,\lambda_2)$ is nothing but
  $\delta(\rho_\lambda\! -\! \lambda_1/\lambda_2)$
  (see also other examples in Ref.~\cite{sampling}).
  Integrating then $f(\rho_\lambda,\lambda_1,\lambda_2\,|\,x_1,x_2)$
  over  $\lambda_1$ and $\lambda_2$ we get 
  Eq.~(\ref{eq:propagazione_delta}).
  \label{fn:uso_delta_come_pdf}
}
and ignoring multiplicative factors,
we can then only focus on 
\begin{eqnarray}
 \tilde f(\ldots) \!\!&\propto& \!\! r_2^{x_2} \cdot e^{-T_2\,r_2}\cdot f_0(r_2)
 \cdot r_1^{x_1} \cdot e^{-T_1\,r_1}\cdot \delta(r_1-\rho\cdot r_2)\cdot f_0(\rho)\,,
 \label{eq:f_tilde}
\end{eqnarray}
having indicated by $\tilde f()$ the unnormalized pdf.

\subsection{Inferred distribution of
  \texorpdfstring{$\rho$}{rho}}\label{ss:inferred_pdf_rho_model_B}
\begin{figure}
\begin{center}
\begin{tabular}{cc}
  \epsfig{file=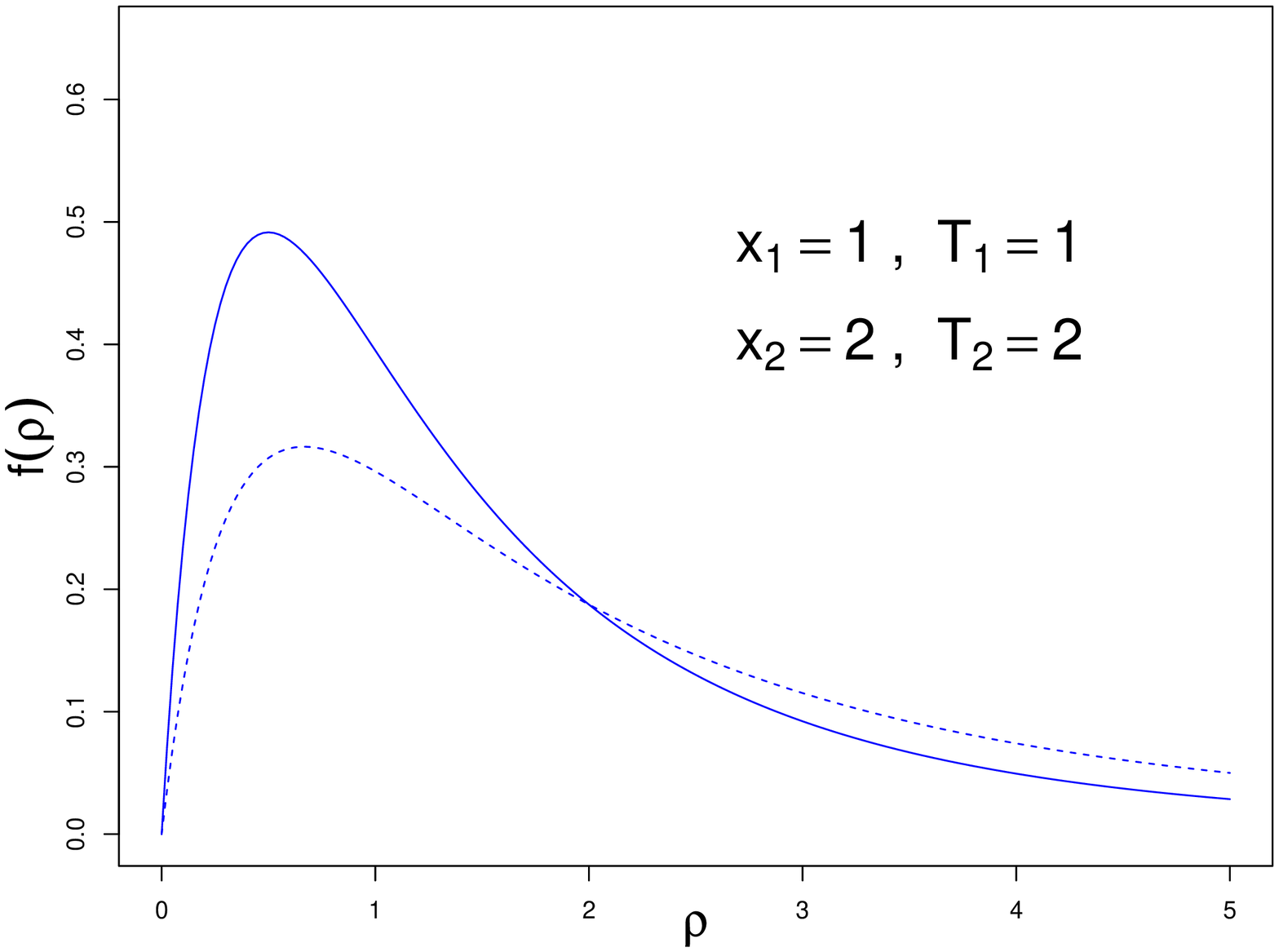,clip=,width=0.5\linewidth}
  &\!\!\!\! \epsfig{file=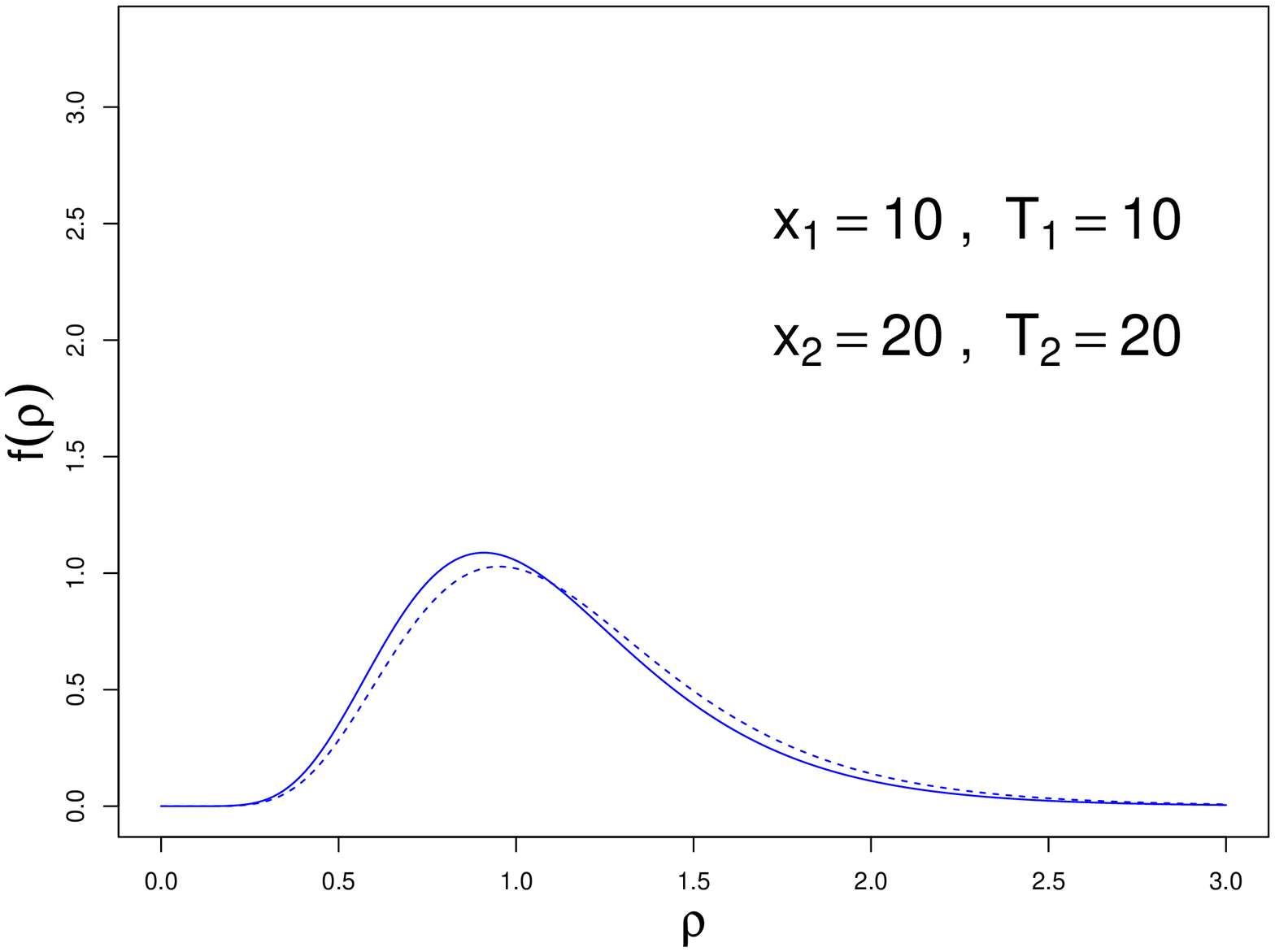,clip=,width=0.5\linewidth}\\
  \epsfig{file=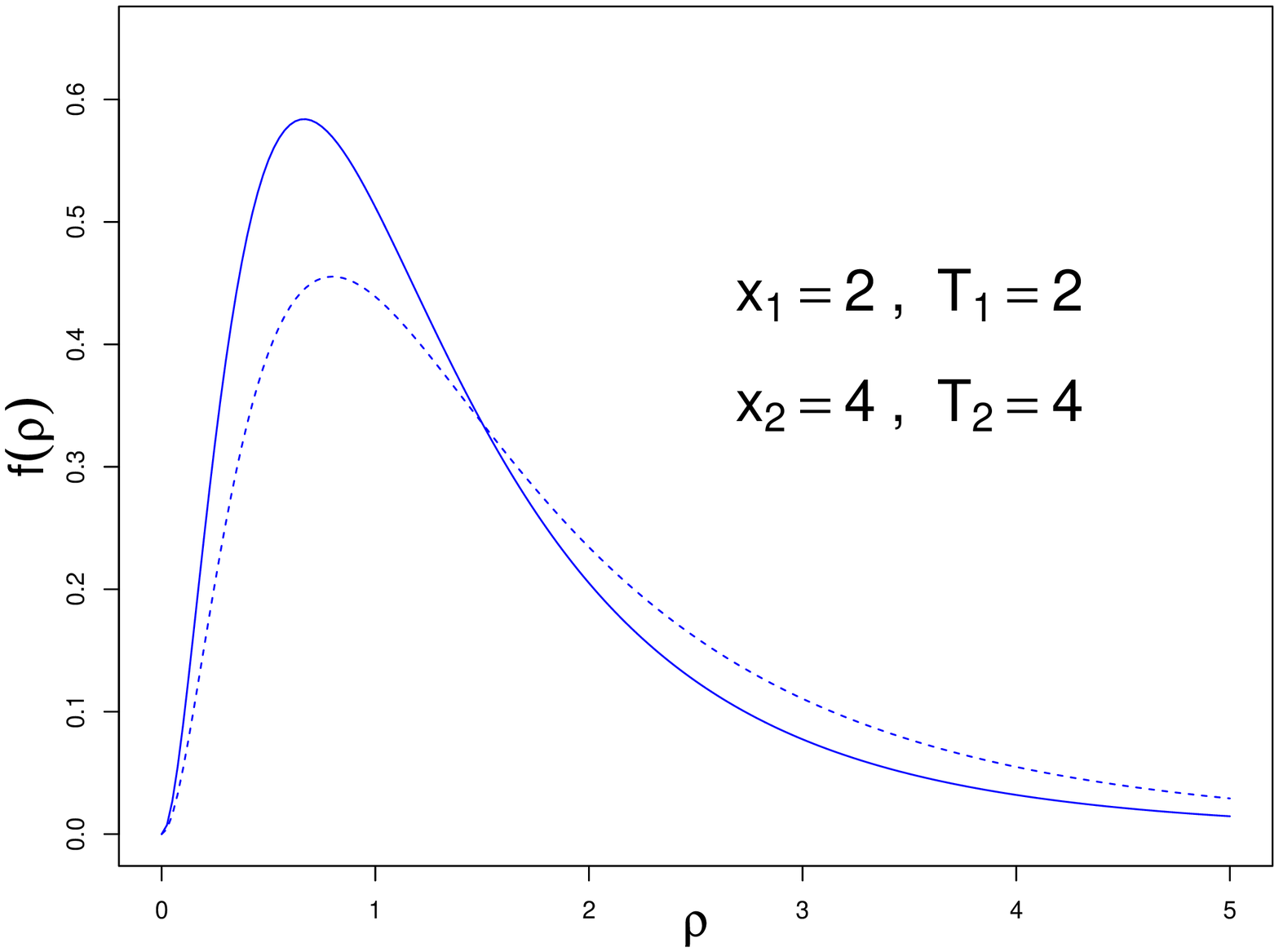,clip=,width=0.5\linewidth}
  &\!\!\!\! \epsfig{file=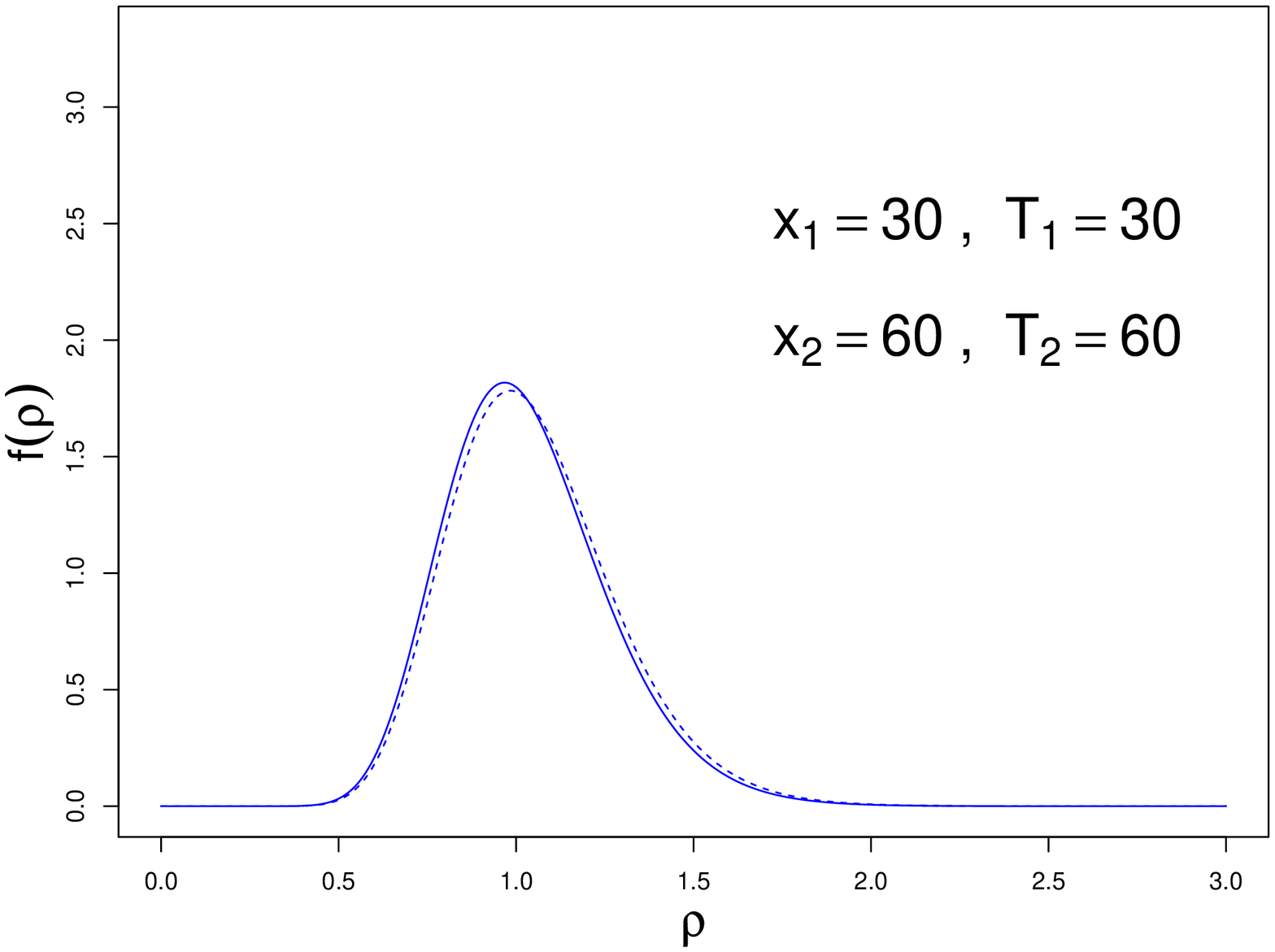,clip=,width=0.5\linewidth} \\
 \epsfig{file=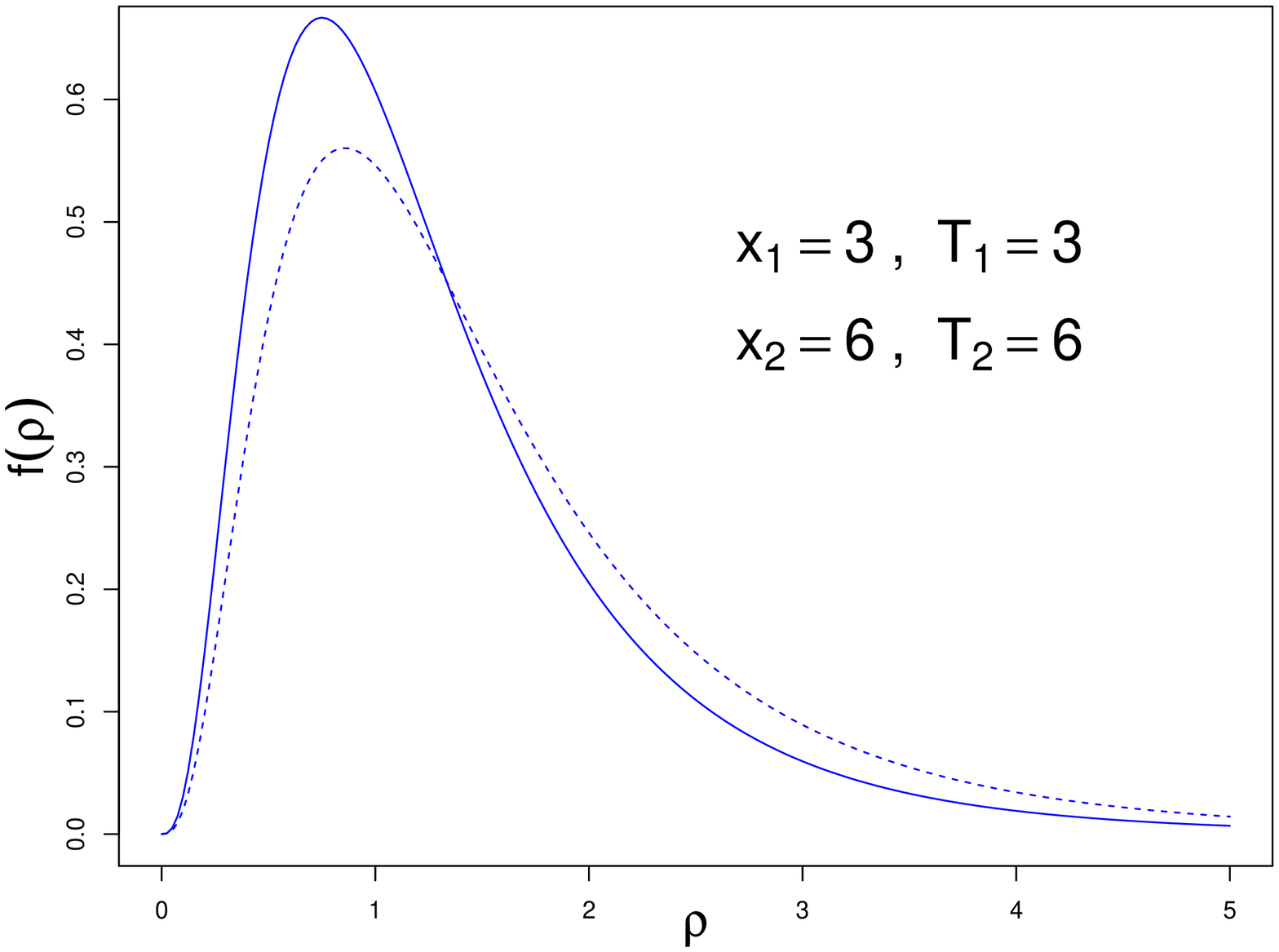,clip=,width=0.5\linewidth}
  &\!\!\!\! \epsfig{file=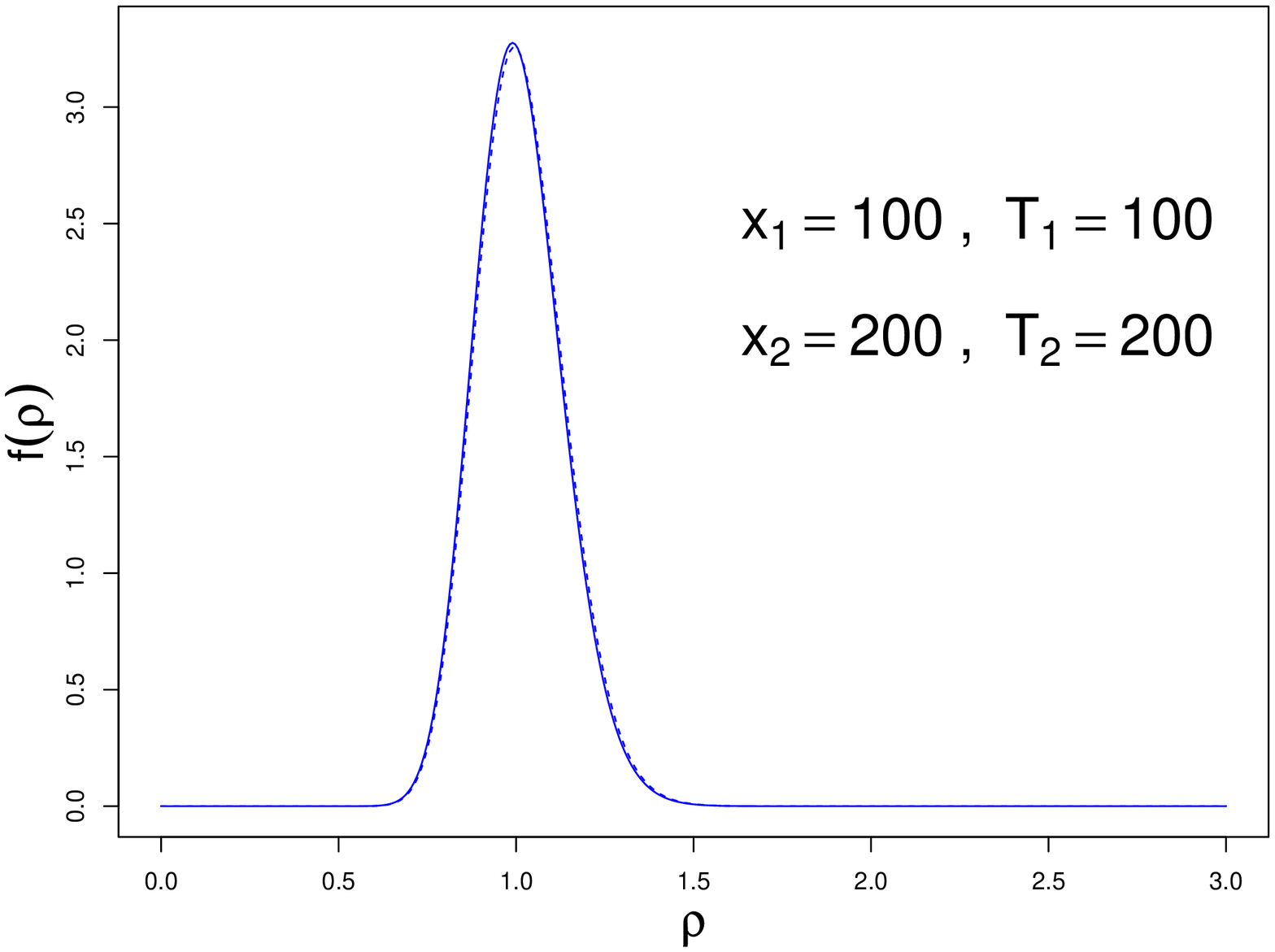,clip=,width=0.5\linewidth} 
\end{tabular}  
\end{center}
\caption{\small \sf Dependence of the inference of $\rho$ from the
  priors. Solid lines: flat priors on $r_1$ and $r_2$, as in
  Figs.~\ref{fig:pdf_rho_r_low} and \ref{fig:pdf_rho_r_high},
  following from the causal model depicted in Fig.~\ref{fig:BN_r_x_rho}. 
  Dashed lines: flat priors on $r_2$ and $\rho$
(causal model of Fig.~\ref{fig:inf_r2_rho}.)} 
\label{fig:pdf_rhp_two_models}
\end{figure}

Let us start with pdf of $\rho$. Our starting point is
\begin{eqnarray}
   f(\rho\,|\,x_1,T_1,x_2,T_2) \!\! &\propto& \!\!\!
   \int_0^\infty\!\!\!\int_0^\infty\!   \tilde f(\ldots)
   \, \mbox{d}r_1\, \mbox{d}r_2\,,  \label{eq:inf_rho_starting} 
\end{eqnarray}
from which it follows
\begin{eqnarray}
    f(\rho\,|\,x_1,T_1,x_2,T_2) \!\!
   &\propto& \!\!\!
  \int_0^\infty\!\!\!\int_0^\infty\!
  r_2^{x_2}  e^{-T_2\,r_2}
  \cdot r_1^{x_1}\! \cdot\! e^{-T_1\,r_1}\cdot \delta(r_1\!-\!\rho\cdot r_2)\!\cdot\!
  f_0(r_2)\!\cdot\! f_0(\rho)\,
  \mbox{d}r_1\, \mbox{d}r_2 \nonumber \\
  &&  \\
  &\propto& 
 \left[ \int_0^\infty\!
  r_2^{x_2}  e^{-T_2\,r_2}
  \cdot (\rho\cdot r_2)^{x_1}\! \cdot\! e^{-T_1\,\rho\,r_2}\cdot
  f_0(r_2)\, \mbox{d}r_2 \right]  \cdot\! f_0(\rho) \\
  &\propto&  \left[\, \rho^{x_1}\cdot
   \int_0^\infty\! r_2^{x_1+x_2}\cdot e^{-(T_2+T_1\cdot\rho)\cdot r_2}\cdot f_0(r_2)
   \, \mbox{d}r_2 \right]  \cdot\! f_0(\rho)\,,
\end{eqnarray}
in which we have explicitly factorized $f_0(\rho)$.  
Again, besides $f_0(r_2)$, we recognize in the integrand 
something proportional to a Gamma pdf. If we then
model also  $f_0(r_2)$ by a Gamma of parameters $\alpha_0$ and $\beta_0$,
and again neglect
irrelevant factors, we get
\begin{eqnarray}
  f(\rho\,|\,x_1,T_1,x_2,T_2)\!\! &\propto & \!\!
   \left[\, \rho^{x_1}\cdot
     \int_0^\infty\! r_2^{x_1+x_2}\cdot e^{-(T_2+T_1\cdot\rho)\cdot r_2}\cdot r_2^{\alpha_0-1}
     \cdot e^{-\beta_0\,r_2}
     \, \mbox{d}r_2 \right]  \cdot\! f_0(\rho)\ \ \ \\
  \!\! &\propto & \!\!
   \left[\, \rho^{x_1}\cdot
     \int_0^\infty\! r_2^{\alpha_0+x_1+x_2-1}\cdot e^{-(\beta_0+T_2+T_1\cdot\rho)\cdot r_2}
     \, \mbox{d}r_2\, \right]  \cdot\! f_0(\rho)\,.
\end{eqnarray}
Indicating, in analogy to what done to obtain
Eq.~(\ref{eq:step_to_pdf_rho_closed_general}), 
the power of $r_2$ as
$\alpha_* - 1 = \alpha_0+ x_1+x_2-1$, and the factor multiplying $r_2$ at the exponent
as $\beta_*=\beta_0+T_2+T_1\cdot\rho$, we get
\begin{eqnarray}
  f(\rho\,|\,x_1,T_1,x_2,T_2)\!\! &\propto & \!\! 
  \left[\, \rho^{x_1}\cdot
    \int_0^\infty\!  r_2^{\alpha_*-1}\cdot e^{-\beta_*\cdot r_2}
    \, \mbox{d}r_2 \right]  \cdot\! f_0(\rho)\\
 \!\!   &\propto & \!\! 
  \left[\, \rho^{x_1}\cdot
    \frac{\Gamma(\alpha_*)}{\beta_*^{\alpha_*}}
    \right]  \cdot\! f_0(\rho) \\
  \!\!   &\propto & \!\! 
  \bigg[\, \rho^{x_1}\!\cdot\!
    (\beta_0\!+\!T_2\!+\!T_1\!\cdot\!\rho)^{-(\alpha_0+x_1+x_2)}
    \bigg]  \cdot\! f_0(\rho)\,.\ \ \ \ \ \
  \label{eq:effectiveLikelihood}
\end{eqnarray}
What is interesting with this result is that we can
consider the term inside the square brackets as an effective likelihood
(remember that multiplicative factors are irrelevant),
and therefore we can rewrite Eq.~(\ref{eq:effectiveLikelihood}) as
\begin{eqnarray}
  f(\rho\,|\,x_1,T_1,x_2,T_2) &\propto &
  {\cal L}(\rho\,;x_1,T_1,x_2,T_2,\alpha_0,\beta_0)\cdot  f_0(\rho)\,.
  \label{eq:factorization_effective_Likelihood}
\end{eqnarray}
For this reason {\em we can serenely proceed assuming 
a  flat prior} about $\rho$, because we can reshape
in a second step the result (see Ref.~\cite{sampling} for details).
So, assuming $f_0(\rho)=k$
and comparing the expression inside the square bracket
of  Eq.~(\ref{eq:effectiveLikelihood})
with
Eq.~(\ref{eq:pdf_rho_closed_general}) we get
the normalization just by analogy, thus getting
\begin{eqnarray}
  f(\rho\,|\,x_1,T_1,x_2,T_2)\!\! & = & \!\!
   \frac{\Gamma(\alpha_0+x_1+x_2)}{\Gamma(x_1+1)\cdot\Gamma(\alpha_0+x_2-1)}
    \cdot T_1^{x_1+1}\cdot (\beta_0+T_2)^{\alpha_0+x_2-1} \cdot \nonumber\\
  &&  \ \  \rho^{x_1} \cdot (\beta_0 + T_2+T_1\cdot\rho)^{-(\alpha_0+x_1+x_2)}
 \,,
\end{eqnarray}
or
\begin{eqnarray}
    f(\rho\,|\,x_1,T_1,x_2,T_2)\!\! & = & \!\!
    \frac{ T_1^{\,x_1+1}\!\cdot\! (\beta_0+T_2)^{\alpha_0+x_2-1}}
         {\mbox{B}(x_1\!+\!1,\alpha_0\!+\!x_2\!-\!1)}
     \cdot 
     \rho^{x_1} \cdot (\beta_0 + T_2+T_1\cdot\rho)^{-(\alpha_0+x_1+x_2)},
     \nonumber \\
     &&
    \label{eq:pdf_rho_f0(rho)=k_priors_r2}
\end{eqnarray}
that, for a flat prior about $r_2$, i.e. $\alpha_0=1$ and $\beta_0=0$, becomes
\begin{eqnarray}
  f(\rho\,|\,x_1,T_1,x_2,T_2)\!\! & = & \!\!
  \frac{\Gamma(x_1+x_2+1)}{\Gamma(x_1+1)\cdot\Gamma(x_2)}
  \cdot T_1^{\,x_1+1}\cdot T_2^{\,x_2} \cdot \rho^{x_1} \cdot ( T_2+T_1\cdot\rho)^{-(x_1+x_2+1)}
  \nonumber \\
  && \\
   \!\! & = & \!\!
  \frac{(x_1+x_2)!}{x_1!\cdot (x_2-1)!}
  \cdot T_1^{\,x_1+1}\cdot T_2^{\,x_2} \cdot \rho^{x_1} \cdot ( T_2+T_1\cdot\rho)^{-(x_1+x_2+1)}
  \nonumber \\
  && \label{eq:pdf_rho_f0(rho)=k}
\end{eqnarray}
The comparison of this result with
Eq.~(\ref{eq_pdf_rho_r}),
obtained using flat priors for $r_1$ and $r_2$,
is at least  surprising: the structures of the pdf's are the same,
but $x_2$ in  Eq.~(\ref{eq_pdf_rho_r}) 
is replaced by $x_2\!-\!1$ in Eq.~(\ref{eq:pdf_rho_f0(rho)=k}).
Obviously, the two results will coincide for large $x_2$,
and also for small $x_2$ there is not a dramatic
difference, as we can see from Fig.~\ref{fig:pdf_rhp_two_models}.
As far as the summaries of the distribution are concerned, we get
\begin{eqnarray}
  \mbox{mode}(\rho) &=& \frac{x_1/T_1}{(x_2+1)/T_2} \\
  && \nonumber \\
  \mbox{E}(\rho) = \mu_\rho &=& 
  \frac{(x_1+1)/T_1}{(x_2-1)/T_2}   \hspace{4.05cm}(\mathbf{x_2>1}) \\
  && \nonumber \\
  \sigma(\rho) &=&  \sqrt{\mu_\rho\cdot
    \left(\frac{T_2}{T_1}\cdot\frac{x_1+2}{x_2-2} - \mu_\rho\right)}
   \hspace{1.5cm}(\mathbf{x_2>2})\,.
\end{eqnarray}  
%
\mbox{}\vspace{-0.5cm}\mbox{}\\
At this point, instead of taking comfort for the fact that
the differences are irrelevant in practical cases,
or {\em tout court} `rejecting Bayesian methods because
of their dependence of priors', it is interesting to try
to understand the origin of this effect,
certainly related to the priors.

But, before proceeding, let
us not forget that Eq.~(\ref{eq:pdf_rho_f0(rho)=k_priors_r2})
was obtained
assuming a flat prior about $\rho$ and that in that model
this prior can be factorized.
Therefore the more general pdf of the rate ratio for the model of
Fig.~\ref{fig:inf_r2_rho}  is
\begin{eqnarray}
  f(\rho\,|\,x_1,T_1,x_2,T_2)\!\! & = & \!\!
   \frac{1}{\mbox{B}(x_1+1,\alpha_0+x_2-1)}
    \cdot T_1^{\,x_1+1}\cdot (\beta_0+T_2)^{\alpha_0+x_2-1} \cdot \nonumber\\
    &&   \rho^{x_1} \cdot (\beta_0 + T_2+T_1\cdot\rho)^{-(\alpha_0+x_1+x_2)}\cdot
    f_0(\rho) \,,
    \label{eq:pdf_rho_f0(rho)=k_priors_r2_f0rho}
\end{eqnarray}
having only the limitation (but in reality almost irrelevant, given
the flexibility of the Gamma distribution) of depending
on the chosen parametrization for $f_0(r_2)$.

\subsection{Cross-influences of priors}
\begin{figure}
\begin{center}
  \epsfig{file=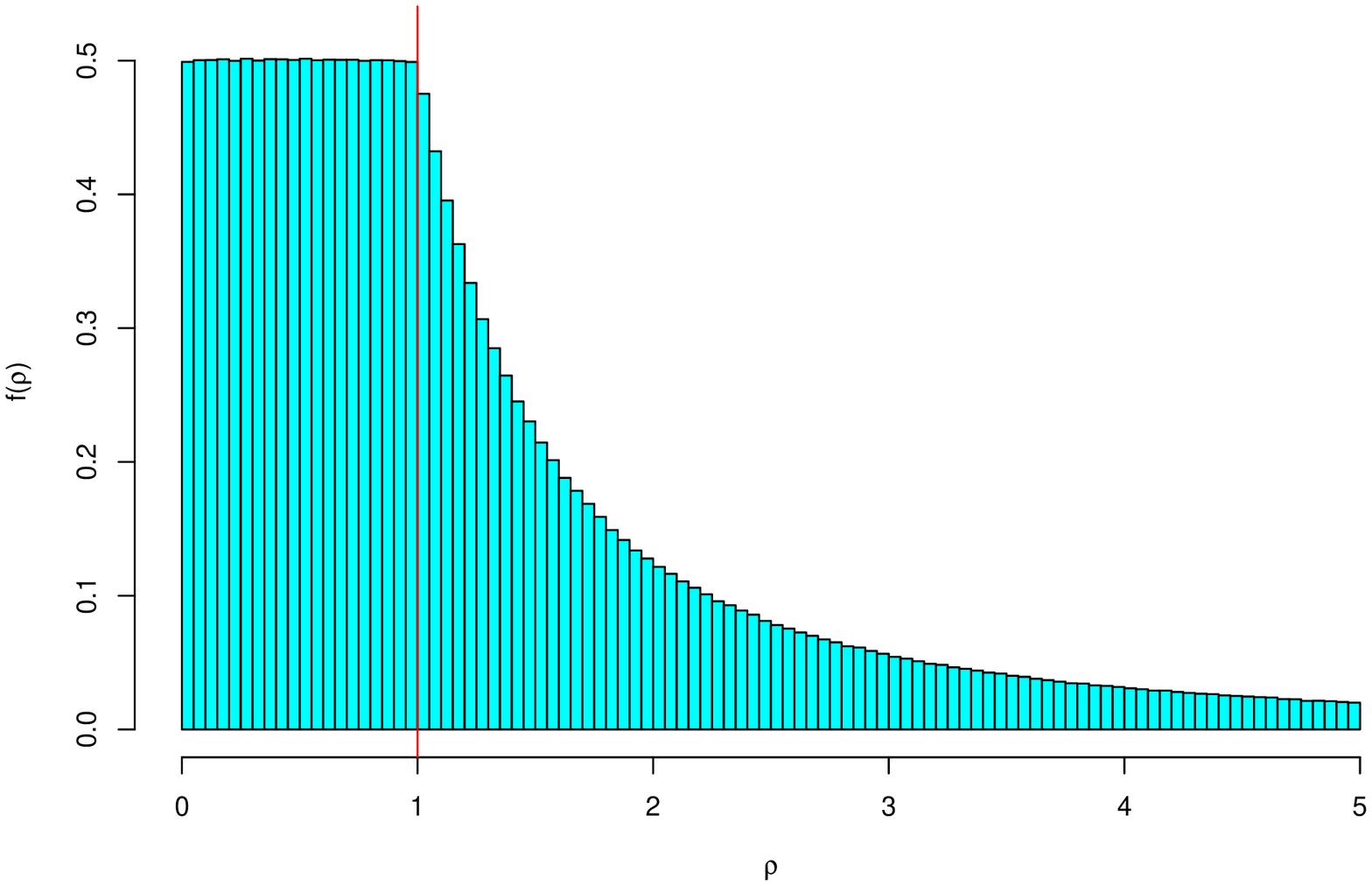,clip=,width=\linewidth}\\
 \epsfig{file=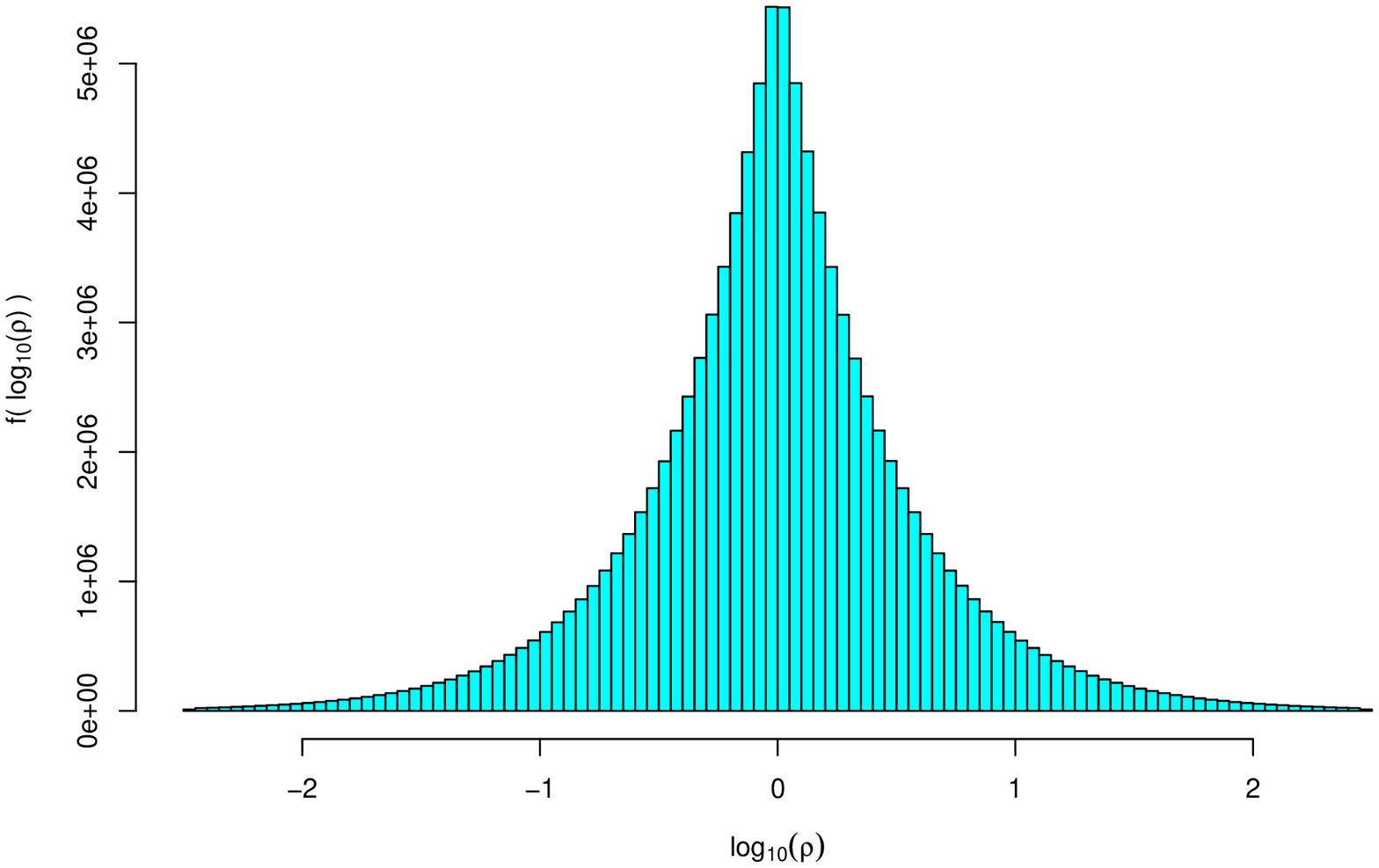,clip=,width=\linewidth}
  \\  \mbox{} \vspace{-1.0cm} \mbox{}
\end{center}
\caption{\small \sf Distribution of $\rho$ implied
  by flat priors on $r_1$ and $r_2$ in linear and log scale. 
  The vertical line in the upper plot
  shows the discontinuity of the distribution at $\rho=1$.}
\label{fig:pdf_rho_r_1_r2_unif}
\end{figure}

One might say that in the first case, that
of Fig.~\ref{fig:BN_r_x_rho}, yielding
Eq.~(\ref{eq_pdf_rho_r}) starting from $f_0(r_1)=f_0(r_2)=k$
there were no priors on $\rho$.
But this is quite not true, because the flat priors
on $r_1$ and $r_2$ impinge on the prior on $\rho$, due
to the relation $\rho=r_1/r_2$. The easiest way to
see what is going on is by Monte Carlo, that is, in R,
\begin{verbatim}
n = 10^7
rM = 100
r1 = runif(n, 0, rM)
r2 = runif(n, 0, rM)
rho = r1/r2
rho.h <- rho[rho<5] 
hist(rho.h, nc=200, col='blue', freq=FALSE)
abline(v=1, col='red')
\end{verbatim}
where the selection of the values below $\rho\!=\!5$ is to
visualize the more interesting region,
shown in the top plot of Fig.~\ref{fig:pdf_rho_r_1_r2_unif}
(a more complete script, which also 
performs the correct normalization of the histogram,
is shown in Appendix B.4). 
The histogram is characterized by
a plateau till $\rho=1$, followed by a slow decreasing.
Curiously, the histogram does not depend on the maximum
value {\tt rM}. 

Although it might be bizarre, this histogram shows in essence
the prior on $\rho$ we have been tacitly assumed,
when flat priors on $r_1$ and $r_2$ were chosen (as a cross check,
the commented instructions of the script of Appendix B.4, executed one by
one, plot the distribution of $r_1$ assuming
a flat prior for $r_2$ and the curious distribution of
the top plot of 
Fig.~\ref{fig:pdf_rho_r_1_r2_unif} for $\rho$).

In order to have a better insight of what is going on,
the bottom plot of the same figure shows the histogram
of $\log_{10}\rho$. The maximum is at $\log\rho=0$
and it decreases symmetrically, exponentially,\footnote{Empirically,
  we can evaluate, taking two points form the
  histogram of Fig.~\ref{fig:pdf_rho_r_1_r2_unif},
  the following exponential:
$f(\log(r))\approx 1.16\times \exp{(-2.30\cdot|\log(r)|)}$.}
as $|\log\rho|$ increases.
This symmetry indicates that the probabilities
to get a value of $\rho$ below or above 1 are the same.
The same conclusion, within the uncertainties
due to sampling, can be drawn from the histogram in linear scale,
since $f(\rho)$ is `about $1/2$' for $0\le \rho \le 1$. Similarly,
from the comparison of the two histograms we can
evaluate, by symmetry arguments, that the probability that
$\rho$ is between 0.1 and 10 is equal to 90\% (exact value, indeed
as we shall see in a while).

It is interesting to get the distribution shown
in the top plot of
Fig.~\ref{fig:pdf_rho_r_1_r2_unif} making a transformation
of variables, as we have done in Eq.~(\ref{eq:propagazione_delta}) and following
equations:\footnote{Perhaps it is worth noticing again
  (see footnote \ref{fn:uso_delta_come_pdf}),
  since this observation seems raised for the first time  in this paper, that
  Eq.~(\ref{eq:propagazione_delta_r1_r2}) can be seen
  not only as an extension to continuous variables of Eq.~(\ref{eq:sommatoria}),
  but also as the joint pdf $f(\rho,r_1,r_2)$ obtained by the
  chain rule, that is
  $f(\rho,r_1,r_2)\! =\! f(\rho\,|\,r_1,r_2)\cdot f(r_1)\cdot f(r_2) $,
  where $f(\rho\,|\,r_1,r_2)\!=\!\delta(\rho\!-\!r_1/r_2)$,
  followed by marginalization.
}
\begin{eqnarray}
  f(\rho) &=& \int_0^{r_M}\!\!\int_0^{r_M}\!\delta(\rho-r_1/r_2)\cdot f(r_1)\cdot
  f(r_2)\,\mbox{d}r_1\,\mbox{d}r_2 \label{eq:propagazione_delta_r1_r2}\\
  &=&  \int_0^{r_M}\!\!\int_0^{r_M}\!r_2\cdot \delta(r_1-\rho\cdot r_2)\cdot
  \frac{1}{r_M}\cdot  \frac{1}{r_M} \,\mbox{d}r_1\,\mbox{d}r_2\,,
\end{eqnarray}
where $r_M$ is the maximum value of $r_1$ and $r_2$.\footnote{If
  you like to reproduce the final result, given by Eq.~(\ref{eq:pdf_rho_integrale}) 
  with {\em Mathematica},
  here are the commands to get it, although the output will appear a
  bit cryptic (but you will recognize the resulting plot):
\mbox{}\vspace{-0.2cm}\mbox{}  
\begin{verbatim} 
  rM = 10
  frho := Integrate[r2*DiracDelta[r1 - rho*r2]/rM^2, {r1, 0, rM}, {r2, 0, rM}]
  frho
  Plot[frho, {rho, 0, 5}]
\end{verbatim}
}

At this point, some care is needed with the limits of the integral
over $r_2$, due to its `natural' upper limit at $r_M$ 
and to that
given by the constraint $\rho\cdot r_2\le 1$, i.e. $r_2\le 1/\rho$.
Therefore, after the trivial integration over $r_1$, we are left with
\begin{eqnarray}
  f(\rho) &=& \frac{1}{r_M^2}\cdot \int_0^{r_2^u}\!r_2 \,\mbox{d}r_2\,, 
\end{eqnarray}
where the upper limit $ r_2^u$ depends on $\rho$ in the following way:
\begin{eqnarray*}
  \rho \le 1 & \longrightarrow & r_2^u = r_M \\
  \rho > 1 & \longrightarrow & r_2^u = r_M/\rho\,.
\end{eqnarray*}
and therefore
\begin{eqnarray*}
  \rho \le 1 & \longrightarrow & f(\rho) =
  \frac{1}{r_M^2}\cdot  \int_0^{r_M}\!\!r_2 \,\mbox{d}r_2 
  =  \frac{1}{r_M^2}\cdot \frac{r_M^2}{2} = \frac{1}{2} \\
  \rho > 1 & \longrightarrow &
   f(\rho) =
   \frac{1}{r_M^2}\cdot  \int_0^{r_M/\rho}\!\!r_2 \,\mbox{d}r_2 
   =  \frac{1}{r_M^2}\cdot \frac{r_M^2}{2\,\rho^2} =
   \frac{1}{2\,\rho^2}\,,  
\end{eqnarray*}
that we summarize as\footnote{We can check
  that $P(1/10\le \rho\le 10)\!=\!9/10$, as previously guessed from symmetry arguments.
}
\begin{eqnarray}
  f\left(\rho\,\,|\,\,f(r_1)\!\!=\!\!\frac{1}{r_M},f(r_2)\!\!=\!\!\frac{1}{r_M}\right)
  &=&  \left\{
 \begin{array}{ll} \frac{1}{2} & \ \ \ \ \ \ (0 \le \rho \le 1) \\
   & \\
                  \frac{1}{2\,\rho^2} &  \ \ \ \ \ \ (\rho >  1) \,,
\end{array}\right.  \label{eq:pdf_rho_integrale}  
\end{eqnarray}
which, indeed, does not depend on the the maximum values
of $r_1$ and $r_2$, as we had already learned playing with
Monte Carlo simulations.\footnote{Curiously, this distribution
  has the property that $f(1/\rho) = f(\rho)$.
  I wonder if there are others.}

For completeness, let also make the game of seeing how
flat priors on $r_2$ and $\rho$ (up to $r_{2_M}$ and $\rho_M$, respectively)
are reflected into $r_1$
in the model of Fig.\ref{fig:inf_r2_rho}:
\begin{eqnarray}
  f(r_1) &=& \int_0^{\rho_M}\!\int_0^{r_{2_M}}\!\delta(r_1-\rho\,r_2)\cdot
  f(\rho)\cdot f(r_2)\,\mbox{d}\rho\,\mbox{d}r_2 \\
  f(r_1) &=& \int_0^{\rho_M}\!\int_0^{r_{2_M}}
  \frac{\delta(\rho-r_1/r_2)}{r_2}\cdot
  \frac{1}{\rho_M}\cdot \frac{1}{r_{2_M}}\,\mbox{d}\rho\,\mbox{d}r_2 \\
  &=& \frac{1}{\rho_M\cdot r_{2_M}}\cdot
  \int_{r_{2_L}}^{r_{2_U}}\frac{1}{r_2}\,\mbox{d}r_2
\end{eqnarray}
where the extremes of integration are $r_{2_L}=r_1/\rho_M$
and  $r_{2_U}=r_{2_M}$.
\newpage
Here is, finally, the pdf of $r_1$, in which we have written
explicitly the conditions:
\begin{eqnarray}
\!f\left(\!r_1\,\,|\,\,f_0(r_2)\!\!=\!\!\frac{1}{r_{2_M}}, 
            f_0(\rho)\!\!=\!\!\frac{1}{\rho_M}\!\right)
  \!\!\!&=&\!\!\!  \frac{1}{\rho_M\cdot r_{2_M}}\!\cdot\!
            \log\left( \frac{r_{2_M}\!\cdot\! \rho_M}{r_1} \right)
            \hspace{0.5cm}(0<r_1\le r_{2_M}\cdot\rho_{M})\ \ \ \
            \nonumber \\
      && \mbox{} \! \! \! \! \!
\end{eqnarray}
\begin{figure}[t]
\begin{center}
  \epsfig{file=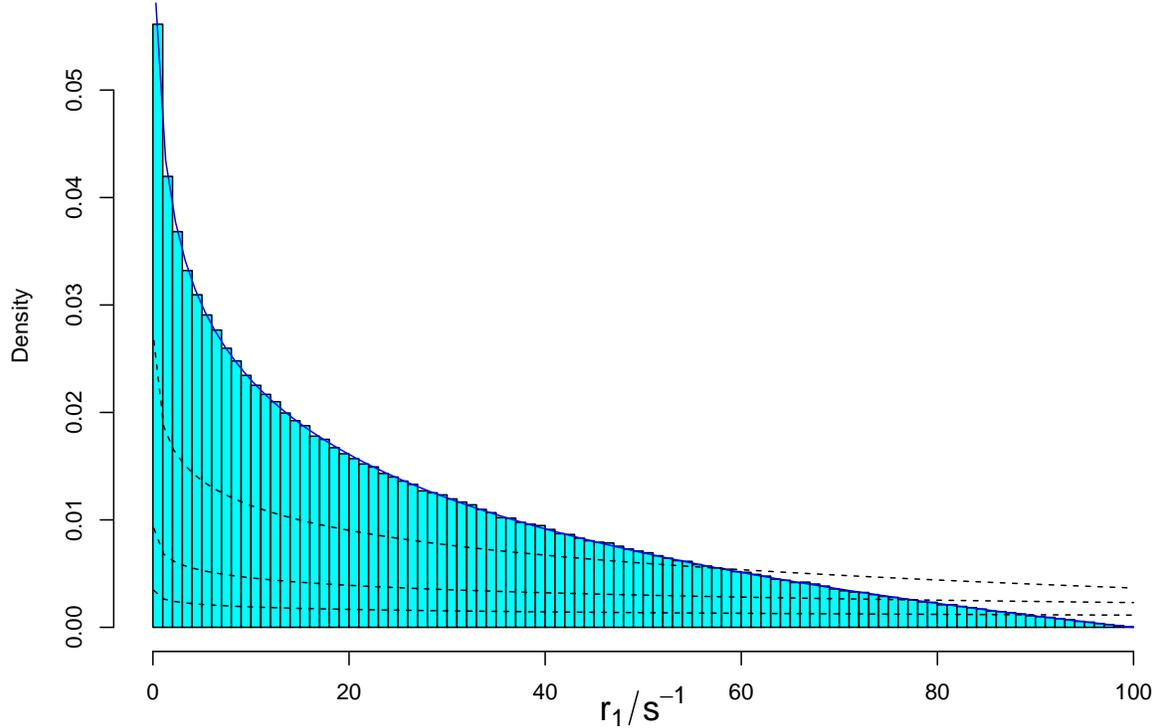,clip=,width=1.025\linewidth}
  \\  \mbox{} \vspace{-1.2cm} \mbox{}
\end{center}
\caption{\small \sf Histogram of $r_1$ implied by 
  priors on $r_2$ and $\rho$ flat up to $\rho_M=r_{2_M}/\mbox{s}^{-1}\!=10$,
  compared to the exact evaluation of the pdf (solid line).
  Dashed lines: pdf of $r_1$ for $\rho_M=r_{2_M}/\mbox{s}^{-1}\!=30, 100, 300$
  (higher to lower, that is from steeper to flatter).}
\label{fig:pdf_r1-flat_rho_r2}
\end{figure}
An example with $\rho_M=1$ and $r_{2_M}=10\,$s$^{-1}$ is 
reported in Fig.~\ref{fig:pdf_r1-flat_rho_r2},
in which the exact pdf (blue solid line)
is compared with the Monte Carlo result. 
The plot also shows the pdf's of $r_1$ for increasing 
maximum values ($\rho_M=r_{2_M}/\mbox{s}^{-1}=30, 100, 300$,
from higher to lower curves).
We see that for $\rho_M\rightarrow \infty$ and
$r_{2_M}\rightarrow \infty$ also the distribution of
$r_1$ becomes flat. 
This is an interesting result, showing that,
contrary to the model of Fig.~\ref{fig:BN_r_x_rho},
the model of Fig.~\ref{fig:inf_r2_rho} can accommodate
in practice flat prior distributions for the three quantities of
interest.\footnote{But when measuring rates, a flat prior
  has more implications than one might think, as discussed in
  chapter 13 of Ref.~\cite{BR}, and therefore
  a full understanding of the physical case is desirable.}

\subsection{Final distributions of
  \texorpdfstring{$r_1$}{r1} and \texorpdfstring{$r_2$}{r2}
  (starting from flat initial distributions of
  \texorpdfstring{$r_2$}{r2} and \texorpdfstring{$\rho$}{rho})}
For completeness, let us also try to get the closed expressions
of  $f(r_1\,|\,x_1,T_1,x_2,T_2)$ and
$f(r_2\,|\,x_1,T_1,x_2,T_2)$, although only under the assumption
of a flat prior of $\rho$. In this case this choice is forced
from the fact that
$f_0(\rho)$ cannot be expressed in term of a conjugate prior
which would then simplify the calculations. For the general case, in fact,
we have to change methods, moving to Markov Chain
Monte Carlo (MCMC), as done e.g.
in Ref.~\cite{sampling} and as it will be sketched in the next section.

In order to get the pdf of $r_1$, we need to restart from
the unnormalized joint distribution (\ref{eq:f_tilde}),
proceeding then like in Eq.~(\ref{eq:inf_rho_starting}),
but this time integrating over $r_2$ and $\rho$
and absorbing the constant priors in the proportionality factor:
\begin{eqnarray}
   f(r_1\,|\,x_1,T_1,x_2,T_2)  &\propto&
   \int_0^\infty\!\!\!\int_0^\infty   \tilde f(\ldots)
   \, \mbox{d}\rho\, \mbox{d}r_2  \label{eq:inf_r1_starting} \\
  &\propto& 
  \int_0^\infty\!\!\!\int_0^\infty\!
  r_2^{x_2} \cdot  e^{-T_2\,r_2}
  \cdot r_1^{x_1}\! \cdot\! e^{-T_1\,r_1}\cdot \delta(r_1\!-\!\rho\cdot r_2)
  \, \mbox{d}\rho\, \mbox{d}r_2 \ \ \  \ \ \ \\
 &\propto& 
  \int_0^\infty\!\!\!\int_0^\infty\!
  r_2^{x_2} \cdot  e^{-T_2\,r_2}
  \cdot r_1^{x_1}  \cdot  e^{-T_1\,r_1}\cdot
  \frac{\delta(\rho\!-\!r_1/r_2)}{r_2}
  \,\mbox{d}\rho\, \mbox{d}r_2 \ \ \  \ \ \ \\
   &\propto&  r_1^{x_1} \cdot  e^{-T_1\,r_1} \cdot \int_0^\infty\!
  r_2^{x_2-1}  e^{-T_2\,r_2}
  \, \mbox{d}r_2\\
    &\propto&  r_1^{x_1} \cdot  e^{-T_1\,r_1}\,,
\end{eqnarray}
thus reobtaining, besides normalization,
Eq.~(\ref{eq:inferenza_r-flat_prior}).

Similarly, we have 
\begin{eqnarray}
   f(r_2\,|\,x_1,T_1,x_2,T_2)  &\propto&
   \int_0^\infty\!\!\!\int_0^\infty   \tilde f(\ldots)
   \, \mbox{d}\rho\, \mbox{d}r_1  \label{eq:inf_r1_starting_2} \\
  &\propto& 
  \int_0^\infty\!\!\!\int_0^\infty\!
  r_2^{x_2} \cdot e^{-T_2\,r_2}
  \cdot r_1^{x_1}\! \cdot\! e^{-T_1\,r_1}\cdot \delta(r_1\!-\!\rho\cdot r_2)
  \, \mbox{d}\rho\, \mbox{d}r_1 \ \ \  \ \ \\
    &\propto& \int_0^\infty\!
  r_2^{x_2} \cdot e^{-T_2\,r_2}
  \cdot (\rho\cdot r_2)^{x_1}\cdot e^{-T_1\,\rho\,r_2}\,  \mbox{d}\rho\\
    &\propto&   r_2^{x_2+x_1} \cdot e^{-T_2\,r_2}
  \cdot  \int_0^\infty\!\! \rho^{x_1}\cdot e^{-T_1\,r_2\,\rho}\,  \mbox{d}\rho \\
     &\propto&   r_2^{x_2+x_1} \cdot e^{-T_2\,r_2}
  \cdot \frac{\Gamma(x_1+1)}{(r_2\,T_1)^{(x_1+1)}} \\
     &\propto&   r_2^{x_2+x_1} \cdot e^{-T_2\,r_2}
 \cdot r_2^{-(x_1+1)} \\
    &\propto&   r_2^{x_2-1} \cdot e^{-T_2\,r_2}\,.
\end{eqnarray}
We see that, differently from $f(r_1\,|\,x_1,T_1,x_2,T_2)$,
the power of $r_2$ is, instead of $x_2$, $x_2\!-\!1$,
that is we get an  effect similar to that found for the
distribution of $\rho$. As a consequence, expected values
and standard deviation of $r_2$ are $x_2/T_2$ and
$\sqrt{x_2}/T_2$, respectively.

\section{Use of MCMC methods to cross-check the closed results
  and to analyze extended models}\label{sec:MCMC}   
So far our models have been rather simple,
missing however several real life complications. For example,
assuming that we do observe the number of counts
due to a Poisson distribution with a given $\lambda=r\cdot T$
clearly implies that we are neglecting {\em efficiency} issues.
In order to include efficiencies
we  need to modify our graphical model of 
Fig.~\ref{fig:inf_r2_rho} (hereafter we stick to this last model),
adding the relevant nodes.

\begin{figure}[t]
\begin{center}
  \epsfig{file=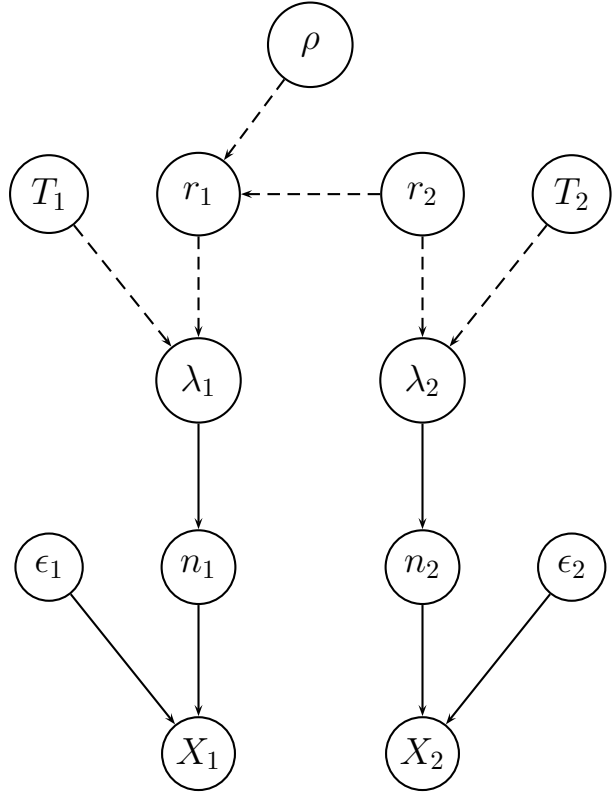,clip=,width=0.53\linewidth}
  \\  \mbox{} \vspace{-1.0cm} \mbox{}
\end{center}
\caption{\small \sf Extension of the model of Fig.~\ref{fig:inf_r2_rho} 
  in order to include efficiencies.}
\label{fig:inf_r2_rho_eff}
\end{figure}

The extended model is shown in Fig.~\ref{fig:inf_r2_rho_eff},
in which we have redefined the symbols, keeping $X_1$
and $X_2$ associated
to the \underline{observed} counts and then calling $n_1$ and $n_2$ those
`produced' by the Poissonians. Each $X_i$ is then binomially
distributed with parameters $n_i$ and $\epsilon_i$. In summary,
listing the `causal relations' from bottom to top, we have 
\begin{eqnarray}
  X_i &\sim& \mbox{Binom}(n_i,\epsilon_i) \label{eq:Xi_sim_B_ni_epsiloni}\\
  n_i &\sim& \mbox{Poisson}(\lambda_i) \label{eq:ni_sim_lambdai} \\
  \lambda_i &=& r_i\cdot T_i \label{eq:lambdai=ri.Ti}\\
  r_1 &=& \rho\cdot r_2  \label{eq:ri=rho.r2}
\end{eqnarray}
At this point we can easily build up the joint distribution
of all quantities in the network, as we have done in the previous section,
and then evaluate the (possibly joint) distribution of the
variables of interest, conditioned by those which are observed
or somehow assumed. Moreover, also the efficiencies $\epsilon_1$
and $\epsilon_2$ are by themselves uncertain, and then
we have to integrate also over them, taking into account
their probability distributions $f(\epsilon_1)$ and $f(\epsilon_2)$.
In fact, their value come from test experiments or, more likely,
from Monte Carlo simulations of the physics process and of the
detector. So we need to enlarge
the model adding four other nodes, taking into account
the probabilistic links
\begin{eqnarray}
  X_i^{(MC)} &\sim& \mbox{Binom}(n_i^{(MC)},\epsilon_i)\,.
\end{eqnarray}
We refrain from adding the four nodes in the network
of Fig.~\ref{fig:inf_r2_rho_eff}, which will become more busy
in a while. Anyway, we can just assign to $\epsilon_1$ and
$\epsilon_2$ the parameter of the probability distribution
resulting from the inferences based on Monte Carlo simulations
(see Ref.~\cite{sampling} for details -- remember that, having
the nodes $\epsilon_1$ and $\epsilon_2$ no parents, they need priors).

\begin{figure}[t]
\begin{center}
  \epsfig{file=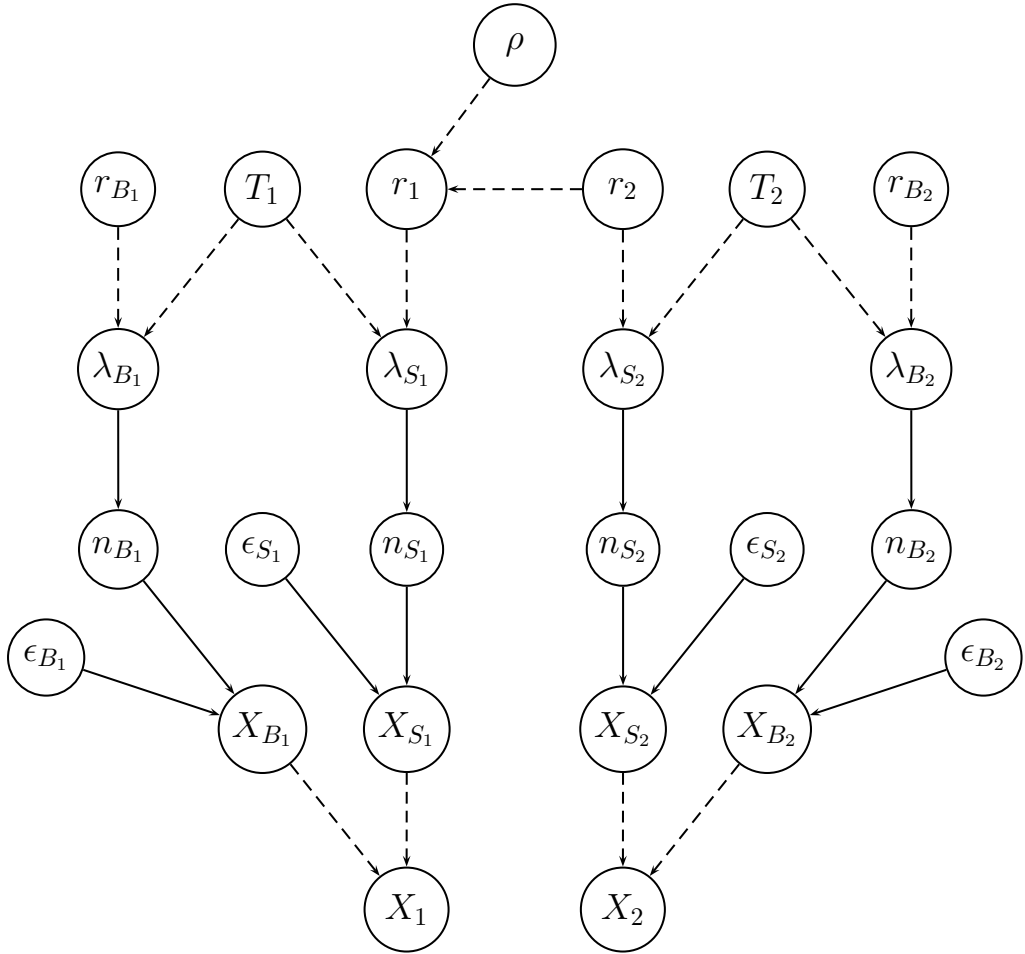,clip=,width=0.89\linewidth}
  \\  \mbox{} \vspace{-1.2cm} \mbox{}
\end{center}
\caption{\small \sf Extended model of Fig.~\ref{fig:inf_r2_rho_eff}
  including also background.}
\label{fig:inf_r2_rho_eff_bkg}
\end{figure}

What is still missing in the model of Fig.~\ref{fig:inf_r2_rho_eff}
is {\em background}. In fact, we do not only lose events
because of inefficiencies, but the  `experimentally defined class'
can get contributions from other `physical class(es)'
(in general there are several physical classes contributing 
as background). Figure \ref{fig:inf_r2_rho_eff_bkg} shows the extension
of the previous model, in which each Poisson process
which describes the {\em signal} has just one background Poisson process.
All variables have subscripts $S$ or $B$, depending if their
are associated to signal or background
(with exception of $r_1$ and $r_2$, which are obviously the two
signal rates). As before, the nodes needed to infer
the efficiencies are not shown in the diagram, which is therefore
missing eight `bubbles'. 

At this point it is clear that trying to achieve closed formulae
is out of hope, and we need to use other methods to perform the
integrals of interest,
namely those based on Markov Chain Monte Carlo. We show here how
to use a powerful package that does the work for
us.  But we do it only for the two cases of which we  already have
closed solutions in hand, that is the models of Figs.~\ref{fig:BN_r_x_rho}
and \ref{fig:inf_r2_rho} starting from uniform priors
for the `top nodes'.
The program we are going to use is {\em JAGS}~\cite{JAGS}
interfaced to R via the package {\tt jrags}~\cite{rjags}.

 $[$\,Introducing  MCMC and related algorithms
  goes well beyond the purpose of this paper
  and we recommend Ref.~\cite{Andrieu} (some examples of application,
  including R scripts, are also provided in
  Ref.~\cite{sampling}).  
  Moreover, mentioning  the Gibbs Sampler algorithm applied to
  probabilistic inference (and forecasting) it is impossible 
  not to refer to the {\em BUGS project} ~\cite{BUGSpaper},
  whose acronym stands
for \underline{B}ayesian inference \underline{u}sing
\underline{G}ibbs \underline{S}ampler, that has
been a kind of revolution in Bayesian analysis,
decades ago limited to simple cases because of computational problems
(see also Sec.~1 of Ref.\cite{JAGS}).
In the BUGS project web site~\cite{BUGSsite}
  it is possible to find packages with excellent Graphical User Interface,
  tutorials and many examples \cite{BUGSexamples}.\,$]$

\subsection{Model A (Fig.~\ref{fig:BN_r_x_rho}), with flat priors
on \texorpdfstring{$r_1$}{r1} and \texorpdfstring{$r_2$}{r2}}
We start from the model of Fig.~\ref{fig:BN_r_x_rho}.
The code that instructs JAGS  about the model is
practically a  transcription
of the expressions to state that a variable follows a given distribution.
Therefore, since we have
\begin{eqnarray}
  X_1 &\sim & \mbox{Poisson}(\lambda_1) \\
  X_2 &\sim & \mbox{Poisson}(\lambda_2) \\
  \lambda_1 &=& r_1\cdot T_1 \\
  \lambda_2 &=& r_2\cdot T_2\\
  \rho &=& r_1/r_2\,,
\end{eqnarray}
we get
\begin{verbatim}
model {
  x1 ~ dpois(lambda1)
  x2 ~ dpois(lambda2)
  lambda1 <- r1 * T1
  lambda2 <- r2 * T2
  r1 ~ dgamma(1, 1e-6) 
  r2 ~ dgamma(1, 1e-6)
  rho <- r1/r2
}
\end{verbatim}
in which are also included the flat priors of
$r_1$ and $r_2$,\footnote{Note that, since
  {\em priors are logically needed}, programs of this kind require
  them, even if they are flat. This can be seen as an annoyance,
  but it is instead a power of these programs: first they can include
  also non trivial priors; second, even if one wants to use flat
  priors, the user is forced to think on the fact that priors are unavoidable,
  instead of following the illusion
  that she is using a prior-free method~\cite{BB},
  sometimes very dangerous, unless one does simple routine measurements
  characterized by a very narrow likelihood~\cite{BR}.}
implemented by Gamma distributions with $\alpha=1$ and $\beta\lll 1$:
\begin{eqnarray}
  r_1 &\sim &  \mbox{Gamma}(1, 10^{-6}) \\
  r_2 &\sim & \mbox{Gamma}(1, 10^{-6})
\end{eqnarray}
The complete R script which calls {\tt rjags} and shows
the results is provided in Appendix B.5
(see Ref.~\cite{sampling} for clarifications about the
structure of the R code). 
\begin{figure}
\begin{center}
  \epsfig{file=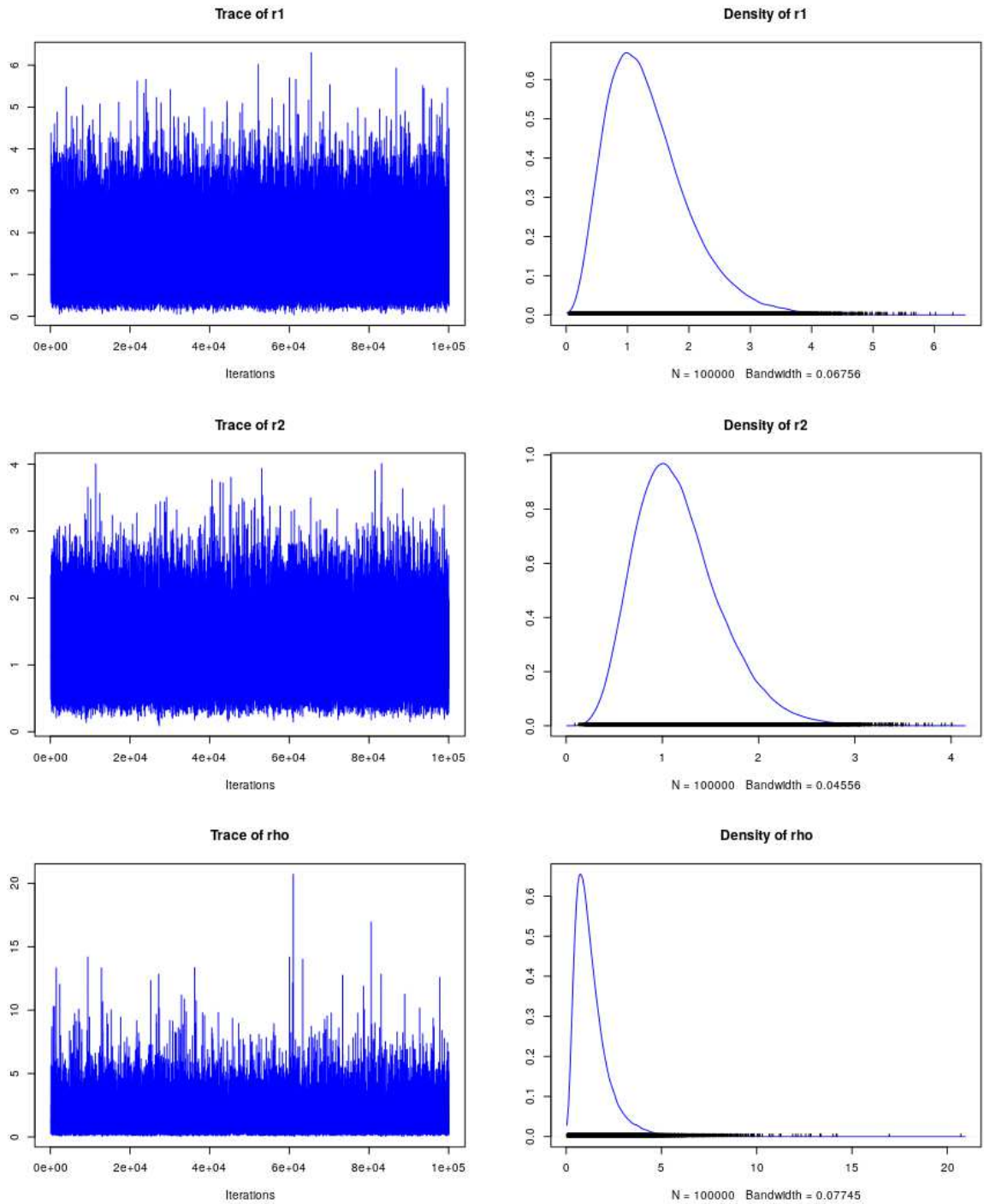,clip=,width=\linewidth}
  \\  \mbox{} \vspace{-1.2cm} \mbox{}
\end{center}
\caption{\small \sf Graphical summary of the chain produced by
  the script of Appendix B.5 implementing the graphical model of
  Fig.~\ref{fig:BN_r_x_rho}.} 
\label{fig:jags_model_1}
\end{figure}
The values of ($x_1\!=\!3,\,T_1\!=\!3\,$s) and ($x_2\!=\!6,\,T_2\!=\!6\,$s)
have been chosen
in order to have small numbers, but with finite expected values
and standard deviation, in order to make a comparison with
the results of the closed formulae.
The parameter determining the `length' of the Markov chain has been
set at $10^5$. 

The summary figure \ref{fig:jags_model_1}, drawn
automatically by R when
the command {\tt plot()} is called with first argument an {\em MCMC
chain object}, shows, for each of the three variables that we have
chosen to {\em monitor},
the `trace' and the 'density'.
The latter is a smoothed representation of the histogram
of the possible occurrences of a variable in the chain.
The former shows the `history'
of a variable during the sampling, and it is important
to understand the quality of the sampling. If the traces appear quite randomic,
as they are in this figure, there is nothing to worry. Otherwise
we have to increase the length of the chain so that it can visit
each `point' (in fact a little volume)
of the space of possibilities with relative frequencies `approximately equal'
to their probabilities (just {\em Bernoulli theorem}, nothing
to do with the `frequentist definition of probability').

Here is the relevant output of the script:
\begin{verbatim}
1. Empirical mean and standard deviation for each variable,
   plus standard error of the mean:

     Mean     SD Naive SE Time-series SE
r1  1.334 0.6671 0.002110       0.002110
r2  1.167 0.4418 0.001397       0.001397
rho 1.333 0.9444 0.002986       0.002986

2. Quantiles for each variable:

      2.5%    25%   50%   75% 97.5%
r1  0.3637 0.8457 1.223 1.700 2.927
r2  0.4701 0.8481 1.111 1.424 2.179
rho 0.2772 0.7048 1.102 1.684 3.770

Exact:
   r1 =  1.333 +- 0.667
   r2 =  1.167 +- 0.441
  rho =  1.333 +- 0.943
\end{verbatim}
As we can see, the agreement between the MCMC and the exact results,
evaluated from Eqs.~(\ref{eq:E_rho_ModelA})
and (\ref{eq:sigma_rho_ModelA}),
is excellent (remember that `{\tt r1 =  1.333 +- 0.667}'
stands for $\mbox{E}(r_1)=1.333\,\mbox{s}^{-1}$ and
 $\sigma(r_1)=0.667\,\mbox{s}^{-1}$).
\mbox{}\\ \mbox{}

\subsection{Model B (Fig.~\ref{fig:inf_r2_rho}), with flat priors
on \texorpdfstring{$\rho$}{rho} and \texorpdfstring{$r_2$}{r2}}
Let us move to the model of Fig.~\ref{fig:inf_r2_rho}, whose
implementation in the JAGS language is the following:
\begin{verbatim}
model {
  x1 ~ dpois(lambda1)
  x2 ~ dpois(lambda2)
  lambda1 <- r1 * T1
  lambda2 <- r2 * T2
  r1 <- rho * r2
  r2  ~ dgamma(1, 1e-6)
  rho ~ dgamma(1, 1e-6)
}
\end{verbatim}
The complete R script, which
uses the same data ($x_1\!=\!3,\,T_1\!=\!3\,$s; $x_2\!=\!6,\,T_2\!=\!6\,$s)
is provided in Appendix B.6. The result is shown in Fig.~\ref{fig:jags_model_2}
and the details are given in the following printouts
\begin{figure}
\begin{center}
  \epsfig{file=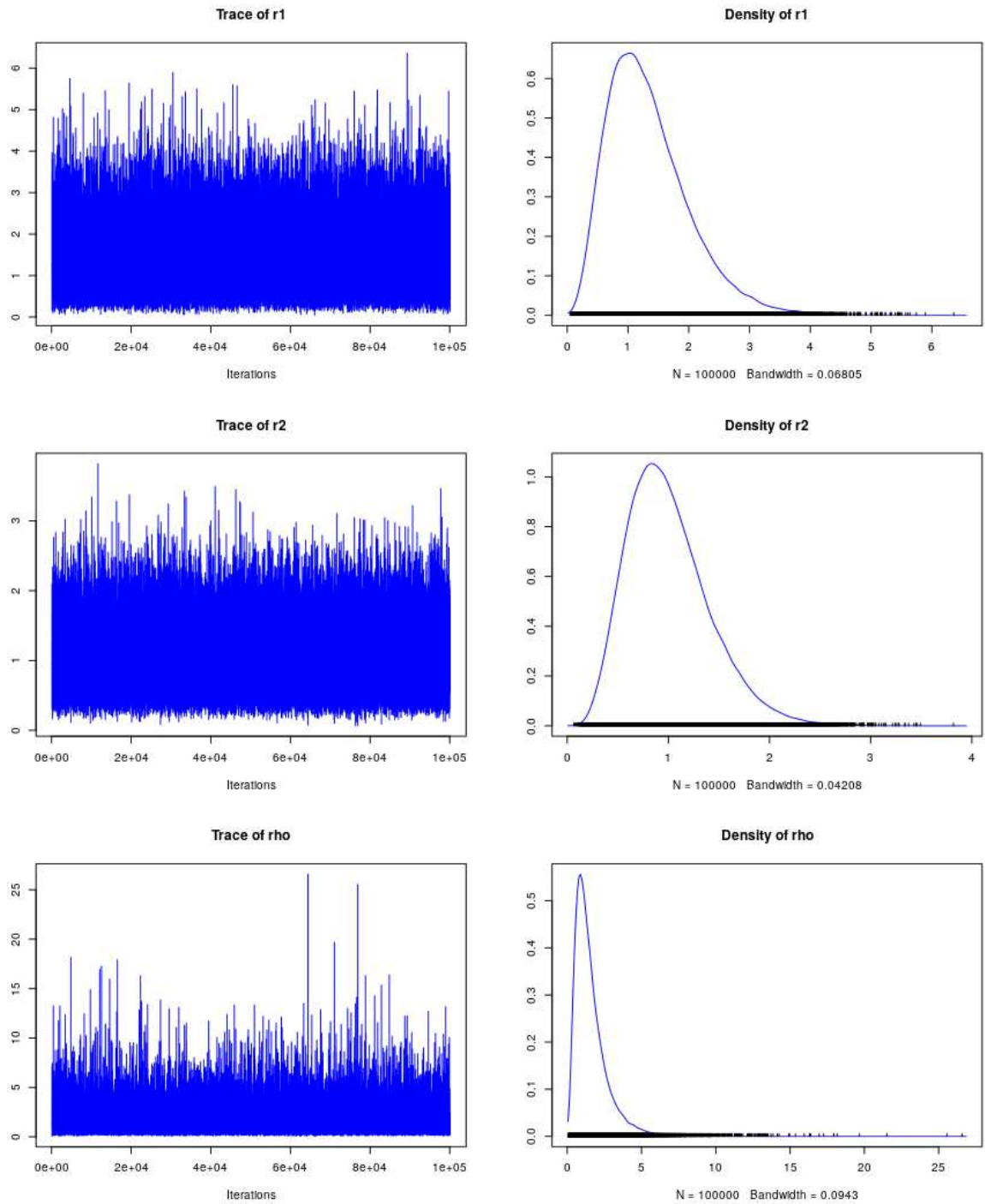,clip=,width=\linewidth}
  \\  \mbox{} \vspace{-1.2cm} \mbox{}
\end{center}
\caption{\small \sf  Graphical summary of the chain produced by
  the script of Appendix B.6 implementing the graphical model of
  Fig.~\ref{fig:inf_r2_rho}.}
\label{fig:jags_model_2}
\end{figure}
\begin{verbatim}
1. Empirical mean and standard deviation for each variable,
   plus standard error of the mean:

     Mean     SD Naive SE Time-series SE
r1  1.334 0.6694 0.002117       0.002117
r2  1.002 0.4068 0.001286       0.001925
rho 1.595 1.1918 0.003769       0.006058

2. Quantiles for each variable:

      2.5%    25%    50%   75% 97.5%
r1  0.3616 0.8438 1.2234 1.704 2.941
r2  0.3671 0.7061 0.9477 1.238 1.940
rho 0.3167 0.8199 1.2923 2.012 4.638

Exact:
   r1 =  1.333 +- 0.667
   r2 =  1.000 +- 0.408
  rho =  1.600 +- 1.200
\end{verbatim}
Again, the agreement between the MCMC and the exact results is excellent.

\subsection{Comparison of the results from the two models}
\begin{figure}[t]
\begin{center}
  \epsfig{file=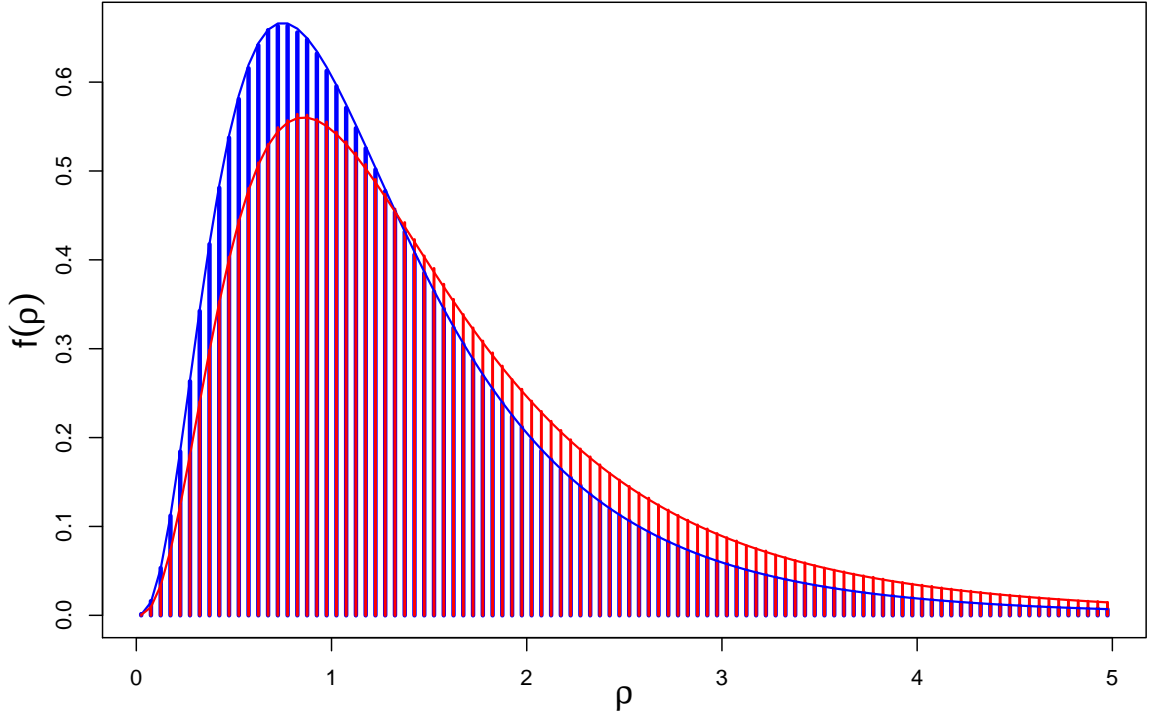,clip=,width=\linewidth}
  \\  \mbox{} \vspace{-1.2cm} \mbox{}
\end{center}
\caption{\small \sf Comparison of the distribution of $\rho=r_1/r_2$ obtained
  by the models of Fig.~\ref{fig:BN_r_x_rho} (blue, slightly
  narrower)
  and Fig.~\ref{fig:inf_r2_rho} (red, slightly wider) in the case of
  ($x_1=3,T_1=3\,$s) and ($x_1=6,T_1=6\,$s) using
  flat priors for the top nodes. The histograms are the JAGS
  results and the lines come from the pdf's in closed form (see text). 
}
\label{fig:model_comparison}
\end{figure}

An overall comparison of the two models,
again based on  the observations
of  3 counts in 3\,s from process 1 and
6 counts in 6\,s from process 2, is shown in
Fig.~\ref{fig:model_comparison},
while  expected values
and standard deviations (separated by `$\pm$') calculated 
from the closed formulae
are summarized in the following table.
\begin{center}
\begin{tabular}{c|c|c}
  & Model A (Fig.~\ref{fig:BN_r_x_rho}) & Model B (Fig.~\ref{fig:inf_r2_rho}) \\
  & $[\,f_0(r_1)\!=\!k\ \& \ f_0(r_2)\!=\!k\,] $  \ \ \ & \ \ \ 
   $[\,f_0(\rho)\!=\!k\ \&\  f_0(r_2)\!=\!k\,] $  \\
  \hline
  $r_1\,(\mbox{s}^{-1})$ & $1.33\pm 0.67$ & $1.33\pm 0.67$ \\
 $r_2\,(\mbox{s}^{-1})$  &   $1.17\pm 0.44$ &  $1.00\pm 0.41$ \\
 $\rho$ & $1.33\pm 0.94$ &  $1.60\pm 1.20$
\end{tabular}  
\end{center}
As we have seen in Fig.~\ref{fig:pdf_rhp_two_models},
the second model produces a distribution of $\rho$ with
higher expected value and higher standard deviation.

\subsection{Dependence of the rate ratio from a physical quantity}
\begin{figure}[t]
\begin{center}
  \epsfig{file=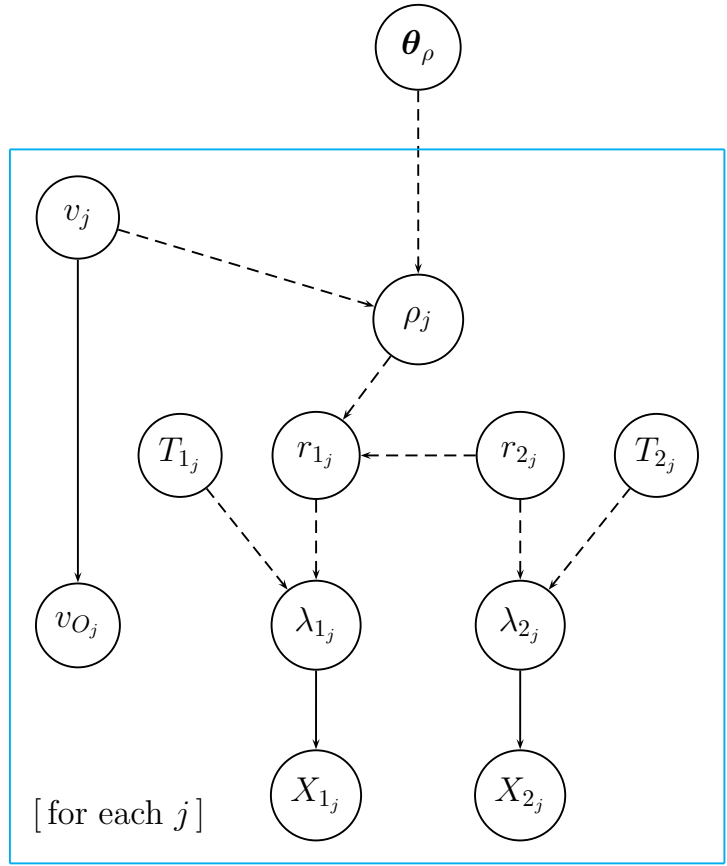,clip=,width=0.63\linewidth}
  \\  \mbox{} \vspace{-0.8cm} \mbox{}
\end{center}
\caption{\small \sf Further extension of the model
  of Fig.~\ref{fig:inf_r2_rho} (neglecting the `complications'
  of the models of Figs.~\ref{fig:inf_r2_rho_eff}
  and \ref{fig:inf_r2_rho_eff_bkg}) to take into account that
  each value $\rho_j$ might depend on the physical quantity
  $v_j$, measured as $v_{m_j}$ via the set of parameters $\bm{\theta}$
}
\label{fig:inf_r2_rho_fit}
\end{figure}

Another interesting question is how to approach the problem
of a ratio of rates that depends on the value of another physical
quantity. 
That is we assume a dependence of $\rho$ from $v$
(symbol for a generic {\em variable}), 
\begin{eqnarray}
  \rho &=& g(v; \bm{\theta}_\rho)\,,
\end{eqnarray}
with $\bm{\theta}_\rho$ the set of parameters of the functional dependence.
The simplest and best understood case is the linear dependence 
\begin{eqnarray}
  \rho &=& m\cdot v + c\,,
\end{eqnarray}
where $\bm{\theta}_\rho=\{m,c\}$, treated
in detail in  Ref.~\cite{fits} where we used the same approach
we are adopting here. In analogy to what done in Fig.~1 there,
we can extend the Model B of Fig.~\ref{fig:inf_r2_rho}
to that of Fig.~\ref{fig:inf_r2_rho_fit}
(we continue to neglect  efficiency and background issues
in order to focus to the core of the problem).
Moreover, as in Fig.~1
of Ref.~\cite{fits}, we have considered the fact that
the physical quantity $v$ is `{\em experimentally observed}' as $v_O$.
In the simple case of a linear dependence  the model
is described by the following relations among the variables
\begin{eqnarray}
  X_{1_j} &\sim & \mbox{Poisson}(\lambda_{1_j}) \\
  X_{2_j} &\sim & \mbox{Poisson}(\lambda_{2_j}) \\
  \lambda_{1_j} &=& r_{j_i}\cdot T_{1_j} \\
  \lambda_{2_j} &=& r_{j_2}\cdot T_{2_j} \\
  r_{1_j} &=& \rho_j\cdot r_{2_j} \\
  \rho_j &=& m\cdot v_j + c \\
   v_{O_j} &\sim & {\cal N}(v_j, \sigma_{E_j})\,,
\end{eqnarray}
in which we have assumed a Gaussian (`normal') error
function of $ v_{O_j}$ around $v_j$,
with standard deviations $\sigma_{E_j}$. 
But the description of the model provided
by the above relations
is not complete (besides the complications
related to inefficiencies and background, that we
continue to neglect).
In fact, we miss priors for $v_j$,  
$r_{2_j}$ and $\bm{\theta}_\rho$,
as they have no parent nodes (instead, we continue to
consider $T_{1_j}$ and $T_{2_j}$ `exactly known',
being their uncertainty usually irrelevant).

The priors which are easier to choose
are those of $v_j$, if their values are `well measured',
that is if $\sigma_{E_j}$ are small enough.
We can then confidently use flat priors, as done e.g.
for the `unobserved' $\mu_{y_i}$ of Fig.~1 in Ref.~\cite{fits}.

Also the priors about $\bm{\theta}_\rho$ can be
chosen quite vague, paying however some care
in order to forbid negative values of $\rho$.
Incidentally,  having mentioned the simple case of linear
dependence, an important sub-case is when $m$
is assumed to be null: the remaining
prior on $c$ becomes indeed the prior on $\rho$,
and the inference of $c$ corresponds to the
inferred value of $\rho$ having taken into account
several instances of $X_1$ and $X_2$ -- this is indeed the question
of the `combination of values of $\rho$' on which we
shall comment a bit more in detail in the sequel.

\begin{figure}[t]
\begin{center}
  \epsfig{file=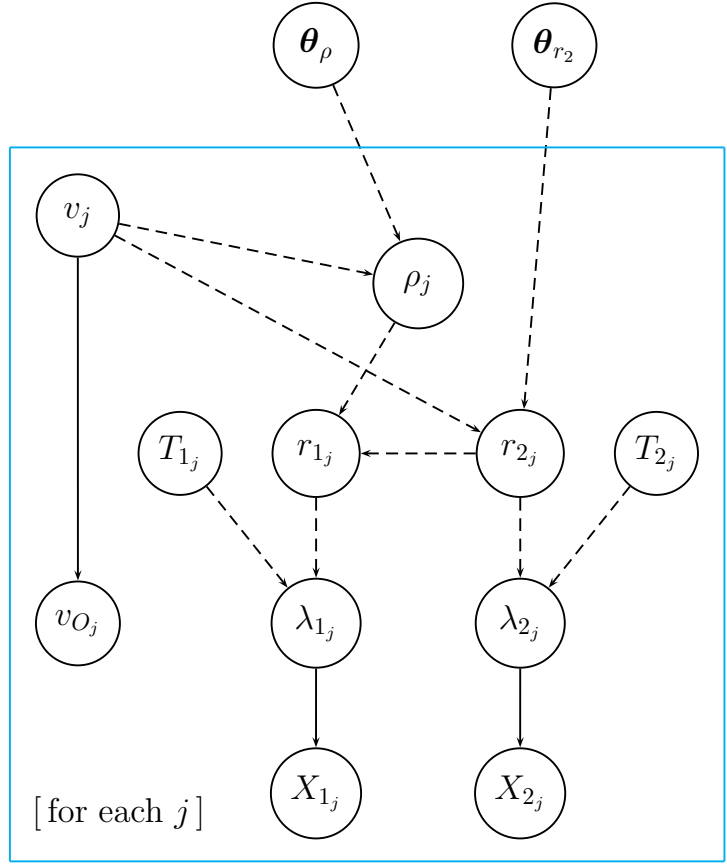,clip=,width=0.63\linewidth}
  \\  \mbox{} \vspace{-0.8cm} \mbox{}
\end{center}
\caption{\small \sf Extension of the model of Fig.~\ref{fig:inf_r2_rho_fit},
  but making also $r_2$ depend on $v$ according to a suited
  law depending on the set of parameters $\bm{\theta}_{r_2}$.
}
\label{fig:inf_r2_rho_fit_2}
\end{figure}

As far as the priors of the rates are concerned,
one could think, a bit naively, that the choice of {\em independent}
flat priors for $r_{2_j}$ could be a reasonable choice.
But we need to understand the physical model underlying this choice.
In fact, most likely, as the ratio $\rho$ might depends on $v$, the same
could be true for $r_2$, but perhaps with a completely
different functional dependence.
For example $r_1$ and $r_2$ could have a strong
dependence on $v$, e.g. they could decrease exponentially,
but, nevertheless, their ratio could be independent of $v$,
or, at most, could just exhibit a small linear dependence.
Therefore we have to add this possibility into the model,
which then becomes as in Fig.~\ref{fig:inf_r2_rho_fit_2},
in which we have included a set
of parameters $\bm{\theta}_{r_2}$, such that
\begin{eqnarray}
  r_2 &=& h(v; \bm{\theta}_{r_2})\,,
\end{eqnarray}
and, needless to say, some priors are required for
$\bm{\theta}_{r_2}$.

At this point, any further consideration goes  beyond the rather
general purpose of this paper, because
we should enter into  details that strongly
depend on the physical case. We hope that the reader could
at least appreciate the level of awareness that
these graphical models provide. The existence of
computing tools in which the models can be implemented
makes then nowadays possible what decades ago
was not even imaginable.

\subsection{Combining ratios of rates}
Let us end the work with the related topic  of
`combining several values' of $\rho$,
a problem we have already slightly touched
above. Let us start phrasing it
in the terms we usually hear about it.
Imagine we have in hand $N$ instances of $(X_1,T_1,X_2,T_2)$.
From each of them we can get a value of $\rho$ with `its uncertainty'.
Then we might be interested in getting a single value, 
combining the individual ones.

The first idea that might come to the mind is to apply the well known
weighted average of the individual values,  using as weights the
inverses of the variances. 
But, before doing it, it is important
to understand the assumptions behind it, that is something
that goes back to none other than Gauss, and for which
we refer to Refs.~\cite{SkepticalJags,CuriousBias}.
The basic idea of Gauss was to get two numbers (let us say
`central value' and standard deviation
-- indeed Gauss used, instead of the standard deviation,
what he called `degree of precision' and
`degree of accuracy'~\cite{CuriousBias}, but
this is an irrelevant detail) such that
{\em they contain the same information of the individual values}.
In practice the rule of combination had to satisfy what
is  currently  known as {\em statistical sufficiency}.
Now it is not obvious at all that the weighted average
using $\mbox{E}(\rho)$ and $\sigma(\rho)$ satisfies sufficiency
(see e.g. the puzzle proposed in the Appendix of Ref.~\cite{CuriousBias}).

\begin{figure}
\begin{center}
  \epsfig{file=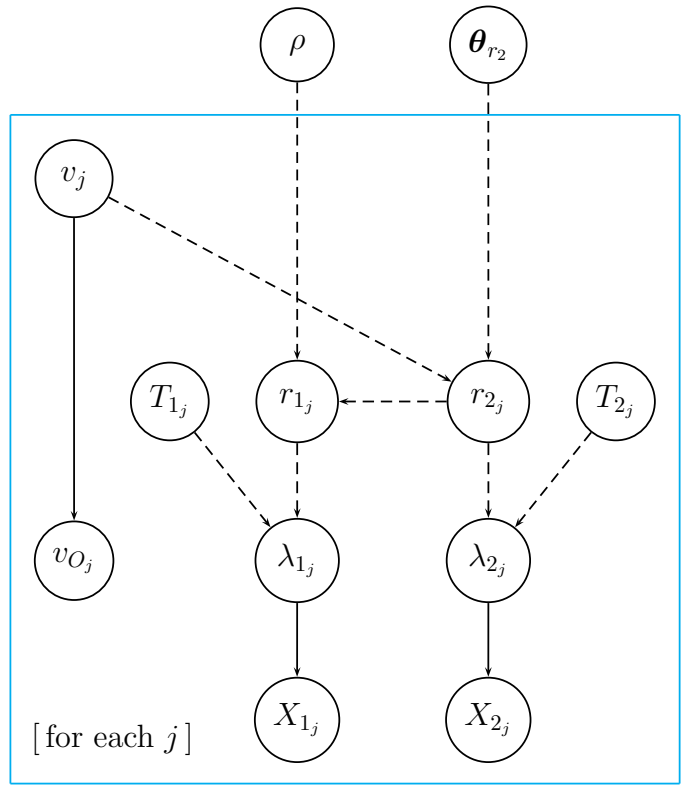,clip=,width=0.59\linewidth} \\
  \mbox{}\\ \mbox{}\\
  \epsfig{file=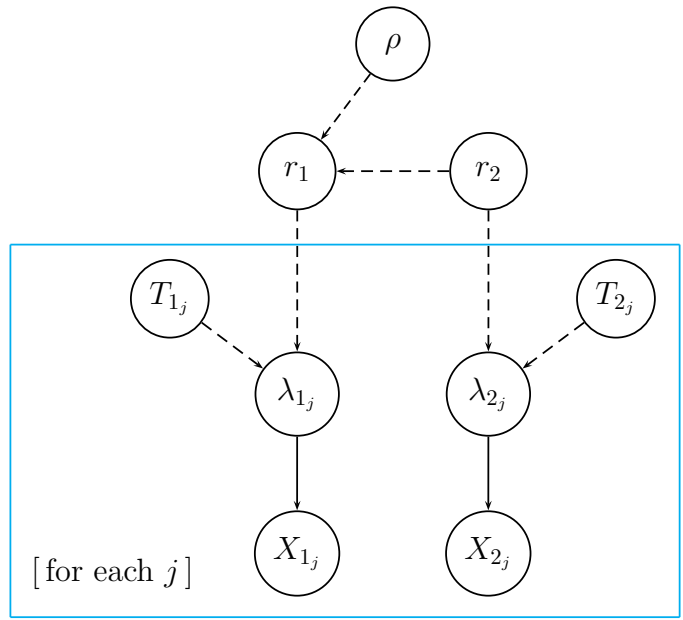,clip=,width=0.59\linewidth}
  \\  \mbox{} \vspace{-1.0cm} \mbox{}
\end{center}
\caption{\small \sf Possible reductions of the model of
  Fig.~\ref{fig:inf_r2_rho_fit_2} for the `combination' of $\rho$
  (see text).
}
\label{fig:combination_rho}
\end{figure}

Therefore, instead of trying to apply the weighted average
as a `prescription', let us see what comes out applying  
consistently the rules of probability on a suitable model,
restarting from that of Fig.~\ref{fig:inf_r2_rho_fit_2}. 
It is clear that if  we consider   meaningful a
combined value of $\rho$ for all
instances of $(X_1,T_1,X_2,T_2)$
it means we assume 
$\rho$ not depending on {\em a} quantity $v$.
However, $r_2$ could. This implies that the values of $r_2$
are strongly correlated to each other.\footnote{At this point a clarification
  is in order. When we make fits and say, again with reference to Fig.~1 of
  Ref.~\cite{fits}, that the observations $y_i$ are independent from
  each other we are referring to the fact that each $y_i$ depends
  only on its $\mu_{y_i}$, e.g. $y_i \sim {\cal N}(\mu_{y_i}, \sigma_Y)$,
  but not on the other $y_{j\ne i}$. Instead,
  the {\em true values  $\mu_{y_i}$ are
  certainly correlated},
  being $\mu_{y} = \mu_{y}( \mu_{x}; \bm{\theta})$. 
  }
Therefore the graphical model
of interest would be that at the top of Fig.~\ref{fig:combination_rho}. 
Again, at this point there is little more to add, because
what would follow depends on the specific physical model.

A trivial case is when both rates, and therefore their ratio, are assumed
to be constant, although unknown, yielding then
the graphical model shown in the bottom diagram of 
Fig.~\ref{fig:combination_rho}, whose related joint pdf, evaluated
by the best suited chain rule, is an extension of
Eqs.~(\ref{eq:chain_rule_model_B})-(\ref{eq:f_tilde})
\begin{eqnarray}
  f(\ldots) \!\!&=&\!\!
  \left[\,\prod_{j=1}^N f(x_{2_j}\,|\,r_2,T_{2_j})\right]\!\cdot\! f_0(r_2)\!\cdot
    \!\left[\,\prod_{j=1}^N f(x_{1_j}\,|\,r_1,T_{1_j})\right]
    \!\cdot\! f(r_1\,|\,r_2,\rho)\!\cdot\! f_0(\rho)\,,\ \ \ \ \ \ \
\end{eqnarray}
from which the unnormalized joint pdf follows:
\begin{eqnarray}
  \tilde f(\ldots) \!\!&\propto& \!\! \left[\,\prod_{j=1}^N r_2^{x_{2_j}}\,
  \cdot e^{-T_{2_j}\,r_2}\right]\!\cdot\! f_0(r_2)
 \! \cdot\! \left[\,\prod_{j=1}^N\,r_1^{x_{1_j}}\! \cdot\! e^{-T_{1_j}\,r_1}\right]\!\cdot
  \delta(r_1-\rho\cdot r_2)\cdot f_0(\rho) \ \ \ \ \ \ \ \ \\
\!\!&\propto& \!\! \left[ r_2^{x_{2_{tot}}}\,
  \cdot e^{-T_{2_{tot}}\,r_2}\right]\!\cdot\! f_0(r_2)
  \cdot \left[r_1^{x_{1_{tot}}}\! \cdot\! e^{-T_{1_{tot}}\,r_1}\right]\!\cdot
  \delta(r_1-\rho\cdot r_2)\cdot f_0(\rho)\,. \ \ \ \ \ \ \ \ 
\end{eqnarray}
We recognize the same structure of Eq.~(\ref{eq:f_tilde}),
with $x_1$ replaced by  $x_{1_{tot}}=\sum_j x_{1_j}$
$T_1$ by  $T_{1_{tot}}=\sum_j T_{1_j}$,
 $x_2$  by  $x_{2_{tot}}=\sum_j x_{2_j}$ and
$T_2$ by  $T_{2_{tot}}=\sum_j T_{2_j}$.
We get then the same result obtained in
Sec.~\ref{ss:inferred_pdf_rho_model_B} if we use the
total numbers of counts in the total times of measurements.
This is a simple and nice result, close to the intuition,
but we have to be aware of the model on which it is based.  

\section{Conclusions}
In this paper we have dealt with the often debated issue
of `ratios of small numbers of events', approaching it
from a probabilistic perspective.
After having shown the difference between predicting
numbers of counts (and their ratios)
and inferring the Poisson parameters (and their ratios) on the base
of the observed numbers of counts,
the attention has been put on the latter, 
{\em  ``a problem in the  probability of causes, \ldots \,
the essential problem of the experimental
  method''}~\cite{Poincare'}.
Having the paper a didactic intent, the basic
ideas of probabilistic inference have been reminded, together with
the use of conjugate priors in order to get closed
results with minimum effort. It has been also shown how to perform
the so called `propagation of uncertainties'
in closed forms, which has required, for the purposes
of this work, to derive
the probability density function of the ratio of
Gamma distributed variables. And, as byproducts,
the `curious' pdf of the ratio of two uniform
variables has been derived and a new derivation
of the formula to get the pdf of a function
of variables has been devised.

The importance of graphical models has been
stressed. In fact, they are not only very useful to
form a global, clearer vision of the problem,
but also to possibly  take into account alternative models.
In the case of rather simple models it has been 
shown how to write down the joint distribution
of all variables, from which
the pdf of the variables of interest follows.
In some cases, thanks to reasonable (or at least
\underline{well stated})
assumptions, closed results have been obtained,
but we have also seen how to use tools
based on MCMC, both to check the closed
results and to tackle more realistic models
(samples of programming code are provided in
Appendix B).

Finally, as far as the issue of `combination of ratios'
is concerned, it has been shown how the solution depends
crucially on the physical model describing the variation
of the rates and/or their ratio in function of
an external variable. Therefore only general indications
on how to approach the problem have been given,
highly recommending the use of MCMC tools
(my preference for small problem with limited amount of data
goes presently to JAGS/rjags, but particle physicists 
might prefer BAT~\cite{BAT}, or perhaps the
more recent, Julia~\cite{Julia} based, BAT.jl~\cite{BAT.jl}).

\mbox{}

I am indebted to Alfredo (Dino) Esposito for many
discussions on the probabilistic and technical
aspects the paper,
some of which admittedly based on Ref.~\cite{sampling},
and for valuable comments on the manuscript.


\newpage
\section*{Appendix A -- From the Bernoulli process to the Poisson process:
  binomial, Poisson and exponential distributions (and more)}
\subsection*{A1. Reminder of basic formulae}
Let us start reminding the well known 
binomial and Poisson distributions, taken verbatim
from Ref.\,\cite{BR}, just to introduce the notation
used in this note.
\begin{description}
\item[Binomial distribution]\ \\
$X\sim \mbox{Binom}(n,p)$ (hereafter ``$\sim$'' stands for ``follows'');
$\mbox{Binom}(n,p)$ stands for {\it binomial} with parameters 
$n$  and $p$:
\begin{equation}
f(x\,|\,n,p) =
\frac{n!}{(n-x)!\,x!}\cdot p^x\cdot (1-p)^{n-x} \, ,
\hspace{1.0 cm}
\left\{ \begin{array}{l}   n = 1, 2, \ldots,  \infty \\
                           0 \le p \le 1 \\
                           x = 0, 1, \ldots, n \end{array}\right.\,.
\nonumber 
\end{equation}                           
Expected value, standard deviation and {\it variation coefficient}
$[$\,$v\equiv\sigma(X)/\mbox{E}(X)$\,$]$:
\begin{eqnarray*}
\mbox{E}(X) &  = &  n\cdot p \\
\sigma(X)   & = & \sqrt{n\cdot p\cdot (1-p)}\\
v  &=&
\frac{\sqrt{n\cdot p\cdot (1-p)}}{n\cdot p} \propto \frac{1}{\sqrt{n}}\, .
\end{eqnarray*} 

\vspace{0.3cm}%

\item[Poisson distribution]\ \\
$X\sim \mbox{Poisson}(\lambda)$:
\begin{equation}
f(x\,|\,\lambda)=\frac{\lambda^x}{x!}\cdot e^{-\lambda}
\hspace{1.0 cm}
\left\{ \begin{array}{l}   0 < \lambda < \infty \\
                           x = 0, 1, \ldots,  \infty\\
                           \end{array} \right.\,.
\nonumber 
\end{equation}
Expected value, standard deviation and variation coefficient
\begin{eqnarray*}
\mbox{E}(X)  & = &  \lambda \\
\sigma(X)  & = &   \sqrt{\lambda} \\
v &=& {1}/{\sqrt{\lambda}}.
\end{eqnarray*}
\item[Binomial $\rightarrow$ Poisson]
\[\mbox{Binom}(n,p)
\xrightarrow 
[\begin{array}{l}n\rightarrow \infty \\ 
p\rightarrow 0 \\
(n\cdot p = \lambda)\end{array}]{}
{\mbox{Poisson}}(\lambda) \,.
\]  
\end{description}

\subsection*{A2. Bernoulli process and related distributions}
A Bernoulli process is characterized by
a probability $p$ of {\em success}, to which
is associated the {\em uncertain number} $X=1$,
and probability $1-p$ of {\em failure},
to which is associated the uncertain number $X=0$. 
Therefore, technically, a {\em Bernoulli distribution}
is just a binomial with $n=1$. But conceptually
it is very important, because it is the basic
process from which other distributions arise:
\begin{itemize}
\item a {\em binomial distribution} describes the probability
  of the total number of
  successes in $n$ {\em independent} Bernoulli trials
  `having' (or more precisely `believed to have')
  the same probability of success $p$;
\item a {\em geometric distribution} describes the
  probability (again assuming independence and constant $p$)
  of the trial
  {\em at which}\footnote{Indeed, it can also be found
    in the literature as the probability of {\em the number of failures
      \underline{before} the first success occurs'}
   (for example, my preferred {\em vademecum} of Probability
  Distributions, that is the
  homonymous {\em app}~\cite{ProbabilityDistributions},
  reports both distributions).}
  the first success occurs;
\item a {\em Pascal distribution} (or {\em negative binomial}
  distribution)
  concerns finally the
  trial at which the $k$-th success occurs.\footnote{Also of
    this distribution there are
    two flavors, the other one describing
    the number of trials \underline{before} the $k$-th
    success\,\cite{ProbabilityDistributions}.} 
\end{itemize}
 \begin{center}
    \epsfig{file=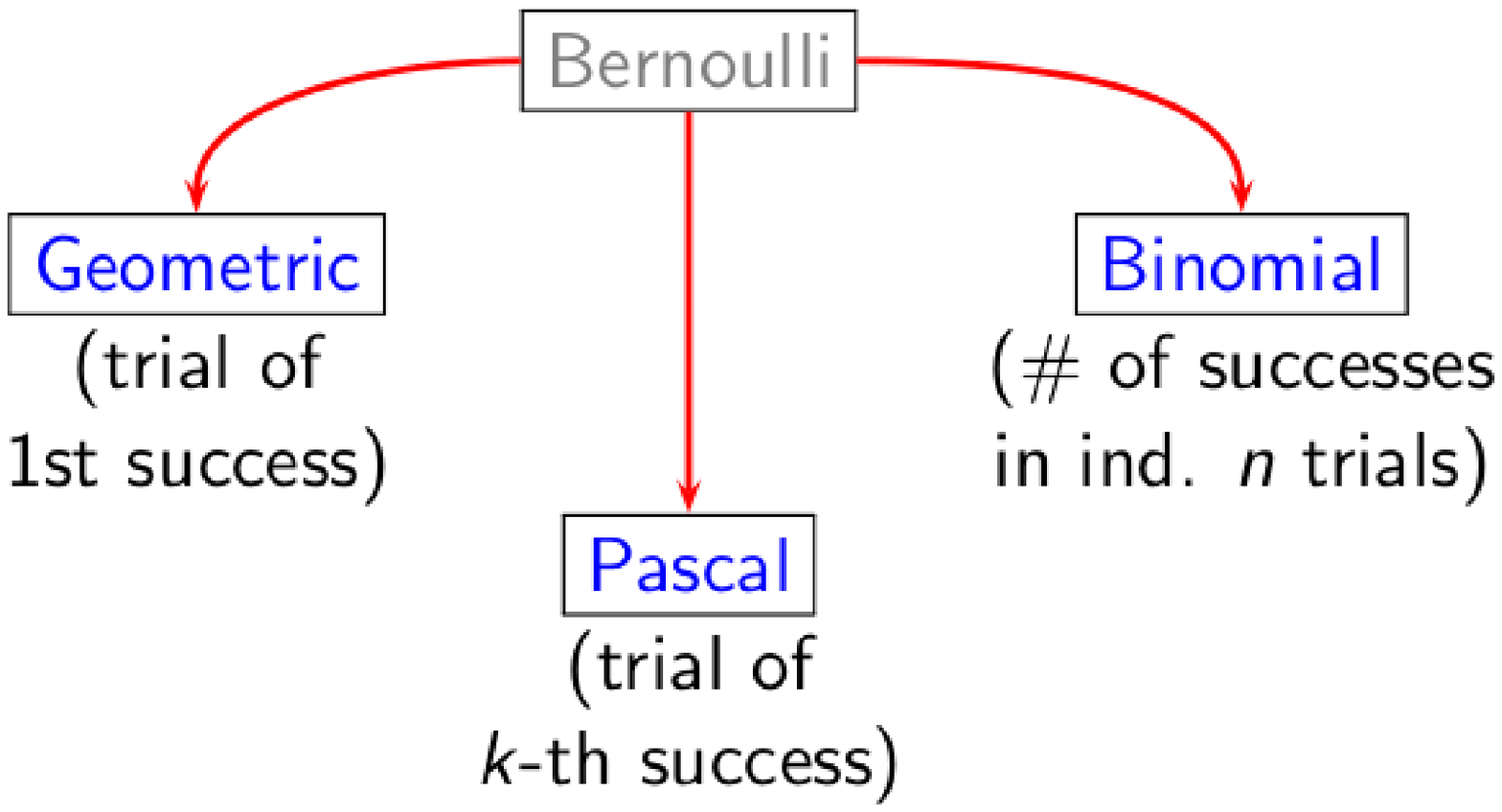,clip=,width=0.6\linewidth}
 \end{center}
\subsection*{A3. Poisson process}
Let us now imagine phenomena that might happen at random
at a given instant\footnote{One could also think at `things'
  occurring in `points' in some different space. All what we are going to
  say in the domain of time can be translated in other domains.} 
\begin{center}
  \mbox{}\\
    \epsfig{file=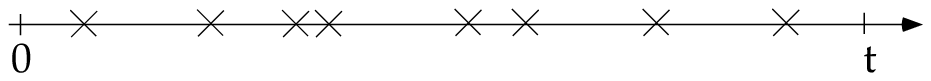,clip=,width=0.55\linewidth}
 \end{center}
such that
\begin{itemize}
\item the probability of one count in
  $\Delta T$ is proportional to $\Delta T$, with $\Delta T$ `small', that is
     $$p=P(\mbox{``1 count in $\Delta T$}'') = r\,  \Delta T$$
  where the proportionality factor $r$ is interpreted as
  the {\em intensity of the process};
    \item the probability that two or more counts occur in
      $\Delta T$ is much smaller than the probability
      of one count (the condition holds if  $\Delta T$ is small enough,
      that will be the case of interest):
      $$P(\ge 2\ \mbox{counts}) \ll P(1\ \mbox{count})\,;$$
\item what happens in one interval does not depend on what
  happened (or `will happen') in other intervals (if disjoint).
\end{itemize}
Let us divide a finite time interval $T$ in $n$ small intervals, i.e. such that
$T=n\,\Delta T$. 
Considering the possible occurrence of a count in each small interval $\Delta T$
as an independent Bernoulli trial, of probability 
   $$p=r\,\Delta T = r\cdot \frac{T}{n}\,,$$
if we are interested in the total number of counts in T
we get a binomial distribution, that is, indicating by $X$
the uncertain number of interest, 
\begin{eqnarray*}
  X &\sim& \mbox{Binom}(n,p)\,.
\end{eqnarray*}
But when $n$ is `very large' (`$n\rightarrow \infty$') we obtain a Poisson
distribution with
\begin{eqnarray*}
  \lambda &=& n\cdot \left( r\cdot\frac{T}{n}\right) = r\cdot T\,,
\end{eqnarray*}
equal to the intensity of the process times the finite time of observation.
In particular, we can see that the physical quantity of interest
is $r$, while the Poisson parameter $\lambda$ is a kind of ancillary
quantity, depending on the measurement time.

\subsection*{A4 -- Waiting time to observe the $k$-event}
It is clear that if we are interested in the probability that
the first count occurs in the $i$-th time interval
of amplitude $\Delta T$, we recover `in principle'
a geometric distribution. But since $\Delta T$ can be
arbitrary small, it makes no sense in numbering the intervals.
Nevertheless, thinking in terms of the $n$ Bernoulli
process can be again very useful. Indeed, the probability
that the first count occurs {\em after} the $x-th$ trial is
equal to the probability that it never occurred in the
trials from 1 to $x$:
\begin{eqnarray*}
  P(X>x) &=& (1-p)^x\,.     
\end{eqnarray*}  
In the domain of time, indicating now by $T$ the
time at which the first event can occur, 
the probability that this variable is larger than the value $t$,
the latter being $n$ times $\Delta T$,
is given by
\begin{eqnarray*}
  P(T>t) &=& (1-p)^n \\
  &=& \left(1-r\cdot \frac{t}{n}\right)^n
  \xrightarrow[\ n\rightarrow\infty\ ]{}\, e^{-r\, t}\,.
\end{eqnarray*}  
As a complement, the cumulative distribution of $T$, from which
the probability density function follows, is given by
\begin{eqnarray*}
  F(t\,|\,r) \equiv P(T\le t) &=& 1 -  P(T>t) = 1-  e^{-r\, t} \\
  && \\
  f(t\,|\,r) \equiv \frac{\mbox{d}F(t\,|\,r)}{\mbox{d}t} &=& r\, e^{-r\, t}\,.
\end{eqnarray*}
The time at which the first count
is recorded is then described by an exponential
distribution
having expected value, standard deviation and variation coefficient equal to
\begin{eqnarray*}
  \mbox{E}(T) &=& 1/r \ [\equiv \tau] \\
  \sigma(T)   &=& 1/r = \tau  \\
  v &=& 1\,,
\end{eqnarray*}
while the {\em mode} (`most probable value')
is always at $T=0$, independently
of $r$.

As we can see, as it is reasonable to be,
the higher is the intensity of the process, the
smaller is the expected time at which the first count occurs
(but note that the distribution extends always rather slowly to
$T\rightarrow\infty$, a mathematical property reflecting the fact that
such a distribution has always a 100\% {\em standard uncertainty},
that is $v=1$).
Moreover, since the choice of the instant at which
we start waiting from the first event is arbitrary
(this is related to the so called `property of no memory' of
the exponential distribution, which has an equivalent in the geometric one),
we can choose it
to be the instant at which a previous count occurred.
Therefore, the same distribution describes the time intervals
between the occurrence of subsequent counts.

Once we have got the probability distribution of $k=1$, using
probability rules we can get that of $k=2$, reasoning
on the fact that the associated variable is the sum
of two exponentials, and so on. We shall not enter
into details,\footnote{It is indeed a useful exercise to
derive the Erlang distribution starting from
$$f(t\,|\,r,k\!=\!2)\ =\ \int_0^\infty\!\int_0^\infty\delta(t-t_1-t_2)\cdot
f(t_1\,|\,r,k\!=\!1)\cdot f(t_2\,|\,r,k\!=\!1)\,\mbox{d}t_1\mbox{d}t_2\,,$$
and going on until the general rule is obtained.
}
but only say that we end with the
{\em Erlang distribution}, given by
\begin{eqnarray*}
  f(t\,|\,r,k) &=&   \frac{r^{k}}{(k-1)!}\cdot t^{k-1}\cdot e^{-r\,t}
  \mbox{}\ \ \ \ \left\{\begin{array}{l}r>0\\
  k:  \mbox{integer}, \ge 1 \end{array} \right.
\end{eqnarray*}
The extension of $k$ to the continuum, indicated for clarity
as $c$, leads to the famous 
{\em Gamma distribution} (here written for our variable $t$)
 \begin{eqnarray*}
  f(t\,|\,r,c) &=&   \frac{r^{c}}{\Gamma(c)}\cdot t^{c-1}\cdot e^{-r\,t}
  \mbox{}\ \ \ \left\{\begin{array}{l}r>0\\
  c > 0 \end{array} \right.
\end{eqnarray*}
 with $r$ the `rate parameter' (and it is now clear the reason
 for the name) and $c$ the `shape parameter'
 (the special cases in which $c$ is integer help to understand its meaning),
 having expected value and standard deviation equal to
 $c/r$ and $\sqrt{c}/r$, both having the dimensions of time
 (this observation helps to remember their expression).

However, since in the text the symbol $r$ is assigned to the
intensity of the physical process of interest, we
are going to use for the Gamma distribution the standard symbols
met in the literature (see e.g. \cite{ProbabilityDistributions}
and \cite{Wiki_Gamma}) applying the following
replacements:
\begin{eqnarray*}
c &\rightarrow& \alpha \\
r &\rightarrow& \beta\,.
\end{eqnarray*}
Using also the usual symbol $X$ for generic variable, here is
a summary of the most important expressions related to the Gamma
distribution (we also add the mode, easily obtained by the condition
of maximum\footnote{Taking the log of $f(x\,|\,\alpha,\beta)$, we get
the condition of maximum by
\begin{eqnarray*}
\frac{\partial}{\partial x}\log f(x\,|\,\alpha,\beta)
&=& \frac{\alpha-1}{x} - \beta = 0\,,
\end{eqnarray*}
resulting in $x=(\alpha-1)/\beta$.
}):
\begin{description}
\item{$X\sim \mbox{Gamma}(\alpha,\beta)$:}
{\small 
 \begin{eqnarray*}
  f(x\,|\,\alpha,\beta) &=&
  \frac{\beta^{\,\alpha}}{\Gamma(\alpha)}\cdot x^{\,\alpha-1}\cdot e^{-\beta\,x}
  \mbox{}\ \ \ \left\{\begin{array}{l}\alpha>0\\
  \beta > 0 \end{array} \right. \\
  \mbox{E}(X) &=& \frac{\alpha}{\beta} \\
  \mbox{Var}(X) &=& \frac{\alpha}{\beta^2} \\
  \sigma(X) &=& \frac{\sqrt{\alpha}}{\beta} \\
  \mbox{mode}(X) &=&  \left\{
\begin{array}{ll} 0 & \mbox{if}\ \alpha < 1 \\ 
                  \frac{\alpha-1}{\beta} & \mbox{if}\ \alpha \ge  1 
\end{array}\right.  
\end{eqnarray*}
}
\end{description}
 Here is, finally, a summary of the distributions
 derived from the `apparently insignificant' Bernoulli process:
\begin{center}
    \epsfig{file=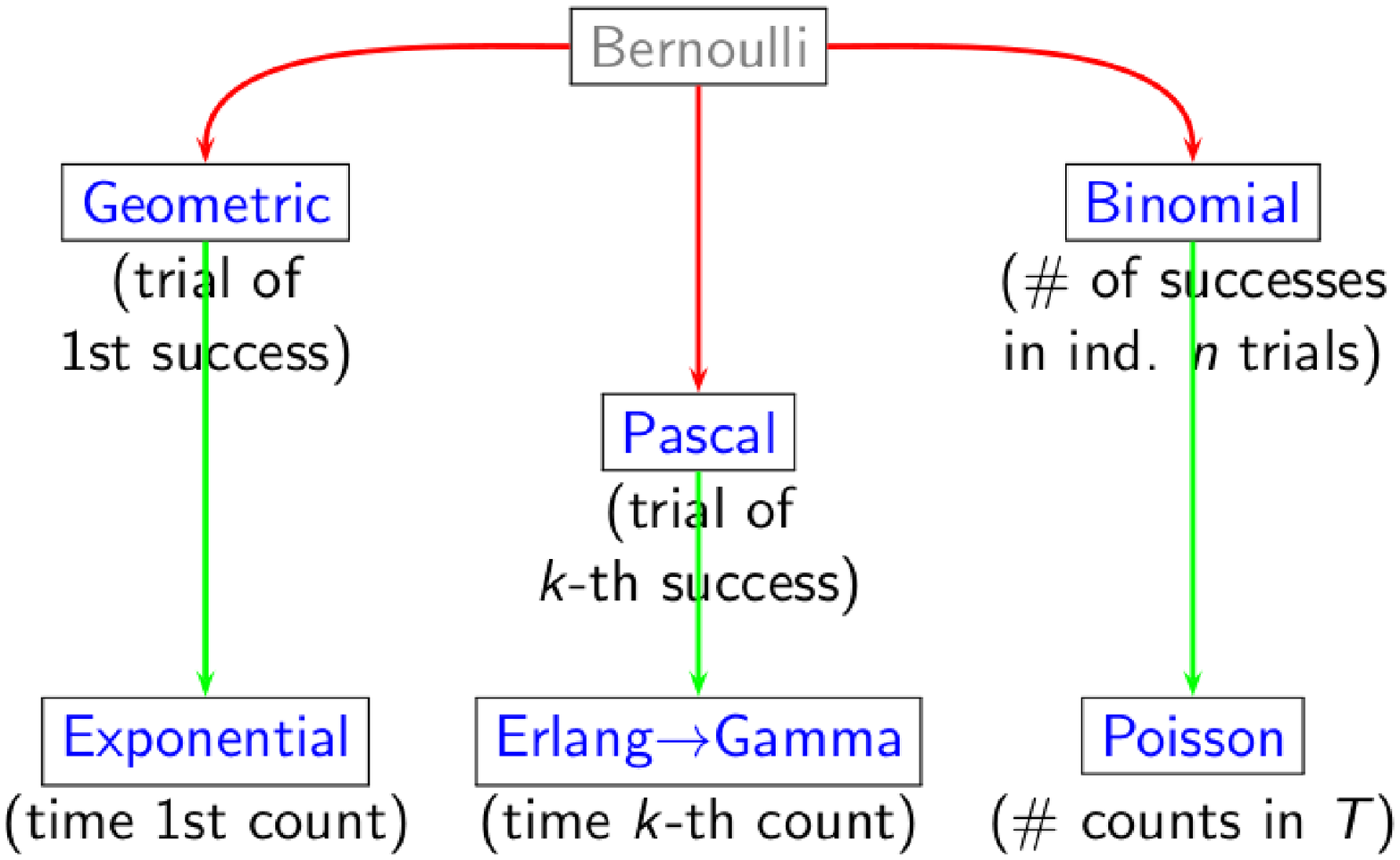,clip=,width=0.725\linewidth}
 \end{center}
For completeness, let us also remind that:
\begin{itemize}
\item the famous {\em $\chi^2$ distribution} is technically a Gamma,
  with $\alpha=\nu/2$ and $\beta=1/2$;
\item most distributions appearing in this scheme,
  with the obvious exception of the geometric and the exponential,
  which have fixed shape, `tend to a Gaussian distribution' for
  some values of the parameters. In particular, for what concerns
  this paper, the Poisson distribution tends to `normality' for `large'
  values of $\lambda$, as well known. However, it is perhaps worth
  remembering that, in general, such a limit applies
  to the cumulative distribution,
  and not to the probability function,  defined for the Poisson
  distribution only for non negative integers:
  $$F(x\,|\,\mbox{Poisson}(\lambda))\, \xrightarrow 
  [\,\ `\lambda\rightarrow \infty'\,\ ]{}
  \,F(x\,|\,{\cal N}(\lambda,\sqrt{\lambda}))\,.$$
\end{itemize}

\newpage
\section*{Appendix B  -- R  and JAGS code}
\subsection*{B.1 -- Distribution of the difference of
  Poisson distributed counts}
{\small 
\begin{verbatim}
dPoisDiff <- function(d, lambda1, lambda2) {
   xmax = round(max(lambda1,lambda2)) + 20*sqrt(max(lambda1,lambda2))
   sum( dpois((0+d):xmax, lambda1) * dpois(0:(xmax-d), lambda2) )
}

l1 = 1
l2 = 1
d = -8:8
fd = rep(0,length(d))
for(i in 1:length(d)) fd[i] = dPoisDiff(d[i], l1, l2)
E.d  <- sum(d*fd)
E.d2 <- sum(d^2*fd)
sigma.d <- sqrt(E.d2 - E.d^2)
cat(sprintf(" d: %.3f +- %.3f ", E.d, sigma.d))
cat(sprintf(" (exact:  %.3f +- %.3f)\n", l1-l2, sqrt(l1+l2)))

barplot(fd, names=d, col='cyan', xlab='d', ylab='f(d)')
\end{verbatim}
(The function {\tt dPoisDiff()} is simple implementation
of the reasoning shown in the text. For a more professional
function see footnote \ref{fn:Skellam}.)

\subsection*{B.2 -- Monte Carlo estimate of the pdf of
$\rho=\lambda_1/\lambda_2$
(flat priors on $\lambda_1$ and $\lambda_2$)}  
\begin{verbatim}
n=10^7
x1 = 1
x2 = 1
lambda1 = rgamma(n, x1+1, 1)
lambda2 = rgamma(n, x2+1, 1)
rho = lambda1/lambda2
E.rho = mean(rho)
sigma.rho = sd(rho)

max.rho.hist = 8
rho = rho[rho<max.rho.hist] 
hist(rho, nc=150, col='cyan', freq=FALSE, xlim=c(0,max.rho.hist), main='',
     xlab=expression(paste(rho, ' = ', lambda[1], '/', lambda[2])) )

# dummy histogram for rough evaluation of the mode
h.rx <- hist(rho, nc=1000, plot=FALSE )
mode = h.rx$mids[which.max(h.rx$density)]

abline(v=mode, col='red')
abline(v=E.rho, col='blue')

p.overflow = (n - length(rho))/n * 100
cat(sprintf("fraction of overflows %.2f%%\n", p.overflow))
text(6,0.63-0.05,expression(paste(x[1], ' =  ', x[2], ' = 1')),
     cex=1.6,col='blue')
text(6,0.54-0.05,sprintf("mean = %.2f;  std = %.2f", E.rho, sigma.rho),
     cex=1.5, col='blue')
text(6,0.46-0.05, sprintf("[ Overflow: %.2f%% ]", p.overflow),
     cex=1.5, col='gray')
text(6,0.37-0.05, sprintf("[ Mode = %.2f ]", mode), cex=1.5, col='red')
\end{verbatim}

\subsection*{B.3 -- Ratio of rates: exact evaluations vs simulation}
\begin{verbatim}
mode.rho <- function(a1,b1, a2,b2) ifelse(a2 >-1, b2/b1* (a1-1)/(a2+1), Inf)
E.rho    <- function(a1,b1, a2,b2) ifelse(a2 > 1, b2/b1*  a1 /(a2-1), Inf)
var.rho  <- function(a1,b1, a2,b2)  ifelse(a2 > 2,
              (b2/b1)^2 * ( a1 /(a2-1) * ((a1+1)/(a2-2) - a1/(a2-1))), Inf)
sigma.rho <- function(a1,b1, a2,b2)  ifelse(a2 > 2,
              sqrt(var.rho(a1,b1, a2,b2)), Inf)
f.rho <- function(rho, a1,b1, a2,b2) {
            lf =   ( a1*log(b1) + a2*log(b2) + (a1-1)*log(rho)
                 + (-a1-a2)*log(b2+rho*b1) - lbeta(a1,a2) )
            return(exp(lf))
}

x1 = 1; T1 = 1
x2 = 2; T2 = 2

a1 = x1+1; b1 = T1
a2 = x2+1; b2 = T2
rho.max = 8
n = 10^6

cat(sprintf("x1,T1 = %.2f, %.2f;  ", x1, T1 ))
cat(sprintf("x2,T2 = %.2f, %.2f;  \n", x2, T2 ))
cat(sprintf("alpha1,beta1 = %d, %d;  ", a1, b1 ))
cat(sprintf("alpha2,beta2 = %d, %d \n", a2, b2 ))
Erho <-  E.rho(a1,b1, a2,b2)
Srho <-  sigma.rho(a1,b1, a2,b2)
cat(sprintf("mode = %.3f;  E() = %.3f ; sigma %.3f\n",
            mode.rho(a1,b1, a2,b2),
            Erho, Srho )) 

x1.r <- rgamma(n, a1, b1)
x2.r <- rgamma(n, a2, b2)
rho.r <- x1.r/x2.r
Mrho  <- mean(rho.r)
SDrho <- sd(rho.r)
cat(sprintf("MC: mean = %.3f ; sigma %.3f\n", Mrho, SDrho ))

rho.r = rho.r[rho.r<rho.max] # only for histogram!!
                 # Warning!! It changes normalization!
norma = length(rho.r)/n
h <- hist(rho.r, nc=150, plot=FALSE)
h$density <- h$density * norma
h$counts <- h$counts * norma
plot(h, col='cyan', freq=FALSE, main='', xlim=c(0,rho.max),
     ylim=c(0,0.52),
     xlab=expression(rho), ylab=expression(paste('f(',rho,')')))

rho = seq(0, rho.max, len=101)
points(rho, f.rho(rho, a1,b1, a2,b2), ty='l', col='blue')
text(6,0.46,bquote(x[1] == .(x1) ~ "," ~  T[1] == .(T1) ), cex=1.5, col='blue')
text(6,0.41,bquote(x[2] == .(x2) ~ "," ~  T[2] == .(T2) ), cex=1.5, col='blue')
Erho.s <- round(Erho, 2)
Srho.s <- round(Srho, 2)
mode.s <- round(mode.rho(a1,b1, a2,b2),2)
Mrho.s <- round(Mrho,2)
SDrho.s <- round(SDrho,2)
text(6,0.35,bquote(E(rho) == .(Erho.s) ~ "," ~
                       sigma(rho) == .(Srho.s) ), cex=1.5, col='blue')
text(6,0.28,bquote(mode(rho) == .(mode.s) ), cex=1.5, col='red')
text(6,0.21,bquote("mean" == .(Mrho.s) ~ "," ~
                       "std" == .(SDrho.s) ), cex=1.5, col='blue')
abline(v=Erho, col='blue')
abline(v=mode.rho(a1,b1, a2,b2), col='red',)
\end{verbatim}

\subsection*{B.4 -- Distribution of $\rho$ implied by uniform
priors on $r_1$ and $r_2$} 
\begin{verbatim}
n = 10^7
rM = 100
r1 = runif(n, 0, rM)
r2 = runif(n, 0, rM)
rho = r1/r2
rho.h <- rho[rho<5]       # for the histogram
norma = length(rho.h)/n   # normalization
h <- hist(rho.h, nc=100, plot=FALSE)
h$density <- h$density * norma
h$counts <- h$counts * norma
plot(h, col='cyan', freq=FALSE, main='', xlim=c(0,5),
     ylim=c(0,0.52),
     xlab=expression(rho), ylab=expression(paste('f(',rho,')')))
abline(v=1, col='red')

# r1.check = rho*r2
# hist(r1.check, nc=200, col='blue', freq=FALSE, xlim=c(0,rM))
\end{verbatim}

\subsection*{B.5 -- Example of JAGS inference of rates and their ratio
for the model of Fig.~\ref{fig:BN_r_x_rho}} 
\begin{verbatim}
#------------  Data  --------------------------------------------
x1 = 3; T1=3
x2 = 6; T2=6
nr = 1e5

#-------------  JAGS model --------------------------------------
library(rjags)
model = "tmp_model.bug"    # name of the model file ('temporary')
write("
model {
  x1 ~ dpois(lambda1)
  x2 ~ dpois(lambda2)
  lambda1 <- r1 * T1
  lambda2 <- r2 * T2
  r1 ~ dgamma(1, 1e-6) 
  r2 ~ dgamma(1, 1e-6)
  rho <- r1/r2
}
", model)

#------------  JAGS call via rjags ------------------------------
data <- list(x1=x1, T1=T1, x2=x2, T2=T2)
jm <- jags.model(model, data)
update(jm, 100)
to.monitor <-  c('r1', 'r2', 'rho')
chain <- coda.samples(jm, to.monitor, n.iter=nr)

#------------  Results -----------------------------------------
print(summary(chain))
plot(chain, col='blue')

cat(sprintf("Exact: \n"))
cat(sprintf("   r1 =  %.3f +- %.3f\n", (x1+1)/T1, sqrt(x1+1)/T1))
cat(sprintf("   r2 =  %.3f +- %.3f\n", (x2+1)/T2, sqrt(x2+1)/T2))
mu.rho    <-  ((x1+1)/T1)/(x2/T2)
sigma.rho <- sqrt(mu.rho*(T2/T1*(x1+2)/(x2-1)-mu.rho)) 
cat(sprintf("  rho =  %.3f +- %.3f\n", mu.rho, sigma.rho))
\end{verbatim}

\subsection*{B.6 -- Example of JAGS inference of rates and their ratio
for the model of Fig.~\ref{fig:inf_r2_rho}} 
\begin{verbatim}
#------------  Data  --------------------------------------------
x1 = 3; T1=3
x2 = 6; T2=6
nr = 1e5

#-------------  JAGS model --------------------------------------
library(rjags)
model = "tmp_model.bug"    # name of the model file ('temporary')
write("
model {
  x1 ~ dpois(lambda1)
  x2 ~ dpois(lambda2)
  lambda1 <- r1 * T1
  lambda2 <- r2 * T2
  r1 <- rho * r2
  r2  ~ dgamma(1, 1e-6)
  rho ~ dgamma(1, 1e-6)
}
", model)

#------------  JAGS call via rjags ------------------------------
data <- list(x1=x1, T1=T1, x2=x2, T2=T2)
jm <- jags.model(model, data)
update(jm, 100)
to.monitor <-  c('r1', 'r2', 'rho')
chain <- coda.samples(jm, to.monitor, n.iter=nr)

#------------  Results -----------------------------------------
print(summary(chain))
plot(chain, col='blue')

cat(sprintf("Exact: \n"))
cat(sprintf("   r1 =  %.3f +- %.3f\n", (x1+1)/T1, sqrt(x1+1)/T1))
cat(sprintf("   r2 =  %.3f +- %.3f\n", x2/T2, sqrt(x2)/T2) )
mu.rho    <-  ((x1+1)/T1)/((x2-1)/T2)
sigma.rho <- sqrt(mu.rho*(T2/T1*(x1+2)/(x2-2)-mu.rho)) 
cat(sprintf("  rho =  %.3f +- %.3f\n", mu.rho, sigma.rho))
\end{verbatim}

\end{document}